\documentclass{aa}  
\usepackage{graphicx}
\usepackage{txfonts}
\usepackage{amsmath}	
\usepackage{amssymb}	
\usepackage{color}
\usepackage{comment}
\usepackage[version=4]{mhchem}
\usepackage{url}
\begin{document} 
\title{X-ray induced chemistry of water and related molecules in low-mass protostellar envelopes}
    \author{Shota Notsu\inst{1,2}\thanks{RIKEN Special Postdoctoral Researcher (SPDR, Fellow)}
          \and
          Ewine F. van Dishoeck\inst{2,3}
          \and
          Catherine Walsh\inst{4}
          \and
          Arthur D. Bosman\inst{5}
          \and 
          Hideko Nomura\inst{6}
          }

   \institute{Star and Planet Formation Laboratory, RIKEN Cluster for Pioneering Research, 2-1 Hirosawa, Wako, Saitama 351-0198, Japan\\
              \email{shota.notsu@riken.jp}
         \and
             Leiden Observatory, Faculty of Science, Leiden University, PO Box 9513, 2300 RA Leiden, The Netherlands            
         \and
            Max-Planck-Institut f\"{u}r Extraterrestrische Physik, Giessenbachstrasse 1, 85748 Garching, Germany
          \and
            School of Physics and Astronomy, University of Leeds, Leeds, LS2 9JT, UK
          \and
            Department of Astronomy, University of Michigan, 1085 South University Avenue, Ann Arbor, MI 48109, USA
          \and
            National Astronomical Observatory of Japan, 2-21-1 Osawa, Mitaka, Tokyo 181-8588, Japan
             }

      \date{Received February 26, 2021; Accepted April 14, 2021}

 
  \abstract
   {Water is a key molecule in star and planet forming regions.
    Recent water line observations toward several low-mass protostars suggest low water gas fractional abundances ($<10^{-6}$ with respect total hydrogen density) in the inner warm envelopes ($r<10^{2}$~au).
    Water destruction by X-rays has been proposed to influence the water abundances in these regions, but the detailed chemistry, including the nature of alternative oxygen carriers, is not yet understood.
   }
   {
   We aim to understand the impact of X-rays on the composition of low-mass protostellar envelopes, focusing specifically on water and related oxygen bearing species. 
   } 
   {
   We compute the chemical composition of two proto-typical low-mass protostellar envelopes using a 1D gas-grain chemical reaction network.
   We vary X-ray luminosities of the central protostars and thus the X-ray ionisation rates in the protostellar envelopes.
    }
   {
   The protostellar X-ray luminosity has a strong effect on the water gas abundances, both within and outside the H$_{2}$O snowline ($T_{\mathrm{gas}}\sim10^{2}$~K, $r\sim10^{2}$~au). Outside, the water gas abundance increases with $L_{\mathrm{X}}$, from $\sim10^{-10}$ for low $L_{\mathrm{X}}$ to $\sim10^{-8}-10^{-7}$ at $L_{\mathrm{X}}>10^{30}$ erg s$^{-1}$.
   Inside, 
   water maintains a high abundance of $\sim 10^{-4}$ for $L_{\mathrm{X}}\lesssim10^{29}-10^{30}$ erg s$^{-1}$, with water and CO being the dominant oxygen carriers.
   For $L_{\mathrm{X}}\gtrsim10^{30}-10^{31}$ erg s$^{-1}$, the water gas abundances significantly decrease just inside the water snowline (down to $\sim10^{-8}-10^{-7}$)
   and in the innermost regions with $T_{\mathrm{gas}}\gtrsim250$~K ($\sim10^{-6}$).
   For these cases, the fractional abundances of O$_{2}$ and O gas reach $\sim 10^{-4}$ within the water snowline, and they become the dominant oxygen carriers.
   In addition, the fractional abundances of \ce{HCO+} and \ce{CH3OH}, which have been used as tracers of the water snowline, significantly increase/decrease within the water snowline, respectively, as the X-ray fluxes become larger.
   The fractional abundances of some other dominant molecules, such as \ce{CO2}, OH, \ce{CH4}, \ce{HCN}, and \ce{NH3}, are also affected by strong X-ray fields, especially within their own snowlines. 
   These X-ray effects are larger in lower density envelope models.
   }
   {X-ray induced chemistry strongly affects the abundances of water and related molecules including O, \ce{O2}, \ce{HCO+}, and \ce{CH3OH}, and can explain the observed low water gas abundances in the inner protostellar envelopes.
   In the presence of strong X-ray fields, gas-phase water molecules within the water snowline are mainly destroyed with ion-molecule reactions and X-ray induced photodissociation.
   Future observations of water and related molecules (using e.g., ALMA and ngVLA) will access the regions around protostars where such X-ray induced chemistry is effective.}


   \maketitle
%

\section{Introduction}
Water is essential for 
habitability of planets, and it is a key molecule in star and planet forming regions.
Water acts as a gas coolant (e.g., \citealt{Neufeld1995}), 
and efficient coagulation of dust grains covered by water ice is a key process in planetesimal and planet formation (e.g., \citealt{Okuzumi2012, Okuzumi2019, Wada2013, Schoonenberg2017, Arakawa2021}).
\\ \\
In diffuse and dense clouds, water gas and ice are important oxygen carriers \citep{Melnick2020, vanDishoeck2021}.
In diffuse and cold gas (gas temperature $T_{\mathrm{gas}}$$\lesssim$ 100 K), water is mainly produced by ion-molecule reactions \citep{Hollenbach2009}.
When such a cloud becomes opaque (extinction $A_{\mathrm{V}}>3$ mag) and cool ( $T_{\mathrm{gas}}$$\lesssim20-30$ K) enough,
water is also efficiently formed by hydrogenation of oxygen atoms sticking onto cold dust grain surfaces where it forms an icy mantle (e.g., \citealt{Cuppen2010}).
Water ice is a dominant oxygen carrier in dark-clouds and pre-stellar cores (e.g., \citealt{Oberg2011, Caselli2012, Marboeuf2014, Boogert2015, Taquet2016, Melnick2020, vanDishoeck2021}).
In warm regions ($T_{\mathrm{gas}}$$>$ 100 K), water ice sublimates from the dust-grain surfaces into the gas phase. 
At temperatures above 250 K, H$_{2}$O is largely produced by gas-phase reactions of O and OH with H$_{2}$ \citep{Baulch1992, Oldenborg1992}.
This high-temperature chemistry route dominates the formation of water in shocks, in the inner envelopes around protostars, and in the warm surface layers of protoplanetary disks.
\\ \\
Recently, water vapor emission from the inner warm envelopes ($T_{\mathrm{gas}}$$>$100 K) of low-mass Class 0 protostars have been investigated using PdBI\footnote[1]{IRAM Plateau de Bure Interferometer, now NOEMA (NOrthern Extended Millimeter Array)} (e.g., \citealt{Jorgensen2010, Persson2012, Persson2014, Persson2016}), ALMA\footnote[2]{Atacama Large Millimeter/submillimeter Array} (e.g., \citealt{Bjerkeli2016, Jensen2019}), and $Herschel$\footnote[3]{Herschel Space Observatory}/HIFI (e.g., \citealt{vanDishoeck2011, vanDishoeck2021, Visser2013}).
The interferometric observations using PdBI and ALMA targeted the para-H$_{2}$$^{18}$O 203 GHz 3$_{13}-2_{20}$ line (upper state energy $E_{\mathrm{up}}$=203.7 K), which is also considered to be a tracer of water emission in the inner warm regions and the position of the water snowline in Class II disks \citep{Notsu2018, Notsu2019}. The velocity-resolved observations using $Herschel$/HIFI targeted several water lines, including the 3$_{12}-3_{03}$ lines of ortho-H$_{2}$$^{16}$O (1097 GHz, $E_{\mathrm{up}}$=249.4 K) and ortho-H$_{2}$$^{18}$O (1096 GHz, $E_{\mathrm{up}}$=248.7 K), and were part of the key program ``Water in star-forming regions with Herschel" (WISH; \citealt{vanDishoeck2011, vanDishoeck2021}), which aimed to study the physics and chemistry of water during star formation across a range of masses and evolutionary stages.
The water abundances in the outer cold envelopes were also investigated, using e.g., the ground-state ortho-H$_{2}$$^{16}$O 557 GHz $1_{10}-1_{01}$ line ($E_{\mathrm{up}}$=61.0 K, e.g., \citealt{Kristensen2010, Kristensen2012, vanDishoeck2011, vanDishoeck2021, Coutens2012, Coutens2013, Mottram2013, Schmalzl2014}).
\\ \\
According to \citet{Persson2012, Persson2014, Persson2016}, and \citet{Visser2013}, the water gas abundance is around 6$\times10^{-5}$ with respect to total H$_{2}$ density in the inner warm envelope and the disk of NGC 1333-IRAS 2A, and this value is similar to the expected value ($\sim10^{-4}$) if water molecules are mostly inherited from the water ice in dark-clouds and pre-stellar cores (e.g., \citealt{Boogert2015}).
In contrast, the water gas abundances in the inner envelopes and disks of NGC 1333 IRAS 4A and 4B are lower by $1-3$ orders of magnitude than the value of NGC 1333-IRAS 2A.
While some of this decrease can be accounted for if the detailed small scale physical structure is considered,
\citet{Persson2016} even found such low water gas abundances when using thin disk$+$envelope models.
Questions on how the water abundance is changed from dense clouds to protostellar envelopes and planet-forming disks and the nature of the main oxygen carrier instead thus arise \citep{vanDishoeck2021}.
Since ALMA has much higher sensitivity and higher spatial and spectral resolution compared with previous instruments, water line surveys toward more Class 0 (and also Class I) protostars are expected using ALMA.
Recently, \citet{Jensen2019} reported ALMA detections of the para-H$_{2}$$^{18}$O 203 GHz ($3_{1,1}-2_{2,0}$) line for the inner warm envelopes around three isolated low-mass Class 0 protostars (L483, B335 and BHR71-IRS1).
The estimated H$_{2}$$^{18}$O column densities in the warm inner envelopes for the three objects are around a few $\times10^{15}$ cm$^{-2}$ in a 0.4" beam, which is similar to that of NGC 1333 IRAS 4B, and around 10 times lower than that of NGC 1333 IRAS 2A \citep{Persson2014}.
According to new observations by \citet{Harsono2020}, water vapor is not abundant in the warm envelopes and disks around Class I protostars, and the upper limits of the water gas abundances averaged over the inner warm disks with $T_{\mathrm{gas}}$$>100$ K are $\sim10^{-7}-10^{-5}$ with respect to H$_{2}$.
\\ \\
There is only limited information on other major oxygen carriers.
In low-mass protostar observations, only one upper limit and a tentative detection are reported for O$_{2}$ lines.
This is partly because O$_{2}$ does not possess electric dipole-allowed rotational transition lines.
\citet{Yildiz2013} observed the O$_{2}$ $3_{3}-1_{2}$ 487.2 GHz line ($E_{\mathrm{up}}=26.4$ K) and 
reported an upper limit O$_{2}$ gas abundance with respect to H$_{2}$ of $6\times10^{-9}$ (3$\sigma$) towards the entire envelope of IRAS 4A using $Herschel$/HIFI, 
and they estimated that the observed O$_{2}$ gas abundance cannot be more than $10^{-6}$ for the inner warm region ($\lesssim10^{2}$ au). 
\citet{Taquet2018} reported the tentative detection ($3\sigma$) of the $^{16}$O$^{18}$O 234 GHz $2_{1}-0_{1}$ line ($E_{\mathrm{up}}=11.2$ K) toward the inner envelope around a low-mass protostar IRAS 16293-2422 B with ALMA.
Assuming that the $^{16}$O$^{18}$O was not detected and using CH$_{3}$OH as a reference species, \citet{Taquet2018} obtained an O$_{2}$/CH$_{3}$OH abundance ratio $<2-5$, which is $3-4$ times lower abundance than that in comet 67P/Churyumov-Gerasimenko.
 \\ \\
The low water gas abundances derived for the inner regions of protostellar envelopes are unexpected because it is assumed that all water ice inherited from the molecular cloud phase would be sublimated in this warm region.   In tandem, observations have failed to identify sufficiently abundant alternative oxygen carriers.  So, what has happened to the water? 
\citet{Stauber2005,Stauber2006} modeled the water gas chemistry including X-ray destruction processes, and suggested that water gas will be destroyed by strong X-ray fluxes 
in the inner warm envelopes of low-mass Class 0 and I protostars on relatively short timescales ($\sim$5000 yr). 
In addition, they suggested that FUV photons from the central source are less effective in destroying water compared with X-ray photons due to extinction.
However, it is not yet understood that the nature of the major oxygen carriers under these conditions.
Moreover, 
it is important to investigate whether HCO$^{+}$ and CH$_{3}$OH are also affected by strong X-ray fluxes, since they have been used as tracers of the water snowline \citep{Visser2015, vantHoff2018, vantHoff2018CH3OH, Leemker2021}.
The chemical model that \citet{Stauber2005,Stauber2006} adopted were limited. Most notably, they did not include detailed gas-grain interactions and grain-surface chemistry (e.g., \citealt{Walsh2015}). 
These additional reactions will be important in considering the abundances of water and related molecules, since major oxygen-bearing molecules including H$_{2}$O, CO$_{2}$, and CH$_{3}$OH are efficiently formed on the grain surfaces.
\\ \\
In this study, we revisit the chemistry of water and related molecules in low-mass Class 0 protostellar envelopes, under various X-ray field strengths.
We adopt a gas-grain chemical reaction network including X-ray induced chemical processes. 
We include gas-phase reactions, thermal and non-thermal gas-grain interactions, and grain-surface reactions, simultaneously.
Through the calculations, we study the radial dependence of the abundance of water and related molecules on the strength of the X-ray field, and identify potential alternative oxygen carriers other than water.
The outline of our model calculations are explained in Section 2. The results and discussion of our calculations are described in Sections 3 and 4, respectively. The conclusions are listed in Section 5.
\section{Protostellar envelope models}
\subsection{Physical structure models}
\subsubsection{Temperature and number density profiles of Class 0 protostellar envelopes}
\begin{figure*}
\begin{center}
\includegraphics[scale=0.75]{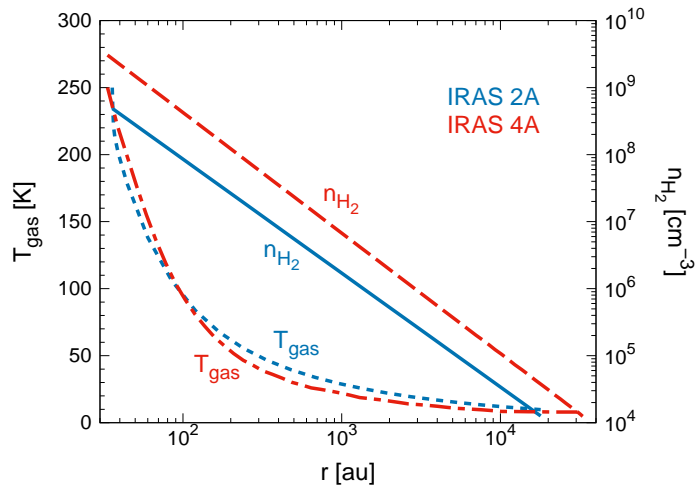}
\end{center}
\caption{
\noindent 
The radial profiles of molecular hydrogen number densities $n_{\mathrm{H}_{2}}$ [cm$^{-3}$] and 
gas temperature $T_{\mathrm{gas}}$ [K] in NGC 1333-IRAS 2A and NGC 1333-IRAS 4A envelope models.
The blue solid line and red dashed line show radial $n_{\mathrm{H}_{2}}$ profiles in IRAS 2A and IRAS 4A envelope models, respectively.
The blue dotted line and red dashed dotted line show radial $T_{\mathrm{gas}}$ profiles in IRAS 2A and IRAS 4A envelope models, respectively.
}\label{Figure1_Tn}
\end{figure*}

For the physical structures of low-mass Class 0 protostellar envelopes, we adopted the radial gas temperature $T_{\mathrm{gas}}$ and molecular hydrogen number density $n_{\mathrm{H}_{2}}$ profiles 
for two sources; NGC 1333-IRAS 2A and NGC 1333-IRAS 4A\footnote[4]{In the remainder of this paper, we define NGC 1333-IRAS 2A and NGC 1333-IRAS 4A as ``IRAS 2A'' and ``IRAS 4A'', respectively.} 
from \citet{Kristensen2012} and \citet{Mottram2013}.
These are the best studied sources with well determined inner and outer water abundances (e.g., \citealt{Persson2012, Persson2014, Persson2016, Mottram2013, Visser2013, vanDishoeck2021}).
The H$^{13}$CO$^{+}$ gas abundance (a good tracer of the water snowline) toward the envelope around IRAS 2A \citep{vantHoff2018} and an upper limit O$_{2}$ gas abundance toward the envelope around IRAS 4A \citep{Yildiz2013} have been also reported.
According to \citet{Jorgensen2007, Jorgensen2009}, 
the differences in luminosities $L_{\mathrm{bol}}$ and envelope masses $M_{\mathrm{env}}$ between these two objects are only a factor of $4-5$ ($L_{\mathrm{bol}}=20L_{\odot}$ and $M_{\mathrm{env}}=1.0M_{\odot}$ for IRAS 2A, and 5.8$L_{\odot}$ and $M_{\mathrm{env}}=4.5M_{\odot}$ for IRAS 4A). 
Thus, they are presumably in similar evolutionary stages of low-mass protostars.
In addition, we used these two profiles in order to examine the effect of density differences on X-ray induced chemistry.
%
\citet{Kristensen2012} derived these $T_{\mathrm{gas}}$ and $n_{\mathrm{H}_{2}}$ profiles using the 1D spherically symmetric dust radiative transfer code DUSTY \citep{Ivezic1997}.
In this procedure, the free model parameters (the radial profile, size, and mass) were fitted to
the spatial extent of the sub-millimeter continuum ($450-850$ $\mu$m) emission and the spectral energy distribution (SED). 
These source models are appropriate on scales of a few 10$^{2}$ - a few 10$^{3}$ au.
Several recent studies (e.g., \citealt{Persson2016, Koumpia2017, vantHoff2018}) also adopted the same models to study the chemistry and line emission in these protostellar envelopes.
In these models. the gas and dust temperatures are taken to be the same ($T_{\mathrm{gas}}=T_{\mathrm{dust}}$), and they are well mixed 
with a gas-to-dust mass ratio of 100:1.
\\ \\
Figure \ref{Figure1_Tn} shows the radial gas temperature and molecular hydrogen number density profiles 
for IRAS 2A and IRAS 4A.
The radial temperature distributions are similar between these two models ($T_{\mathrm{gas}}\sim$250 K in the innermost region and $T_{\mathrm{gas}}\sim$10 K at the outer edge).
At the same radii, the density in IRAS 4A is around $3-6$ times larger than that in IRAS 2A. The differences in densities between these two objects gradually increase as the radii decrease.
In the inner edge at $T_{\mathrm{gas}}=250$ K ($r\sim35$ au), $n_{\mathrm{H}_{2}}$ in IRAS 2A is $4.9\times10^{8}$ cm$^{-3}$ and $n_{\mathrm{H}_{2}}$ in IRAS 4A is $3.1\times10^{9}$ cm$^{-3}$.
The effects of the small scale structures such as disks are neglected, but they will lower the temperature for some fraction of the gas.  
\subsubsection{X-ray fields}
\begin{figure*}
\begin{center}
\includegraphics[scale=1.4]{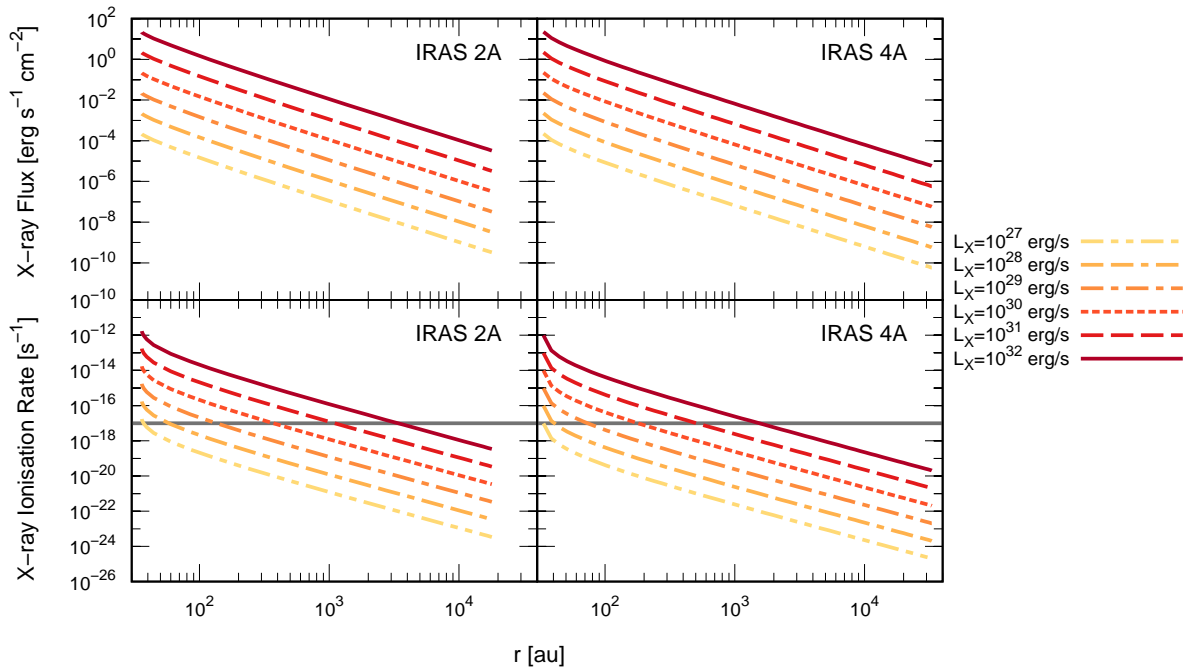}
\end{center}
\caption{
\noindent 
[Top panels] The radial profiles of the X-ray flux $F_{\mathrm{X}}(r)$ [erg s$^{-1}$ cm$^{-2}$] in NGC 1333-IRAS 2A (top left panel) and NGC 1333-IRAS 4A (top right panel) envelope models.
[Bottom panels] The radial profiles of the ``secondary'' X-ray ionization rate $\xi_{\mathrm{X}}(r)$ [s$^{-1}$] in NGC 1333-IRAS 2A (bottom left panel) and NGC 1333-IRAS 4A (bottom right panel) envelope models.
The horizontal gray solid lines show the assumed constant cosmic-ray ionisation rate $\xi_{\mathrm{CR}}(r)=$$1.0\times10^{-17}$ s$^{-1}$.
In all panels, the different line styles and colors in the radial $F_{\mathrm{X}}(r)$ and $\xi_{\mathrm{X}}(r)$ profiles denote models with different central star X-ray luminosities $L_{\mathrm{X}}$. 
%
}\label{Figure2_FX}
\end{figure*}

The observed X-ray spectra from YSOs are usually fitted with the emission spectrum of a thermal plasma (e.g., \citealt{Hofner1997,Stauber2005, Bruderer2009}).
The thermal X-ray spectrum can be approximated with
\begin{equation}
F_{\mathrm{X, in}}(E,r)=F_{0}(r)\exp(-E/kT_{\mathrm{X}}) \ \ \mathrm{[photons \ s^{-1} \ cm^{-2} \ eV^{-1}],}
\end{equation}
where $r$ is the radius in the envelope from the central protostar, $F_{\mathrm{X, in}}(E,r)$ is the incident X-ray flux per unit energy, $k$ is the Boltzmann constant, and $T_{\mathrm{X}}$ is the temperature of the X-ray emitting plasma.
The factor $F_{0}(r)$ can be calculated from the following equation,
\begin{eqnarray}
L_{\mathrm{X}}&=&4\pi r^{2}\int_{E_{\mathrm{min}}}^{E_{\mathrm{max}}} F_{\mathrm{X, in}}(E,r)E \mathrm{d}E \ \ \mathrm{[erg \ s^{-1}]} \\
&=&4\pi r^{2}\int_{E_{\mathrm{min}}}^{E_{\mathrm{max}}} F_{0}(r)\exp(-E/kT_{\mathrm{X}}) E \mathrm{d}E \ \ \mathrm{[erg \ s^{-1}]},
\end{eqnarray}
where $L_{\mathrm{X}}$ is the X-ray luminosity of the central protostar.
The local (attenuated) X-ray flux per unit energy $F_{\mathrm{X}}(E,r)$ is given by the following equation,
\begin{equation}
F_{\mathrm{X}}(E,r)=F_{\mathrm{X, in}}(E,r)\exp(-\tau(E,r)),
\end{equation}
where $\tau(E,r)$ is the total optical depth from the central protostar position to $r$. 
The energy-integrated total attenuated X-ray flux $F_{\mathrm{X}}(r)$ at radius $r$ of the envelope is given by the following equation,
\begin{equation}
F_{\mathrm{X}}(r)=\int_{E_{\mathrm{min}}}^{E_{\mathrm{max}}}F_{\mathrm{X}}(E,r)E \mathrm{d}E  \ \ \mathrm{[erg \ s^{-1} \ cm^{-2}]}.
\end{equation}
\\
$\tau(E,r)$ is determined by the following equation,
\begin{equation}
\tau(E,r)=\tau_{\mathrm{p}}(E,r)+\tau_{\mathrm{c}}(E,r),
\end{equation}
where $\tau_{\mathrm{p}}(E,r)$ and $\tau_{\mathrm{c}}(E,r)$ are the optical depths determined by photoabsorption and incoherent Compton scattering of hydrogen \citep{Nomura2007}.
We note that the attenuation of the X-rays is mainly determined by photoabsorption especially at $E<10$ keV, 
and the influence of Compton scattering of hydrogen on the chemistry is negligible \citep{Stauber2005, Bruderer2009}. 
\\ \\
Assuming that the photoabsorption cross section of an atom is equal to its photoionization cross section, $\tau_{\mathrm{p}}(E,r)$ is obtained by the following equation, 
\begin{equation}
\tau_{\mathrm{p}}(E,r)=N_{\mathrm{H}}(r)\sigma_{\mathrm{tot,p}}(E)=N_{\mathrm{H}}(r)\sum_{i} x(i)\sigma_{i\mathrm{,p}}(E),
\end{equation}
where $N_{\mathrm{H}}$(r) is the total hydrogen column density from the central protostar position to $r$, $\sigma_{\mathrm{tot,p}}(E)$ is the total photoabsorption cross section given by the sum of the photoionization cross sections for each element $\sigma_{i\mathrm{,p}}(E)$ multiplied by its fractional abundance $x(i)$.
We calculate the values of $\sigma_{i\mathrm{,p}}(E)$ using the analytical method in \citet{Verner1993}, as done in \citet{Walsh2012}.
$\tau_{\mathrm{c}}(E,r)$ is obtained by the following equation,
\begin{equation}
\tau_{\mathrm{c}}(E,r)=N_{\mathrm{H}}(r)\sigma_{\mathrm{c}}(E),
\end{equation}
where $\sigma_{\mathrm{c}}(E)$ is the incoherent Compton scattering cross section of hydrogen. 
We have adopted the values of $\sigma_{\mathrm{c}}(E)$ from the NIST/XCOM database \citep{Berger1999}.
\\ \\
In Class I and II protostars, the values of observed X-ray luminosities are typically around $L_{\mathrm{X}}$
$\sim10^{28}-10^{31}$ $\mathrm{erg \ s^{-1}}$
\citep{Imanishi2001, Preibisch2005, Gudel2009}.
However, the values of $L_{\mathrm{X}}$ in low-mass Class 0 protostars have not yet been well determined (e.g., \citealt{Hamaguchi2005, Forbrich2006, Giardino2007, Gudel2009, Kamezaki2014, Grosso2020}), since the X-rays from the central Class 0 protostars are absorbed by their surrounding dense envelopes.
Recently, \citet{Grosso2020} reported a powerful X-ray flare from the Class 0 protostar HOPS 383 with $L_{\mathrm{X}}\sim4\times10^{31}$ erg s$^{-1}$ in the $2-8$ keV energy band.
\citet{Takasao2019} discussed from their simulations that protostar X-ray flares occur repeatedly (e.g., once in around 10 days) even in Class 0 protostars without magnetospheres.
These flares are thought to occur when a portion of the large-scale magnetic fields, which are transported by accretion, are removed from the protostar as a result of magnetic reconnection.
\citet{Stauber2007} discussed the X-ray strengths from CN, CO$^{+}$ and SO$^{+}$ abundances, and they estimated that values of $L_{\mathrm{X}}$ in Class 0 low-mass protostars are around $10^{29}-10^{32}$ $\mathrm{erg \ s^{-1}}$, which are comparable to those in low-mass Class I protostars.
However, \citet{Benz2016} discussed that the abundances of CN and CO$^{+}$ obtained by $Herschel$/HIFI observations can also be explained by FUV irradiation of outflow cavity walls (see also \citealt{Bruderer2010}), and suggested that
the spatial resolution at scales of a few $\times10^{3}$ au is not sufficient to detect molecular tracers of X-rays.
\citet{Benz2016} also estimated the X-ray luminosities from the upper limits of H$_{3}$O$^{+}$ line fluxes obtained with $Herschel$/HIFI towards some low-mass protostars ($L_{\mathrm{X}}<10^{30}$ erg s$^{-1}$ in the Class 0 object IRAS16293-2422 and $L_{\mathrm{X}}\gtrsim10^{31}$ erg s$^{-1}$ in the Class I object TMC1).
\\ \\
In order to investigate the dependence of the chemical evolution on the strength of the X-ray field, we take values of $L_{\mathrm{X}}=$\ 0, $10^{27}$,  $10^{28}$, $10^{29}$,  $10^{30}$, $10^{31}$, and $10^{32}$ $\mathrm{erg \ s^{-1}}$.
We adopt $kT_{\mathrm{X}}=$\ 2.6 keV ($=3\times10^{7}$ K), which is similar to \citet{Stauber2006}, and is also consistent with typical Class I protostars \citep{Imanishi2001, Preibisch2005}.
We set $E_{\mathrm{min}}=0.1$ keV and 
$E_{\mathrm{max}}=100$ keV 
to cover a sufficient range of X-rays in our calculations.
According to \citet{Maloney1996}, \citet{Stauber2005}, and \citet{Bruderer2009},
the shape of the X-ray spectrum will vary for different values of $kT_{\mathrm{X}}$, with e.g., $10^{7}-10^{8}$ K.  
However, they discussed that the calculated abundances differ only a factor of a few at most for the different X-ray temperatures, and that the influence of the X-ray luminosities on the chemistry is dominant.
Note that we assume a constant value of X-ray luminosity during $10^{5}$ year, since protostellar X-ray flares repeatedly occur and as a first step we would like to know the overall influence of X-ray fields on chemistry (see also Section 4.7).
\\ \\
In our calculations, the FUV radiation field from the central protostar is neglected.
According to \citet{Stauber2007}, X-rays are suggested to be more effective for chemistry than FUV fields in the low-mass protostellar envelopes.
Low-mass protostars ($L_{\mathrm{bol}}\sim10^{1-2}L_{\odot}$, $T_{\mathrm{eff}}<10^{4}$ K) emit much less UV photons than high-mass protostars ($L_{\mathrm{bol}}\sim10^{4-5}L_{\odot}$, $T_{\mathrm{eff}}\gtrsim$ a few $\times 10^{4}$ K) due to their lower surface temperatures.
Thus, FUV photons from the central source are not effective in destroying molecules in Class 0 protostellar envelopes \citep{Stauber2005, Stauber2006}.
Some FUV radiation from the disk-star boundary can escape through outflow cavities, but only affects a narrow layer along the cavity walls \citep{Visser2012}.
\\ \\
The top panels of Figure \ref{Figure2_FX} show the radial profiles of $F_{\mathrm{X}}(r)$ in the IRAS 2A and IRAS 4A envelope models.
In both models, the values of $F_{\mathrm{X}}(r)$ in the innermost region are around $2\times10^{-4}$ erg s$^{-1}$ cm$^{-2}$ in the case of $L_{\mathrm{X}}=10^{27}$ $\mathrm{erg \ s^{-1}}$, and around 
20 erg s$^{-1}$ cm$^{-2}$ 
in the case of $L_{\mathrm{X}}=10^{32}$ $\mathrm{erg \ s^{-1}}$.
In the outer envelopes, the values of $F_{\mathrm{X}}(r)$ reduce because of the increasing values of $N_{\mathrm{H}}(r)$.
Compared with the IRAS 2A model, $F_{\mathrm{X}}(r)$ of the IRAS 4A model is lower in the outer regions due to higher densities (see also Figure \ref{Figure1_Tn}).
The values of $F_{\mathrm{X}}(r)$ at $r\sim10^{3}$ au are $\sim1\times10^{-7}$ erg s$^{-1}$ cm$^{-2}$ (IRAS 2A) and $\sim6\times10^{-8}$ erg s$^{-1}$ cm$^{-2}$ (IRAS 4A) in the case of $L_{\mathrm{X}}=10^{27}$ $\mathrm{erg \ s^{-1}}$, and $\sim1\times10^{-2}$ erg s$^{-1}$ cm$^{-2}$ (IRAS 2A) and $\sim6\times10^{-3}$ erg s$^{-1}$ cm$^{-2}$ (IRAS 4A) in the case of $L_{\mathrm{X}}=10^{32}$ $\mathrm{erg \ s^{-1}}$.
%
%
%
\subsection{Calculations of chemical evolution}
We calculate the chemical evolution of low-mass Class 0 protostellar envelopes using a detailed gas-grain chemical reaction network including X-ray induced chemical processes \citep{Walsh2012, Walsh2015}.
Note that \citet{Stauber2005, Stauber2006} focused on gas-phase water chemistry only.
In order to investigate the radial dependence of the abundances of both gas and ice molecules on X-ray fields, we include gas-phase reactions, thermal and non-thermal gas-grain interactions, and grain-surface reactions, simultaneously.
\\ \\
The chemical network adopted in this work is based on the chemical model from \citet{Walsh2015}, as also used in \citet{Eistrup2016, Eistrup2018}, and \citet{Bosman2018}.
The detailed background theories and procedures are also discussed in these papers and our previous works (e.g., \citealt{Walsh2010, Walsh2012, Walsh2014, Walsh2014F}, \citealt{Heinzeller2011}, \citealt{Notsu2016, Notsu2017, Notsu2018}), although there are some differences between those studies and our paper.
Here we provide a summary and describe important update of our adopted chemical network in this paper.
Consistent with \citet{Stauber2006}, the chemical evolution in envelopes is run for $10^{5}$ years, which is the typical age of Class 0 protostars.
\subsubsection{Gas-phase reactions}
Our gas-phase chemistry is the complete network from the recent release of the UMIST Database for Astrochemistry (UDfA), termed ``RATE12'' which is publicly available\footnote[5]{\url{http://udfa.ajmarkwick.net}} \citep{McElroy2013}.
RATE12 includes gas-phase two-body reactions, photodissociation and photoionisation, direct cosmic-ray ionisation, and cosmic-ray-induced photodissociation and photoionisation.
Since the FUV radiation fields from the central protostar is neglected in our calculations (see also Section 2.1.2), the photodissociation and photoionisation by FUV radiation is not included.
In contrast, we have supplemented this gas-phase network with direct X-ray ionisation reactions, and X-ray-induced photoionisation and photodissociation processes (see \citealt{Walsh2012, Walsh2015}, and Section 2.2.4 in this paper).
In these X-ray induced photoreaction, UV photons are generated internally via the interaction of secondary electrons (produced by X-rays, see also Section 2.2.4) with H$_{2}$ molecules \citep{Gredel1987, Gredel1989}. 
As in \citet{Walsh2015}, we also add a set of three-body reactions and ``Hot" H$_{2}$ chemistry, although they are not expected to be important at the densities and temperatures calculated in this study.
Moreover, the gas phase chemical network is supplemented with reactions for important species, for example the CH$_{3}$O radical, that are not included in RATE12. 
The gas-phase formation and destruction reactions for these species are taken from the Ohio State University (OSU) network \citep{Garrod2008}.
\subsubsection{Gas-grain interactions}
\begin{table}
\caption{UV photodesorption yields}              
\label{Table:1}      
\centering                                      
\begin{tabular}{c c c}          
\hline\hline                        
Species $j$ & $Y_{\mathrm{des}}$($j$)& References\\    
&[molecules photon$^{-1}$]&\\
\hline                                   
    H$_{2}$O & $^*$$1.300\times10^{-3}$ & $a$\\
    CO & $2.700\times10^{-3}$ & $b, c$ \\ 
    CO$_{2}$ & $2.300\times10^{-3}$ & $c, d$\\
    N$_{2}$ & $1.800\times10^{-4}$ & $b, c$ \\
    CH$_{3}$OH & $^*$$2.475\times10^{-4}$ & $e$ \\
    All other species & $1.000\times10^{-3}$ & $f, g$ \\
\hline                                             
\end{tabular}
\tablebib{$^{a}$\citet{Oberg2009b}; $^{b}$\citet{Oberg2007}; $^{c}$\citet{Oberg2009a}; $^{d}$\citet{Fillion2014}; $^{e}$\citet{Bertin2016}; $^{f}$\citet{Walsh2015}; $^{g}$\citet{Cuppen2017};
\\
$^*$ For H$_{2}$O and CH$_{3}$OH, the values of $Y_{\mathrm{des}}$($j$) are the sum of all of their fragmentation pathways (see texts in Section 2.2.2).
}
\end{table}
In our calculations, we consider the freezeout of gas-phase molecules on dust grains, and the thermal and non-thermal desorption of molecules from dust grains \citep{Hasegawa1992,Walsh2010,Walsh2012,Walsh2014,Walsh2015,Notsu2016}.
The adopted non-thermal desorption mechanisms are cosmic-ray-induced (thermal) desorption \citep{Leger1985, Hasegawa1993, Hollenbach2009}, reactive desorption (see Section 2.2.3), and photodesorption.
We note that the direct cosmic-ray-induced desorption have no significant impact on chemistry, since its reaction timescale is typically much longer ($>>10^{7}$ years) than the age of protostars \citep{Hollenbach2009}. 
\\ \\
We include photodesorption by both external X-ray photons and UV photons generated internally via the interaction of secondary electrons produced by cosmic-rays with H$_{2}$ molecules.
Following \citet{Walsh2015}, we assume compact spherical grains with a radius $a$ of 0.1 $\mu$m and a fixed density of $\sim10^{-12}$ relative to the gas number density. 
We adopt a value for the integrated cosmic-ray-induced UV photon flux as $10^{4}$ photons cm$^{-2}$ s$^{-1}$ \citep{Prasad1983, Walsh2014}.
We scale the internal UV photon flux by the cosmic-ray ionisation rate.
\\ \\
We use experimentally determined photodesorption yields, $Y_{\mathrm{des}}$($j$), where available (e.g., \citealt{Oberg2007, Oberg2009a, Oberg2009b, Bertin2016, Cuppen2017}).
Such experiments were conducted by using UV lamps that mimic well the FUV radiation field (such as 100$-$200 nm) produced locally by H$_{2}$ emission excited by cosmic-rays or X-rays.
For all species without experimentally determined photodesorption yields, a value of $10^{-3}$ molecules photon$^{-1}$ is used.
The values of photodesorption yields adopted in our work are the same as those in \citet{Walsh2015}, except the value of CH$_{3}$OH.
Recent studies into methanol ice photodesorption showed that methanol does not desorb intact at low temperatures (e.g., \citealt{Bertin2016, Cruz-Diaz2016}), and the value of intact photodesorption yield for CH$_{3}$OH are considered to be much lower ($\sim$$10^{-6}-10^{-5}$) than that in the previous estimates ($\sim$$10^{-3}$, \citealt{Oberg2009c}).
The values of photodesorption yields adopted in this work, $Y_{\mathrm{des}}$($j$), are listed in Table \ref{Table:1}.
On the basis of \citet{Oberg2009b}, \citet{Arasa2010, Arasa2015}, \citet{Bertin2016}, \citet{Cruz-Diaz2016}, and \citet{Walsh2018}, we include the fragmentation pathways for water ice (50\% H$_{2}$O and 50\% OH$+$H) and methanol photodesorption (e.g., 85.0\% CO+H$_{2}$+H$_{2}$, 6.1\% CH$_{3}$OH, 4.85\% H$_{2}$CO$+$H$_{2}$, 3.0\% CH$_{3}$+OH).
The values of $Y_{\mathrm{des}}$($j$) for H$_{2}$O and CH$_{3}$OH listed in Table \ref{Table:1} are the sum of all of these fragmentation pathways.
The adopted value of intact photodesorption yield for CH$_{3}$OH is 1.5$\times10^{-5}$ [molecules photon$^{-1}$] and 6.1\% of $Y_{\mathrm{des}}$($j$) for CH$_{3}$OH.
\\ \\
As in \citet{Walsh2014}, we treat X-ray induced photodesorption as we treat UV photodesorption, and assume the same photodesorption yields for both X-ray induced photodesorption and UV photodesorption.
In addition, following \citet{Walsh2014}, we do not include the photodesorption by UV photons generated internally via the interaction of secondary electrons produced by X-rays with H$_{2}$ molecules.
This is because experimental constraints for X-ray induced photodesorption are limited, and the interaction of X-ray photons with ice is still not well understood (for more details, see e.g., \citealt{Andrade2010, Walsh2014}).
In Sections 4.2 and 4.3, we discuss the rates of X-ray induced photodesorption in detail, with conducting additional test calculations.
We note that we also allow X-rays to photodissociate grain mantle material (see also Section 2.2.3 and \citealt{Walsh2014}), in which UV photons are generated internally via the interaction of secondary electrons (produced by X-rays, see Section 2.2.4) with H$_{2}$.
Recently, \citet{Dupuy2018} and \citet{Basalgete2021a, Basalgete2021b} experimentally investigated X-ray induced photodesorption rates of H$_{2}$O, O$_{2}$, CH$_{3}$OH, and other related molecules (for more details, see Section 4.3).
\\ \\
The sticking coefficient is assumed to be 1 for all species, except for H that leads to H$_{2}$ formation (for more details, see Appendix B.2 of \citealt{Bosman2018}).
Compared with \citet{Walsh2015}, the values of molecular binding energies, $E_{\mathrm{des}}$($j$), are updated on the basis of the recent extensive literature review performed by \citet{Penteado2017} and grain-surface chemistry review by \citet{Cuppen2017}.
The values of binding energies for several important molecules, $E_{\mathrm{des}}$($j$), are listed in Table \ref{Table:2}.
%
%
\subsubsection{Grain-surface reactions}
For the grain-surface reactions, we use the reactions included in the Ohio State University (OSU) network \citep{Garrod2008}.
In addition to grain-surface two-body reactions and reactive desorptions, grain-surface cosmic-ray-induced and X-ray-induced photodissociations are also included in our calculations \citep{Garrod2008, Walsh2014, Walsh2015}. 
In these X-ray induced photodissociation reactions, UV photons are generated internally via the interaction of secondary electrons (produced by X-rays, see also Section 2.2.4) with H$_{2}$ molecules \citep{Gredel1987, Gredel1989}.
In addition, as \citet{Walsh2018} adopted, we include an extended grain-surface chemistry network for methanol and its related compounds from \citet{Woods2013} and \citet{Chuang2016}.
Moreover, we have also added the hydrogenation abstraction pathway during hydrogenation from HNCO to NH$_{2}$CHO \citep{Noble2015}. 
As in \citet{Walsh2015} and \citet{Bosman2018}, the additional water formation routes studied by \citet{Cuppen2010} and \citet{Lamberts2013} are also included.
The grain-surface two-body reaction rates are calculated assuming the Langmuir-Hinshelwood mechanism only, and using the rate equation method as described in \citet{Hasegawa1992}.
Only the top two monolayers of the ice mantle are chemically ``active".
We assume that the size of the barrier to surface diffusion is $0.3\times$$E_{\mathrm{des}}$($j$) \citep{Walsh2015}.
For the lightest reactants, H and H$_{2}$, we adopt either the classical diffusion rate or the quantum tunnelling rate depending on which is fastest \citep{Hasegawa1992,Bosman2018}.
For the latter quantum tunnelling rates, we adopt a rectangular barrier of width 1.0 $\AA$ \citep{Hasegawa1992, Bosman2018}.
As in \citet{Bosman2018}, reaction-diffusion competition for grain-surface reactions with a reaction barrier \citep{Garrod2011} is not included.
\\ \\
We note that grain-surface reactions take place on finite grain surfaces, where the populations of certain chemical species can become very small, i.e., $<<$1. 
If surface reactions occur very quickly in such a regime (the stochastic limit situation), the reaction rates might be overestimated compared with the actual values \citep{Garrod2008b, Garrod2011, Cuppen2017}.
Such stochastic effects would be more important on the smaller dust grains (such as $a\lesssim0.1$ $\mu$m), since the number of surface sites per grain is smaller \citep{Barzel2007, Garrod2008b}.
\citet{Stantcheva2004} and \citet{Vasyunin2009} showed that the stochastic effects are most important on chemical evolution in moderately warm regions ($T_{\mathrm{dust}}\sim30$ K), and that the abundances of molecules such as H$_{2}$O and CO$_{2}$ can differ by more than an order of magnitude.
In contrast, they also showed that such effects are not important in the regions with low ($T_{\mathrm{dust}}\lesssim10$ K) and high ($T_{\mathrm{dust}}\gtrsim50$ K) temperatures (see also \citealt{Caselli1998}).
Comparing with the physical structures shown in Figure \ref{Figure1_Tn}, the molecular abundances just outside the water snowline ($r\sim$(1-a few)$\times10^{2}$ au) will not be strongly influenced by such effects.
In addition, sizes of dust grains in protostellar envelopes are on average larger than 0.1 $\mu$m \citep{Ormel2009, Miotello2014, Li2017}, and thus the effects would be smaller than those in diffuse clouds.
The micro- and macroscopic Monte Carlo techniques would be helpful for much more precise treatment of the grain-surface chemistry (e.g., \citealt{Tielens1982, Vasyunin2009, Vasyunin2013, Garrod2009, Cuppen2017}).
\subsubsection{X-ray ionisation rates}
We include a set of gas-phase and grain-surface X-ray-induced reactions which we duplicate from the existing set of cosmic-ray-induced reactions contained in RATE12 \citep{McElroy2013, Walsh2015}.
The reaction rates are estimated by scaling the cosmic-ray-induced reaction rates by the ratio of the local X-ray ionization rate $\xi_{\mathrm{X}}(r)$ and cosmic-ray ionisation rate $\xi_{\mathrm{CR}}(r)$.
\\ \\
In this study, we calculate the ``secondary'' X-ray ionization rate at each radius $\xi_{\mathrm{X}}(r)$ by the following equation (see also \citealt{Glassgold1997,Walsh2012}),
\begin{equation}
\xi_{\mathrm{X}}(r)=\sum_{i} \int_{E_{i}}^{E_{\mathrm{max}}} x(i)\sigma_{i\mathrm{,p}}(E)F_{\mathrm{X}}(E,r)\left[\frac{E-E_{i}}{\Delta\epsilon}\right] \mathrm{d}E \ \ \mathrm{[s^{-1}]},
\end{equation}
where $E_{i}$ is the ionization potential for each element $i$.
$x(i)$, $\sigma_{i\mathrm{,p}}(E)$, and $F_{\mathrm{X}}(E,r)$ are determined as described in Section 2.1.2.
The number of secondary ionizations per unit energy produced by primary photoelectrons is given by the expression $(E-E_{i})/\Delta\epsilon$, where $\Delta\epsilon=37$ eV is the mean energy required to make an ion pair.
X-rays interact only with atoms, regardless of whether an atom is bound within a molecule or free \citep{Glassgold1997}.
According to \citet{Maloney1996}, these ``secondary''  ionization rates $\xi_{\mathrm{X}}(r)$ dominate the total ionization rates in X-ray dissociation regions.
For atoms heavier than Li, inner-shell ionization is followed by the Auger effect, in which the excited, photo-produced ion undergoes two- or even three-electron decay \citep{Glassgold1997}.
Our calculations do not include the Auger effect. According to \citet{Igea1999} and \citet{Stauber2005}, Auger electrons, as well the primary photoelectron, are negligible compared to the secondary electrons for the ionization of the gas.
\\ \\
In this study, we adopt a constant value for the cosmic-ray ionisation rate of $\xi_{\mathrm{CR}}(r)=$$1.0\times10^{-17}$ s$^{-1}$ at all radii \citep{Umebayashi2009}.
The bottom panels of Figure \ref{Figure2_FX} show the radial profiles of the X-ray ionisation rate $\xi_{\mathrm{X}}(r)$ in the IRAS 2A and IRAS 4A envelope models.
In both models, the values of $\xi_{\mathrm{X}}(r)$ in the innermost region are around $10^{-17}$ s$^{-1}$ in the case of $L_{\mathrm{X}}=10^{27}$ $\mathrm{erg \ s^{-1}}$, and around $10^{-12}$ s$^{-1}$ in the case of $L_{\mathrm{X}}=10^{32}$ $\mathrm{erg \ s^{-1}}$.
In the outer envelopes, the values of $\xi_{\mathrm{X}}(r)$ are reduced because of increasing values of $N_{\mathrm{H}}(r)$.
Compared with the IRAS 2A model, $\xi_{\mathrm{X}}(r)$ of the IRAS 4A model is lower in the outer regions due to higher densities (see also Figure \ref{Figure1_Tn}).
The values of $\xi_{\mathrm{X}}(r)$ at $r\sim10^{3}$ au are $\sim10^{-21}$ s$^{-1}$ (IRAS 2A) and $\sim10^{-22}$ s$^{-1}$ (IRAS 4A) in the case of $L_{\mathrm{X}}=10^{27}$ $\mathrm{erg \ s^{-1}}$, and $\sim10^{-16}$ s$^{-1}$ (IRAS 2A) and $\sim10^{-17}$ s$^{-1}$ (IRAS 4A) in the case of $L_{\mathrm{X}}=10^{32}$ $\mathrm{erg \ s^{-1}}$.
In regions with $\xi_{\mathrm{X}}(r)$$>$$\xi_{\mathrm{CR}}(r)$ ($=1.0\times10^{-17}$ s$^{-1}$), X-ray induced photoionisation and photodissociation processes are considered to be dominant compared with cosmic-ray induced photoionisation and photodissociation processes.
In the cases of $L_{\mathrm{X}}\gtrsim10^{31}$ $\mathrm{erg \ s^{-1}}$, the values of $\xi_{\mathrm{X}}(r)$ are larger than that of $\xi_{\mathrm{CR}}(r)$ at $r\lesssim10^{3}$ au in the IRAS 2A model and $r\lesssim5\times10^{2}$ au in the IRAS 4A model.
Inside the water snowline ($r<10^{2}$ au), the values of $\xi_{\mathrm{X}}(r)$ are larger than that of $\xi_{\mathrm{CR}}(r)$ in the cases of $L_{\mathrm{X}}>10^{29}$ $\mathrm{erg \ s^{-1}}$ for IRAS 2A and $L_{\mathrm{X}}>10^{30}$ $\mathrm{erg \ s^{-1}}$ for IRAS 4A.
\\ \\
We have also included the direct (``primary'') X-ray ionization of elements. The reaction rate $\xi_{\mathrm{PX,} i}(r)$ for each element $i$ is given by the following equation \citep{Verner1993, Walsh2012},
\begin{equation}
\xi_{\mathrm{PX,} i}(r)=\int_{E_{i}}^{E_{\mathrm{max}}}\sigma_{i\mathrm{,p}}(E)F_{\mathrm{X}}(E,r) \mathrm{d}E \ \ \mathrm{[s^{-1}]}.
\end{equation}
\subsubsection{Initial abundances}
\begin{table}
\caption{Initial abundances for dominant molecules in our protostellar envelope models and their binding energies}              
\label{Table:2}      
\centering                                      
\begin{tabular}{c c c c}          
\hline\hline                        
Species $j$ & $n_{j, \mathrm{gas}}$/$n_{\mathrm{H}}$ &  $n_{j, \mathrm{ice}}$/$n_{\mathrm{H}}$ & $E_{\mathrm{des}}$($j$) [K]\\    
\hline                                   
    H &$3.807\times10^{-5}$&$4.458\times10^{-17}$&650\tablefootmark{a}\\
    H$_{2}$ &$4.997\times10^{-1}$&$4.140\times10^{-5}$&430\tablefootmark{b}\\
    H$_{2}$O &$7.080\times10^{-7}$&$1.984\times10^{-4}$&4880\tablefootmark{c}\\      
    O &0.0& $2.073\times10^{-13}$&1660\tablefootmark{d}\\
    O$_{2}$ &0.0&$4.035\times10^{-12}$&898\tablefootmark{e}\\
    OH & $5.164\times10^{-8}$&$6.019\times10^{-14}$&3210\tablefootmark{d}\\   
    C &$2.571\times10^{-8}$&$1.310\times10^{-16}$&715\tablefootmark{f}\\
    CO &$7.532\times10^{-5}$&$2.946\times10^{-5}$&855\tablefootmark{g}\\
    CO$_{2}$ &$7.487\times10^{-7}$&$2.856\times10^{-7}$&2267\tablefootmark{e}\\
    HCO$^{+}$ &$3.553\times10^{-9}$& --- &---\\   
    CH$_{4}$ &$1.120\times10^{-6}$&$7.384\times10^{-6}$&1252\tablefootmark{h}\\
    CH$_{3}$OH &$3.558\times10^{-9}$&$6.027\times10^{-7}$&3820\tablefootmark{i}\\            
    H$_{2}$CO &$1.108\times10^{-7}$&$8.437\times10^{-6}$&3260\tablefootmark{e}\\   
    C$_{2}$H &$1.776\times10^{-10}$&$5.537\times10^{-17}$&1330\tablefootmark{f}\\   
    C$_{2}$H$_{2}$ &$7.440\times10^{-8}$&$3.291\times10^{-10}$&2090\tablefootmark{i}\\   
    N &$2.105\times10^{-5}$&$5.531\times10^{-14}$&715\tablefootmark{f}\\   
    N$_{2}$ &$9.765\times10^{-6}$&$5.411\times10^{-6}$&790\tablefootmark{g}\\
    NH$_{3}$ &$2.933\times10^{-7}$&$1.327\times10^{-5}$&2715\tablefootmark{i}\\
    CN &$3.016\times10^{-9}$&$1.406\times10^{-15}$&1355\tablefootmark{f}\\
    HCN &$7.718\times10^{-8}$&$2.772\times10^{-6}$&3610\tablefootmark{f}\\
\hline                                             
\end{tabular}
\tablebib{$^{a}$\citet{Al-Halabi2007}; $^{b}$\citet{Acharyya2014}; $^{c}$\citet{Dulieu2013}; $^{d}$\citet{He2014}, \citet{He&Vidali2014}; 
$^{e}$\citet{Noble2012}; $^{f}$Average between \citet{Hasegawa1993} and \citet{Aikawa1996} values; $^{g}$\citet{Oberg2005}; $^{h}$\citet{Smith2016}; 
$^{i}$Estimated from \citet{Colling2004}
}
\end{table}
To generate a set of initial abundances for input into protostellar envelope models, we run a dark cloud model ($T_{\mathrm{gas}}=T_{\mathrm{dust}}=10$ K, $n_{\mathrm{H}_{2}}=10^{4}$ cm$^{-3}$, $\xi_{\mathrm{CR}}(r)=1.0\times10^{-17}$ s$^{-1}$).
As \citet{Walsh2015} adopted, 
the values of volatile elemental abundances for O, C, and N are $3.2\times10^{-4}$, $1.4\times10^{-4}$, and $7.5\times10^{-5}$ relative to total hydrogen nuclei density, respectively.
These values are based on diffuse cloud observations \citep{Cardelli1991, Cardelli1996, Meyer1998}.
For other elements, we use the low-metal elemental abundances from \citet{Graedel1982}.
In this way, we begin the envelope calculations with an ice reservoir on the grain mantle built up in the dark-cloud and pre-stellar core phases.
We use initial abundances at a time of 3.2$\times10^{5}$ years on the basis of \citet{Walsh2015} and \citet{Drozdovskaya2016}, except for the values of O gas, O$_{2}$ gas, and H$_{2}$O ice, which allow we treat as free parameters in our study but such that elemental oxygen abundance is preserved of $3.2\times10^{-4}$.
This time scale of 3.2$\times10^{5}$ years is consistent with the observed pre-stellar core lifetime of
$\sim(2-5)\times10^{5}$
years \citep{Enoch2008}.
\\ \\
In the above calculation under the dark cloud condition, the abundances with respect to total hydrogen nuclei density of O gas, O$_{2}$ gas, and H$_{2}$O ice at a time of 3.2$\times10^{5}$ years are 
$8.5\times10^{-5}$, $2.2\times10^{-6}$, and $1.1\times10^{-4}$.
If we consider longer time evolution ($\gtrsim10^{6}$ years), however, the abundances of O gas and O$_{2}$ gas become much smaller ($<<10^{-6}$, see also \citealt{Yildiz2013, Taquet2018})
and the abundance of H$_{2}$O ice becomes larger ($\sim2\times10^{-4}$, see also \citealt{Schmalzl2014}).
\\ \\
Previous chemical calculations (e.g., \citealt{Walsh2015, Eistrup2016, Eistrup2018, Drozdovskaya2016}) adopted a similarly high abundance for H$_{2}$O ice
($\sim(1-3)\times10^{-4}$), and low or zero abundances for O and O$_{2}$ gas as initial conditions. 
Thus here we assume that all oxygen atoms in these three species are incorporated into H$_{2}$O ice ($=1.984\times10^{-4}$).
\\ \\
Observations show that H$_{2}$O ice is indeed a major oxygen carrier in dark clouds and pre-stellar cores, although measured water ice
abundances are consistently a factor of 2--4 below the expected value of $2\times10^{-4}$ if all volatile oxygen that is not contained in CO
is in water ice \citep{Oberg2011,Boogert2015}.  
Chemical modeling of \citet{Schmalzl2014} and \citet{Furuya2016} show that the water ice
abundance in pre-stellar cores increases with pre-collapse time (see
also \citealt{vanDishoeck2021}), and that such a low water ice
abundance can only be obtained for a short pre-stellar period.  At
pre-collapse times of $t_{\mathrm{pre}}<10^{6}$ years, a considerable
amount of oxygen is also found in other oxygen bearing species (mainly
O besides CO).  At $t_{\mathrm{pre}}\gtrsim10^{6}$ years, oxygen
returns into the water network and water ice then becomes dominant
oxygen reservoir (up to $\sim2\times10^{-4}$) with CO.
\\ \\
The observed low water ice abundances with respect to hydrogen nuclei of
low-mass protostellar envelopes of $\sim(3-8)\times10^{-5}$ would require short
pre-collapse lifetimes of $t_{\mathrm{pre}}\lesssim10^{5}$ years \citep{Schmalzl2014}, less
than the observed pre-stellar core lifetimes of
$\sim(2-5)\times10^{5}$ years \citep{Enoch2008}.
In addition, this shorter pre-collapse phase is inconsistent with the discussions in
\citet{Yildiz2013} who argued for a long pre-collapse phase of at
least $10^{6}$ years to explain the lower upper limit of gas-phase cold
O$_{2}$ abundances ($<<10^{-6}$) towards IRAS 4A (see also
\citealt{Taquet2018}). 
Possible mitigations of this conundrum include
the possibility that a fraction of water ice is locked up in larger
micron-sized grains that do not contribute to the infrared water ice
bands, or the presence of some amount of ``Unidentified Depleted
Oxygen (UDO)'' which has also been invoked to explain the oxygen budget
in diffuse clouds \citep{Whittet2010, Schmalzl2014, vanDishoeck2021}. 
Here we do not consider either of these two options.
\\ \\
The fractional abundances with respect to total hydrogen nuclei density for dominant and important molecules, which are used as initial abundances in our protostellar envelope models, are listed in Table \ref{Table:2}.
\section{Results}
\subsection{Water fractional abundances}
\begin{figure*}
\begin{center}
\includegraphics[scale=0.67]{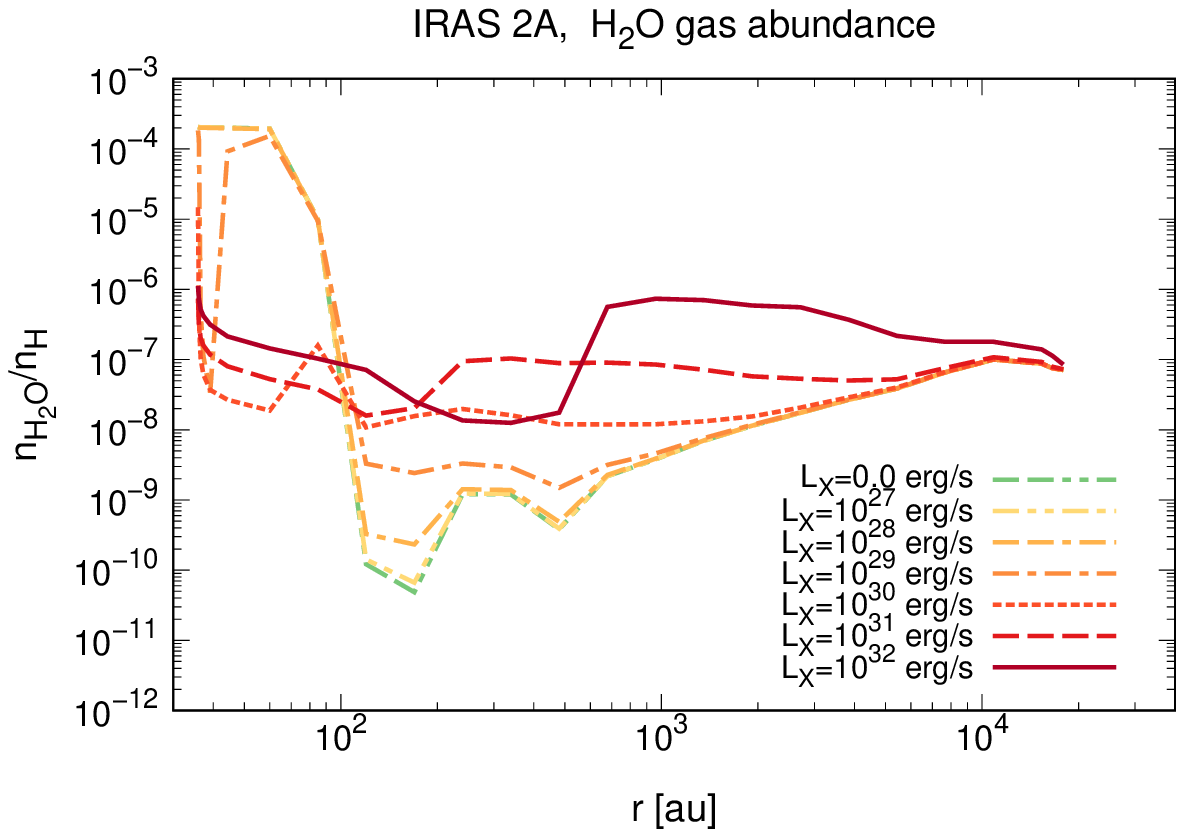}
\includegraphics[scale=0.67]{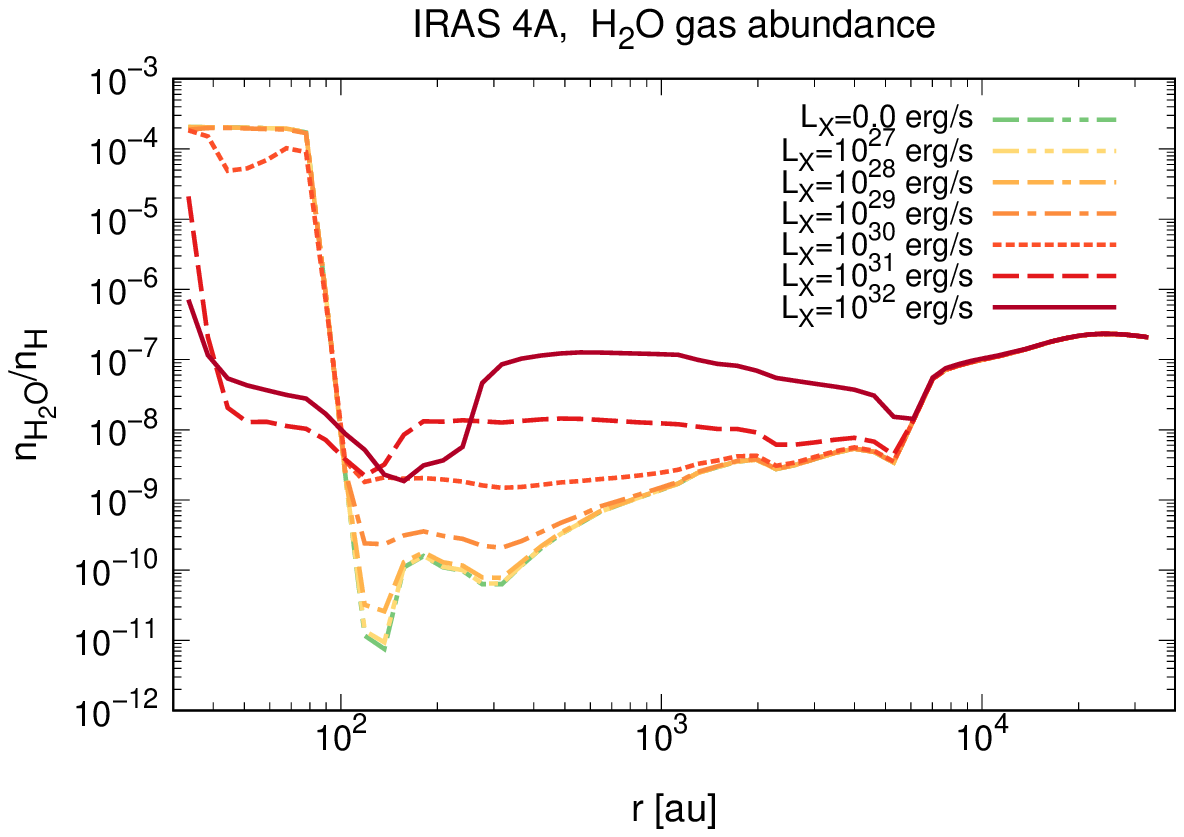}
\includegraphics[scale=0.67]{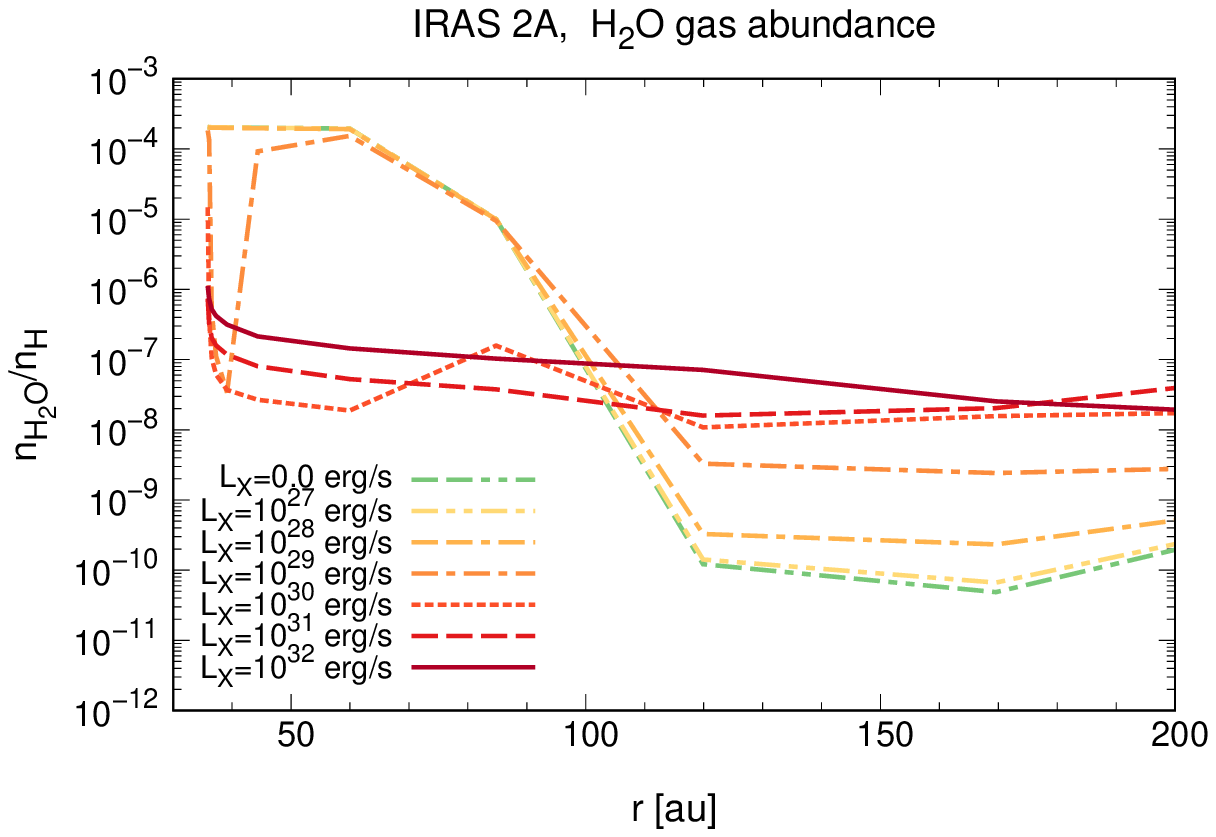}
\includegraphics[scale=0.67]{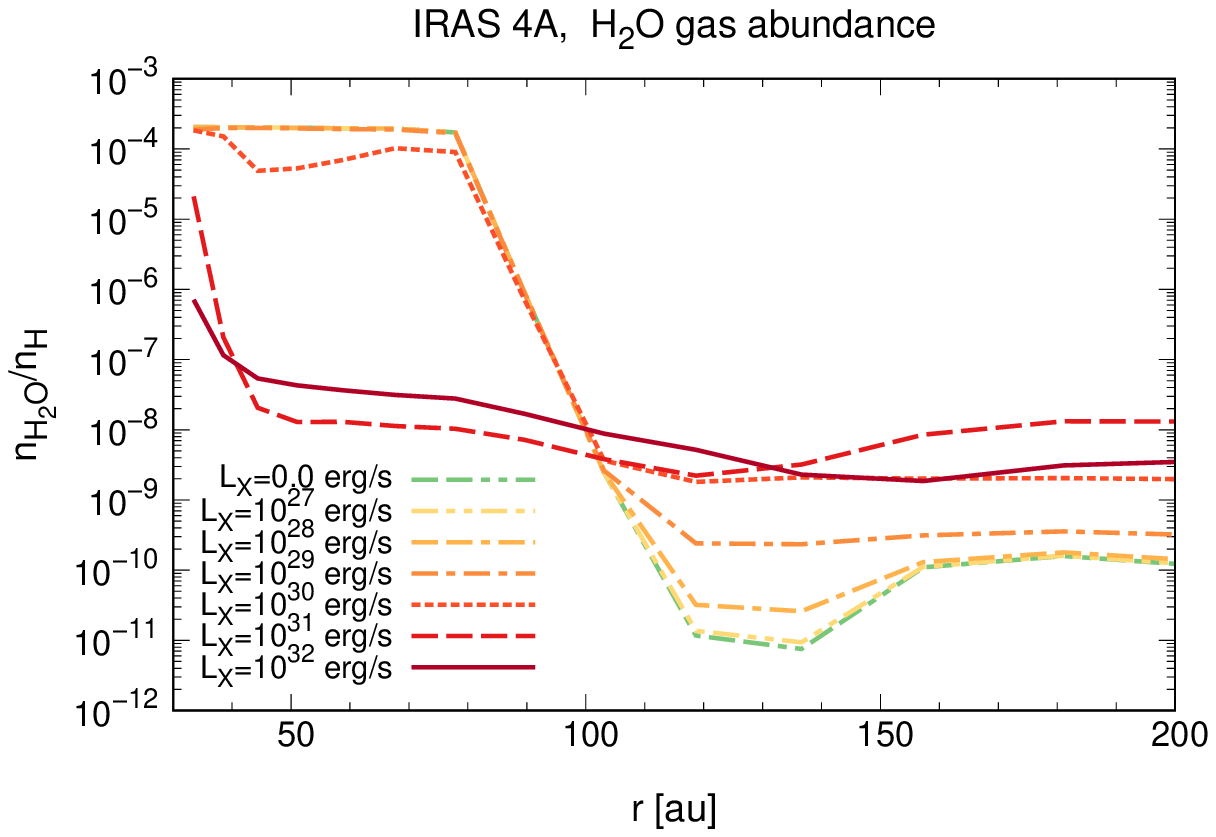}
\end{center}
\caption{
\noindent 
[Top panels]: The radial profiles of water gas fractional abundances with respect to total hydrogen nuclei densities $n_{\mathrm{H}_{2}\mathrm{O}}$/$n_{\mathrm{H}}$ in NGC 1333-IRAS 2A (left panel) and NGC 1333-IRAS 4A (right panel) envelope models.
The double-dashed double-dotted lines, the dashed double-dotted lines, the double-dashed dotted lines, the dashed dotted lines, the dotted lines, the dashed lines, and the solid lines 
show radial $n_{\mathrm{H}_{2}\mathrm{O}}$/$n_{\mathrm{H}}$ profiles for values of central star X-ray luminosities $L_{\mathrm{X}}$=0, $10^{27}$, 10$^{28}$, 10$^{29}$, 10$^{30}$, 10$^{31}$, and 10$^{32}$ erg s$^{-1}$, respectively. The line colors gradually change from green to yellow, orange, red, and brown as the values of $L_{\mathrm{X}}$ increase.
In Figures \ref{Figure5_O2&Ogas}$-$\ref{Figure7_CH3OHgas&ice}, \ref{FigureC1_CO$_{2}$}, \ref{FigureC2_CO}, \ref{FigureD1_CH4&C2H&HCNgas}, \ref{FigureE1_NH3&N2gas}, \ref{FigureH1_H2O&O2&O&OHgas_reaction-rev}, and \ref{FigureH2_H2O&O2&O&OHgas_Edes-O-rev}, 
we adopt the same line type and color patterns of the calculated radial fractional abundance profiles of gas-phase molecules for different values of $L_{\mathrm{X}}$.
 [Bottom panels]: The same radial profiles as shown in the top panels, but we enlarge the inner regions ($r<200$ au) on a linear scale.
%
}\label{Figure3_H2Ogas}
\end{figure*} 
\begin{figure*}
\begin{center}
\includegraphics[scale=0.67]{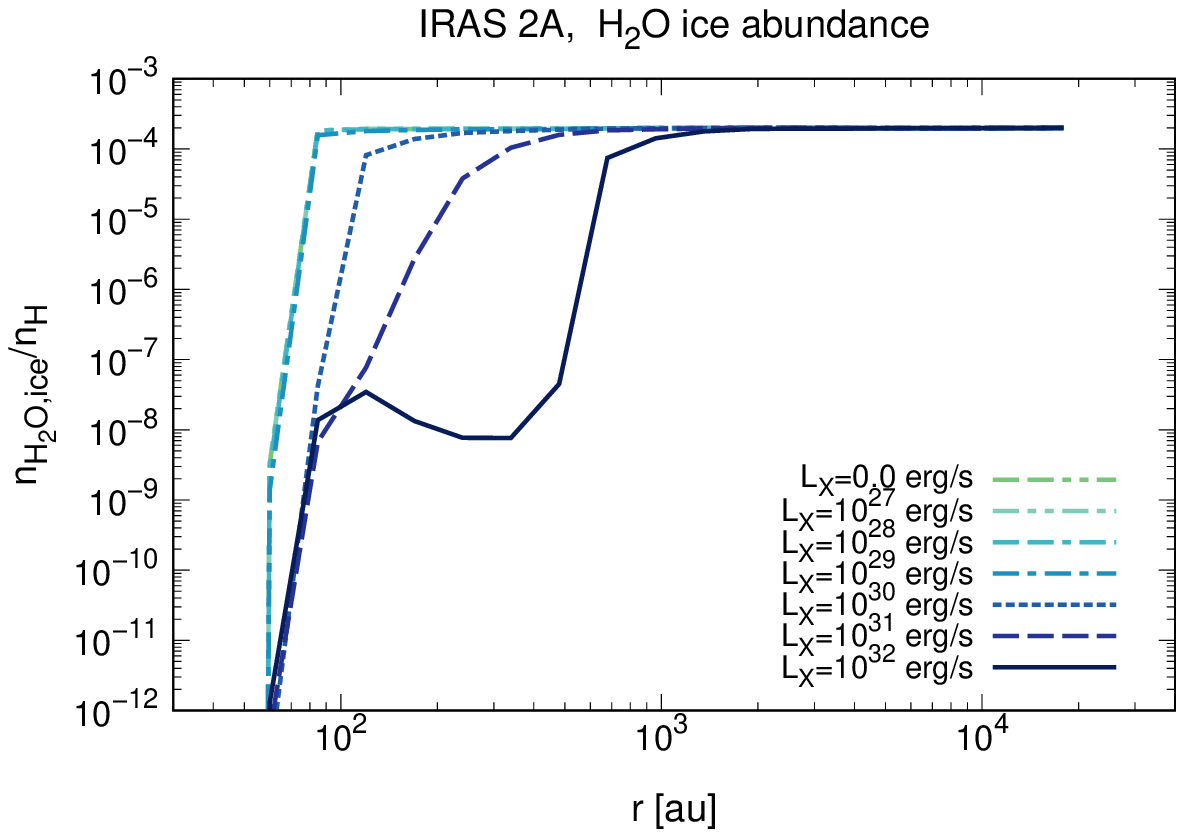}
\includegraphics[scale=0.67]{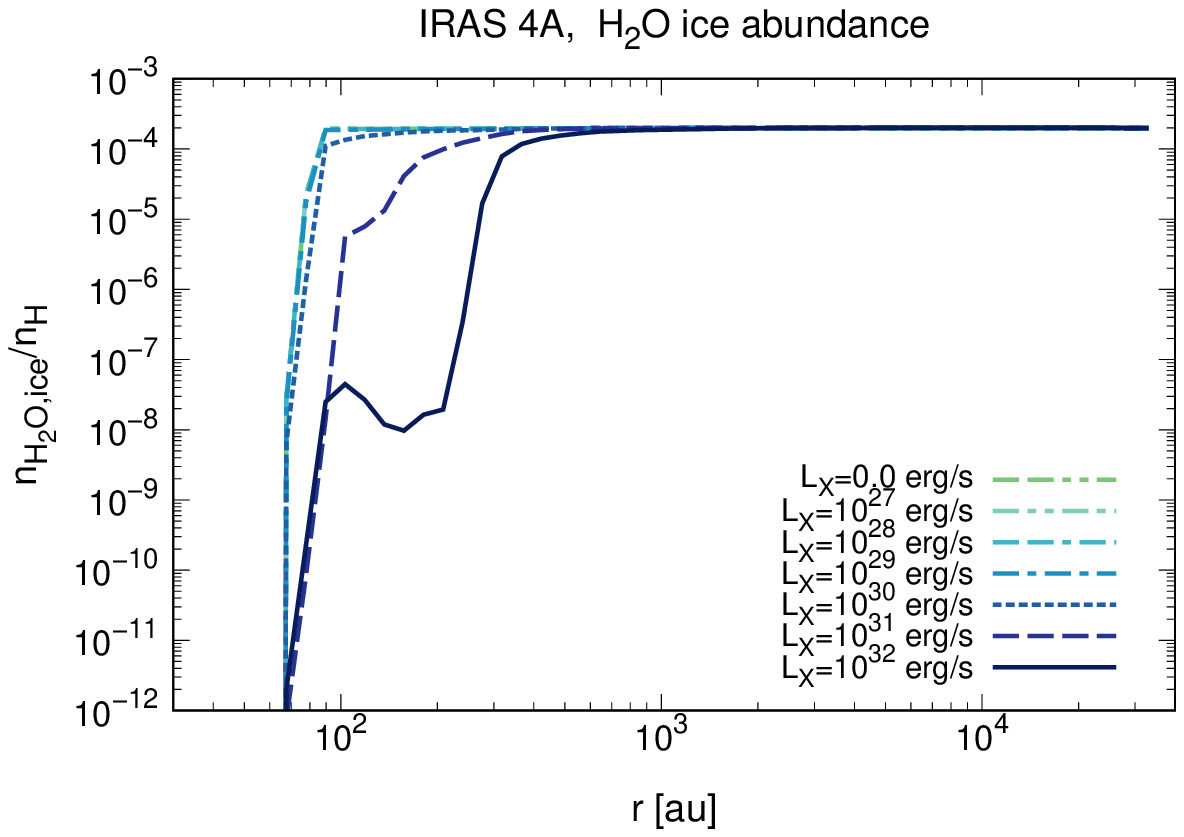}
\end{center}
\caption{
\noindent 
The radial profiles of water ice fractional abundances $n_{\mathrm{H}_{2}\mathrm{O}, \mathrm{ice}}$/$n_{\mathrm{H}}$ in NGC 1333-IRAS 2A (left panel) and NGC 1333-IRAS 4A (right panel) envelope models.
%
The line colors are gradually change from green to cyan, blue, and indigo as the values of $L_{\mathrm{X}}$ become large.
The profiles for $L_{\mathrm{X}}=0-10^{29}$ erg s$^{-1}$ are almost overlapped.
In Figures \ref{Figure7_CH3OHgas&ice}, \ref{FigureC1_CO$_{2}$}, and \ref{FigureC2_CO}, we adopt the same line type and color patterns of the radial fractional abundance profiles of icy-phase molecules for different values of $L_{\mathrm{X}}$.
}\label{Figure4_H2Oice}
\end{figure*} 
Figure \ref{Figure3_H2Ogas} shows the radial profiles of the water gas fractional abundances with respect to total hydrogen nuclei densities $n_{\mathrm{H}_{2}\mathrm{O}}$/$n_{\mathrm{H}}$ in IRAS 2A (left panels) and IRAS 4A (right panels) envelope models, for the various X-ray luminosities ($L_{\mathrm{X}}$=0, $10^{27}$, 10$^{28}$, 10$^{29}$, 10$^{30}$, 10$^{31}$, and 10$^{32}$ erg s$^{-1}$).
Figure \ref{Figure4_H2Oice} shows the radial profiles of water ice fractional abundances $n_{\mathrm{H}_{2}\mathrm{O}, \mathrm{ice}}$/$n_{\mathrm{H}}$ in the same models.
In both models, the water snowline positions are at $r\sim10^{2}$ au, where $T_{\mathrm{gas}}$ is around 100 K. 
\\ \\
For $L_{\mathrm{X}}=0$ erg s$^{-1}$, water gas abundances are around $2\times10^{-4}$ inside the water snowline ($T_{\mathrm{gas}}>10^{2}$ K, $r<10^{2}$ au), and sharply decrease to $\lesssim10^{-10}$ just outside the water snowline ($T_{\mathrm{gas}}<10^{2}$ K, $r>10^{2}$ au).
The water gas abundances increase in the outer low density envelopes ($n_{\mathrm{H}_{2}\mathrm{O}}$/$n_{\mathrm{H}}$$\sim10^{-7}$ at $r\sim10^{4}$ au), since 
in this region the water abundances are mainly determined by the balance between freeze-out of water vapor and cosmic-ray induced photodesorption of water ice, which maintains an approximately constant number density of gas phase water (for more details, see \citealt{Schmalzl2014}).
\\ \\
Outside the water snowline,
for $L_{\mathrm{X}}\gtrsim10^{30}$ erg s$^{-1}$, water gas abundances become higher (up to $n_{\mathrm{H}_{2}\mathrm{O}}$/$n_{\mathrm{H}}$$\sim10^{-8}-10^{-7}$), compared with the values ($n_{\mathrm{H}_{2}\mathrm{O}}$/$n_{\mathrm{H}}$$\sim10^{-10}$) for $L_{\mathrm{X}}\lesssim10^{27}$ erg s$^{-1}$ in IRAS 2A and $L_{\mathrm{X}}\lesssim10^{28}$ erg s$^{-1}$ in IRAS 4A. 
In addition, water ice abundances (see Figure \ref{Figure4_H2Oice}) are around $2\times10^{-4}$ outside the water snowline for $L_{\mathrm{X}}\lesssim10^{30}$ erg s$^{-1}$, and they become much lower (below to $n_{\mathrm{H}_{2}\mathrm{O}}$/$n_{\mathrm{H}}$$\sim10^{-8}$ at a few $\times10^{2}$ au) for $L_{\mathrm{X}}\gtrsim10^{31}$ erg s$^{-1}$.
We conclude that photodesorption by external X-ray photons (e.g., \citealt{Walsh2015, Cuppen2017, Dupuy2018}) is important in this region (see also Sections 4.2 and 4.3). 
This X-ray effect is stronger in the IRAS 2A model, since it has around a $3-6$ times lower density and thus higher X-ray fluxes (see Figure \ref{Figure2_FX}) than the IRAS 4A model.
The lower density in the IRAS 2A also decreases the efficacy of two-body ion-molecule reactions (see Appendix A where we demonstrate the effect of density only on the chemistry).
%
Water gas abundances at $r\gtrsim10^{3}$ au are also affected by strong X-ray fluxes, although $\xi_{\mathrm{X}}(r)$ is smaller than $\xi_{\mathrm{CR}}(r)$ in these regions. 
This is because at $r\gtrsim10^{3}$ au the rates of the X-ray induced photodesorption of water ice are around 10$^{3}$ times larger than those of cosmic-ray induced photodesorption,
and much larger ($>10^{20}$ times) than that of thermal desorption and cosmic-ray induced (thermal) desorption.
The chemical model adopted by \citet{Stauber2005,Stauber2006} did not include non-thermal desorption processes, 
and thus they did not find this dependence of the gaseous water abundances on X-ray fluxes outside the water snowline.
\\ \\
Inside the water snowline ($T_{\mathrm{gas}}>10^{2}$ K, $r<10^{2}$ au), for $L_{\mathrm{X}}\lesssim10^{29}$ erg s$^{-1}$ in IRAS 2A 
and $L_{\mathrm{X}}\lesssim10^{30}$ erg s$^{-1}$ in IRAS 4A, the gas maintains high water abundances of 10$^{-4}$, and they are the dominant oxygen carrier along with CO.
On the other hand, for $L_{\mathrm{X}}\gtrsim10^{30}$ erg s$^{-1}$ in IRAS 2A and $L_{\mathrm{X}}\gtrsim10^{31}$ erg s$^{-1}$ in IRAS 4A,
water gas abundances become much smaller just inside the water snowline ($T\sim100-250$ K, below to $n_{\mathrm{H}_{2}\mathrm{O}}$/$n_{\mathrm{H}}$$\sim$$10^{-8}-10^{-7}$) and in the innermost regions ($T\sim250$ K, $n_{\mathrm{H}_{2}\mathrm{O}}$/$n_{\mathrm{H}}$$\sim10^{-6}$).
\\ \\
Within the water snowline, water sublimates from the dust-grain surfaces into the gas phase.
According to \citet{Stauber2005, Stauber2006}, \citet{vanDishoeck2013, vanDishoeck2014} and \citet{Walsh2015}, in the presence of X-rays, gas-phase water in this region is mainly destroyed with X-ray induced photodissociation (H$+$OH), and ion-molecule reactions (with, for example, HCO$^{+}$, H$^{+}$, H$_{3}^{+}$, and He$^{+}$).
For $L_{\mathrm{X}}\gtrsim10^{29}$ erg s$^{-1}$ in IRAS 2A and $L_{\mathrm{X}}\gtrsim10^{30}$ erg s$^{-1}$ in IRAS 4A, the X-ray ionization rates $\xi_{\mathrm{X}}(r)$ are larger than 
our adopted cosmic ray ionization rate $\xi_{\mathrm{CR}}(r)$ ($=$$1.0\times10^{-17}$ s$^{-1}$) within the water snowline (see Figure \ref{Figure2_FX}).
Thus, these processes are important to explain the dependence on X-ray fluxes and number densities within the water snowline.
\citet{Stauber2006} discussed that these ion-molecule reactions are more effective than X-ray-induced photodissociation, with resulting water gas abundances varying by less than 15\% if they ignored the X-ray induced photodissociation.
In our calculations, we also confirm that the reaction rates of these ion-molecule reactions are larger than those of the X-ray induced photodissociation leading to H+OH, and that the former reactions become more important compared with the latter reaction as the gas densities become larger (see also Appendix A).
\\ \\
At $r\sim60$ au and for $L_{\mathrm{X}}=10^{32}$ erg s$^{-1}$ in the IRAS 2A model,
the water gas abundance and $n_{\mathrm{H}_{2}\mathrm{O}}$
are $1.4\times10^{-7}$ and 5.9$\times10^{1}$ cm$^{-3}$ at $t=10^{5}$ years, respectively, and 
the HCO$^{+}$ gas abundance and $n_{\mathrm{HCO}^{+}}$\footnote[6]{$n_{\mathrm{HCO}^{+}}$ is the number denisity of HCO$^{+}$ and we obtain the value in Section 3.3 and Figure \ref{Figure6_HCO+&OHgas}.} are
$9.8\times10^{-9}$ and $4.0$ cm$^{-3}$ at $t=10^{5}$ years, respectively. 
On the basis of these values, the rate coefficient of the ion-molecule reaction with H$_{2}$O+HCO$^{+}$$\rightarrow$CO+H$_{3}$O$^{+}$, $k_{1}$, is
$\sim3.7\times10^{-9}$ cm$^{3}$ s$^{-1}$ \citep{Adams1978}, and the reaction rate,
$R(1)=$$k_{1}$$n_{\mathrm{H}_{2}\mathrm{O}}$$n_{\mathrm{HCO}^{+}}$, is $\sim8.7\times10^{-7}$ cm$^{-3}$ s$^{-1}$.
In contrast, the rate coefficient of X-ray-induced photodissociation leading to H+OH, $k_{2}$, is $\sim8.6\times10^{-11}$ s$^{-1}$ \citep{Gredel1989}, and the reaction rate, $R(2)=$$k_{2}$$n_{\mathrm{H}_{2}\mathrm{O}}$,
 is $\sim5.1\times10^{-9}$ cm$^{-3}$ s$^{-1}$.
 \\ \\
We note that for $L_{\mathrm{X}}\gtrsim10^{31}$ erg s$^{-1}$, abundances of HCO$^{+}$ within the water snowline ($\gtrsim10^{-9}$, see Section 3.3) are larger than 
those of other molecular ions which are important to water gas destruction such as He$^{+}$ ($\lesssim10^{-10}$).
This makes HCO$^{+}$ the most important destructor of H$_{2}$O in highly ionized regions.
%
\\ \\
In the innermost high temperature region ($T_{\mathrm{gas}}\sim250$~K), the following two-body reaction with the reaction barrier of 1736 K \citep{Oldenborg1992},
 \begin{equation}
 \mathrm{OH}+\mathrm{H}_{2}\rightarrow\mathrm{H}+\mathrm{H}_{2}\mathrm{O},
 \end{equation}
becomes more efficient, and thus water gas abundances become relatively large ($n_{\mathrm{H}_{2}\mathrm{O}}$/$n_{\mathrm{H}}$$\sim10^{-6}$) even in the highest X-ray flux cases ($L_{\mathrm{X}}=10^{32}$ erg s$^{-1}$).
As \citet{Stauber2005, Stauber2006} noted, X-ray destruction processes are more effective in lower density models.
%
%
\subsection{Molecular and atomic oxygen fractional abundances}
\begin{figure*}
\begin{center}
\includegraphics[scale=0.67]{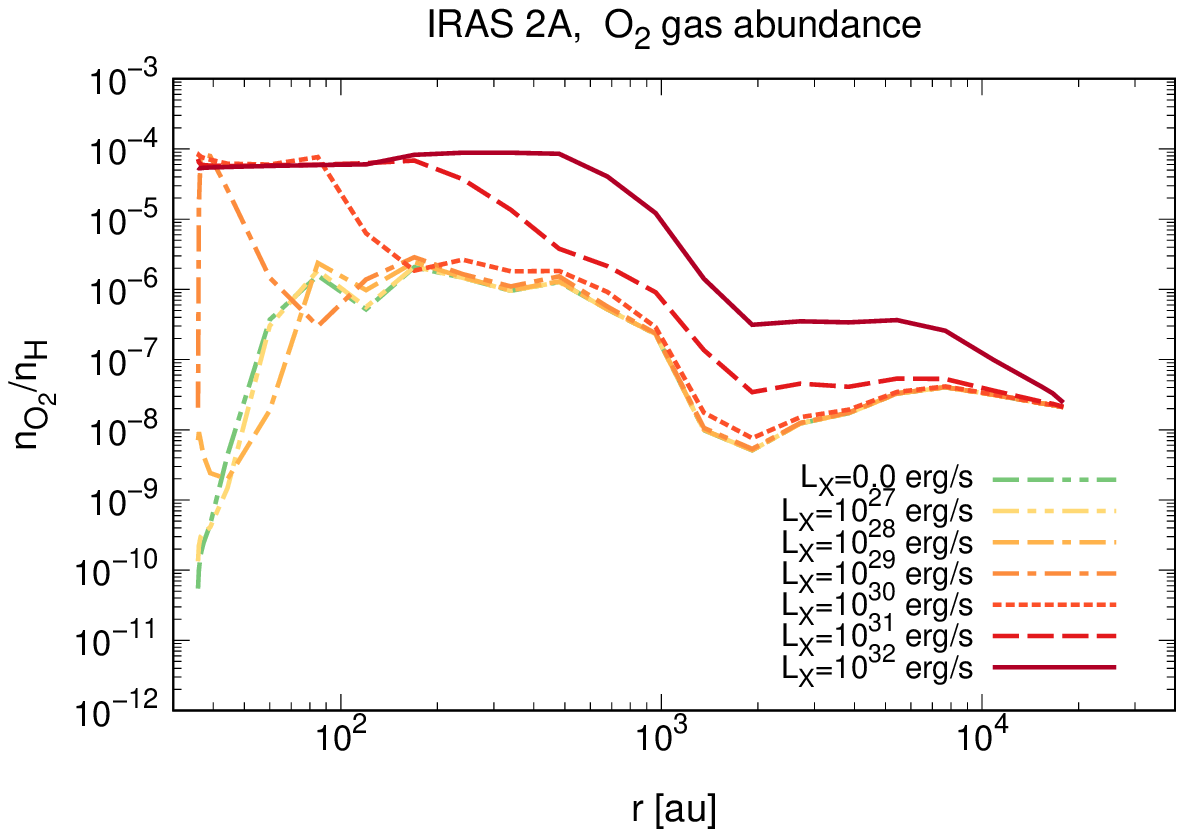}
\includegraphics[scale=0.67]{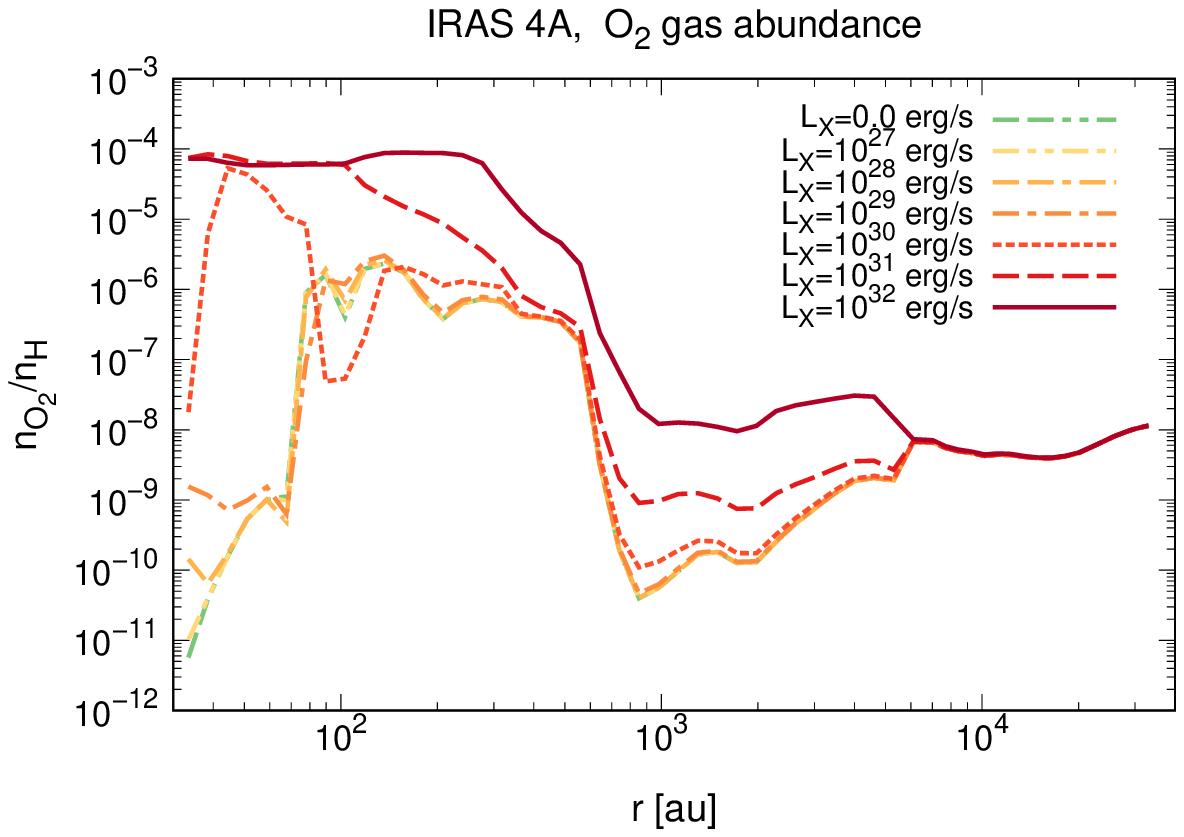}
\includegraphics[scale=0.67]{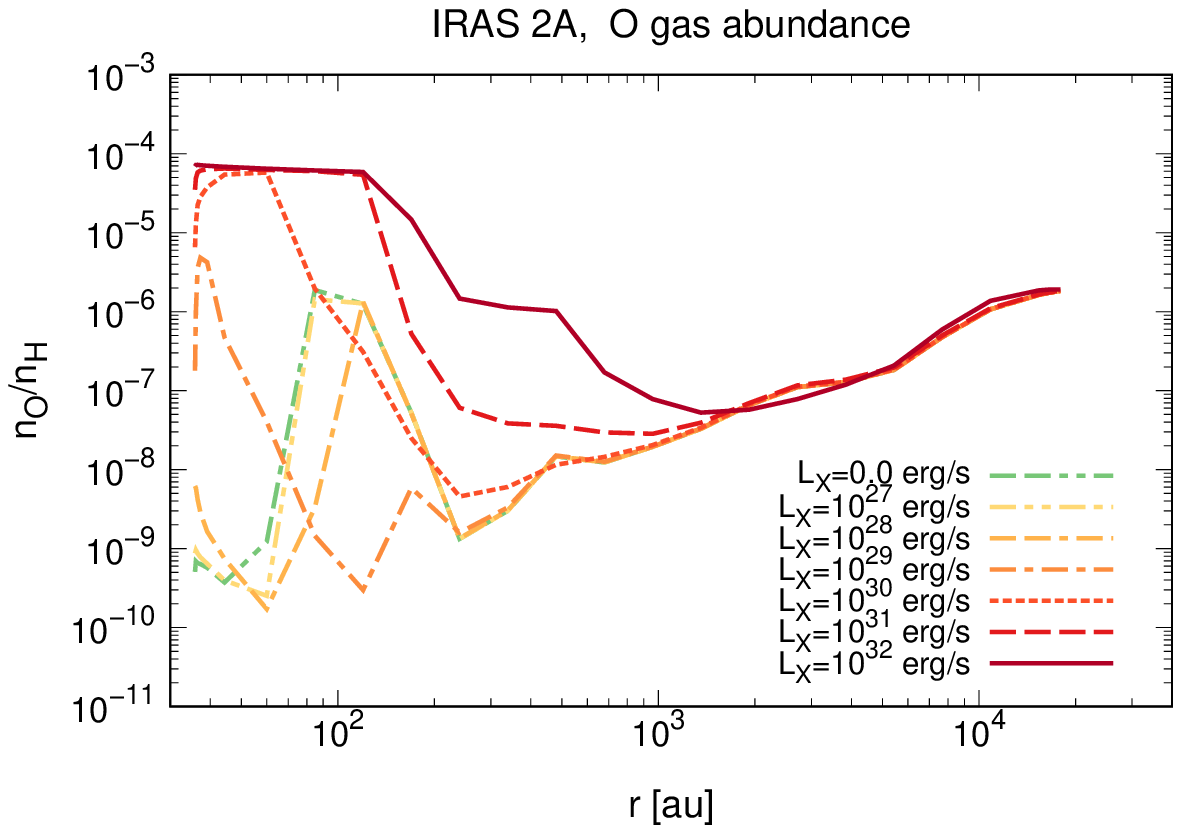}
\includegraphics[scale=0.67]{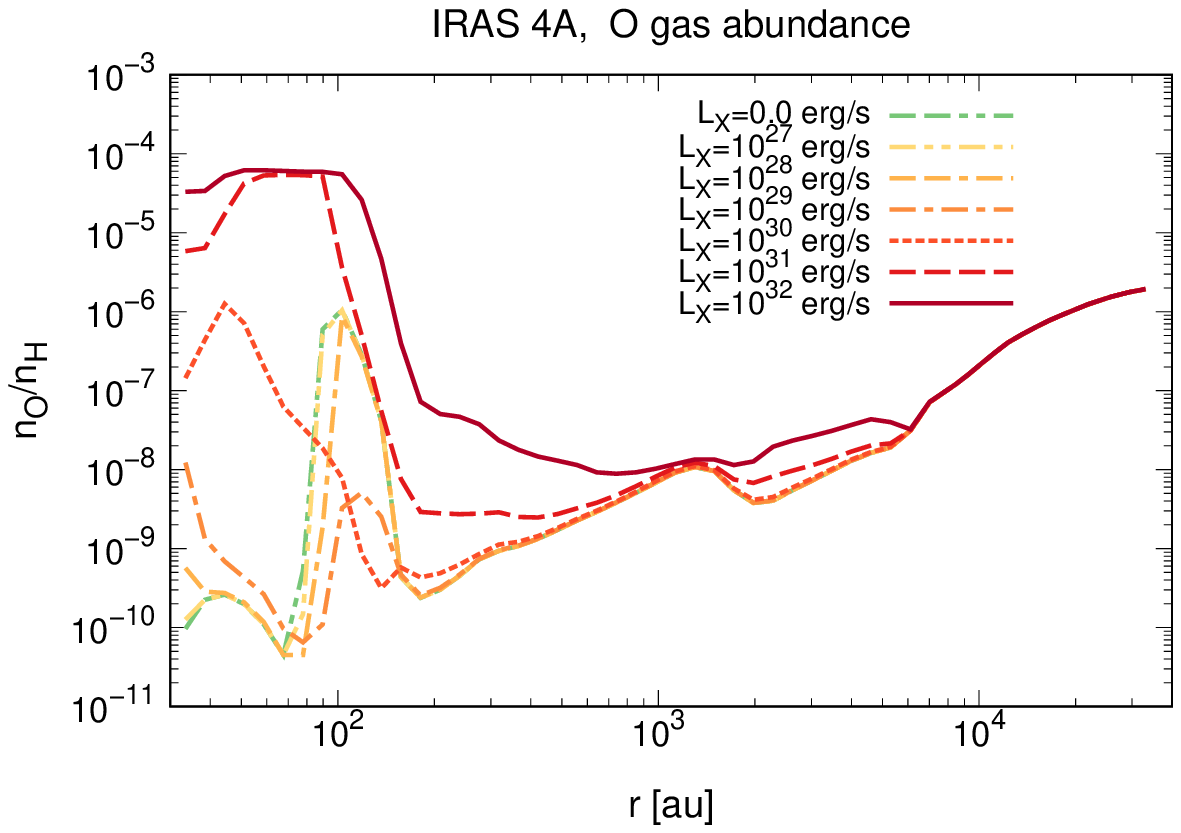}
\end{center}
\caption{
\noindent 
[Top panels]: The radial profiles of molecular oxygen gas fractional abundances $n_{\mathrm{O}_{2}}$/$n_{\mathrm{H}}$ in NGC 1333-IRAS 2A (left panel) and NGC 1333-IRAS 4A (right panel) envelope models.
%
[Bottom panels]: 
The radial profiles of atomic oxygen gas fractional abundances $n_{\mathrm{O}}$/$n_{\mathrm{H}}$ in the IRAS 2A (left panel) and IRAS 4A (right panel) envelope models.
%
}\label{Figure5_O2&Ogas}
\end{figure*} 
Figure \ref{Figure5_O2&Ogas} presents the radial profiles of the fractional abundance of gaseous molecular oxygen $n_{\mathrm{O}_{2}}$/$n_{\mathrm{H}}$ (top panels) and atomic oxygen $n_{\mathrm{O}}$/$n_{\mathrm{H}}$ (bottom panels) in IRAS 2A (left panels) and IRAS 4A (right panels) envelope models, for the various X-ray luminosities.
It is seen that the O$_{2}$ abundances at $r\lesssim10^{4}$ au (IRAS 2A) and $r\lesssim6\times10^{3}$ au (IRAS 4A), and the O abundance at $r\lesssim10^{3}$ au increase (within each snowline position) as X-ray luminosities become larger.
Both molecular and atomic oxygen are very volatile ($E_{\mathrm{des}}$(O)=1660 K and $E_{\mathrm{des}}$(O$_{2}$)=898 K) compared with
H$_{2}$O ($E_{\mathrm{des}}$(H$_{2}$O)=4880 K), thus their snowline positions are located in the outer envelopes ($r>5\times10^{2}$ au).
\\ \\
Inside the water snowline, both molecular and atomic oxygen abundances are much lower ($<10^{-8}$) in the cases of low X-ray luminosities ($L_{\mathrm{X}}\lesssim10^{28}$ erg s$^{-1}$ in IRAS 2A and $L_{\mathrm{X}}\lesssim10^{29}$ erg s$^{-1}$ in IRAS 4A).
In contrast, for moderate X-ray luminosities ($L_{\mathrm{X}}\sim10^{29}$ erg s$^{-1}$ in IRAS 2A and $L_{\mathrm{X}}\sim10^{30}$ erg s$^{-1}$ in IRAS 4A) and high X-ray luminosities ($L_{\mathrm{X}}\gtrsim10^{30}$ erg s$^{-1}$ in IRAS 2A and $L_{\mathrm{X}}\gtrsim10^{31}$ erg s$^{-1}$ in IRAS 4A),
their abundances become larger, and reach about $\sim5\times10^{-5}-10^{-4}$ with $L_{\mathrm{X}}\gtrsim10^{31}$ erg s$^{-1}$.
Compared with the water gas abundances, both molecular and atomic oxygen have opposite dependence on X-ray fluxes.
Thus, the identity of the main volatile oxygen carrier in the inner regions is very sensitive to the X-ray flux from the central protostars (see also Section 4.1).
In Sections 3.2-3.7 and 4.1, and Appendix C, D, and E, we adopted the same definition for the values of low, moderate and high X-ray luminosities.
\\ \\
According to \citet{Woitke2009} and \citet{Walsh2015}, in the presence of X-rays, atomic oxygen is mainly produced by X-ray induced photodissociation of OH and CO.
OH is efficiently produced by X-ray induced photodissociation and fragmental photodesorption of H$_{2}$O (see Section 3.4), and thus the O abundance becomes larger as X-ray fluxes become larger.
In addition, as also discussed in \citet{Walsh2015} and \citet{Eistrup2016}, molecular oxygen is formed in the gas-phase via the following reaction, 
\begin{equation}
\label{reaction12}
\mathrm{O}+\mathrm{OH}\rightarrow\mathrm{O}_{2}+\mathrm{H}, 
\end{equation}
and is destroyed via photodissociation and reactions with C and H to yield CO and OH, respectively.
Note that reaction (\ref{reaction12}) is a barrierless neutral-neutral reaction and has a negligible temperature dependance \citep{Carty2006, Taquet2016}.
Both the O and OH abundances become larger as X-ray fluxes become larger, and thus the O$_{2}$ abundances become larger, especially in the inner warm envelope where water is sublimated from dust grains.
\subsection{HCO$^{+}$ fractional abundances}
\begin{figure*}
\begin{center}
\includegraphics[scale=0.67]{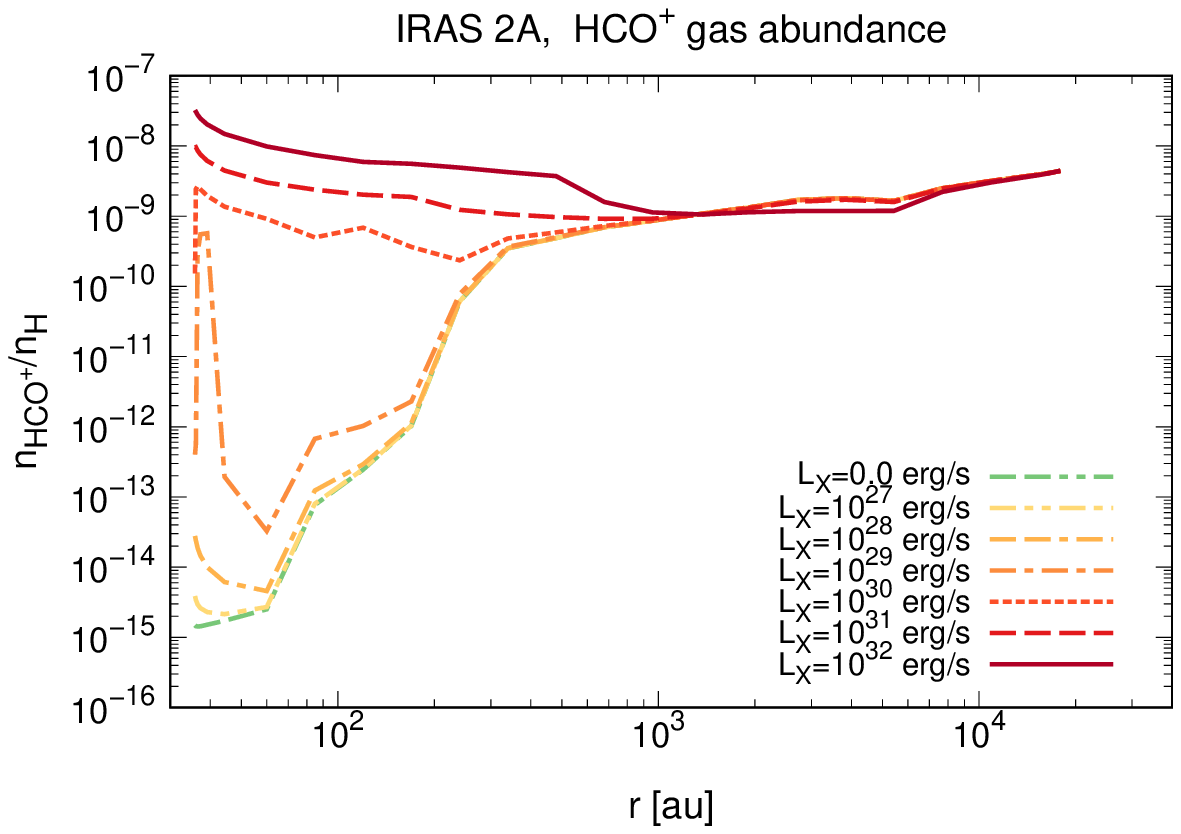}
\includegraphics[scale=0.67]{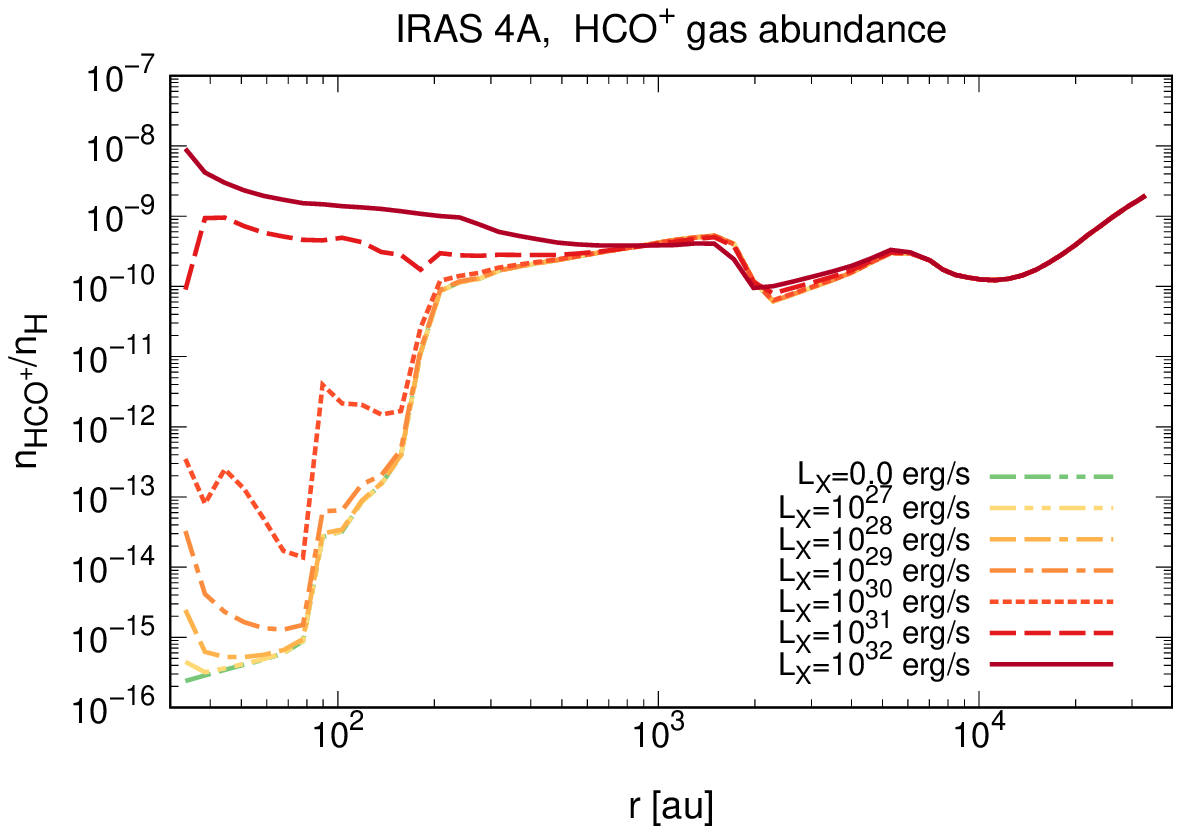}
\includegraphics[scale=0.67]{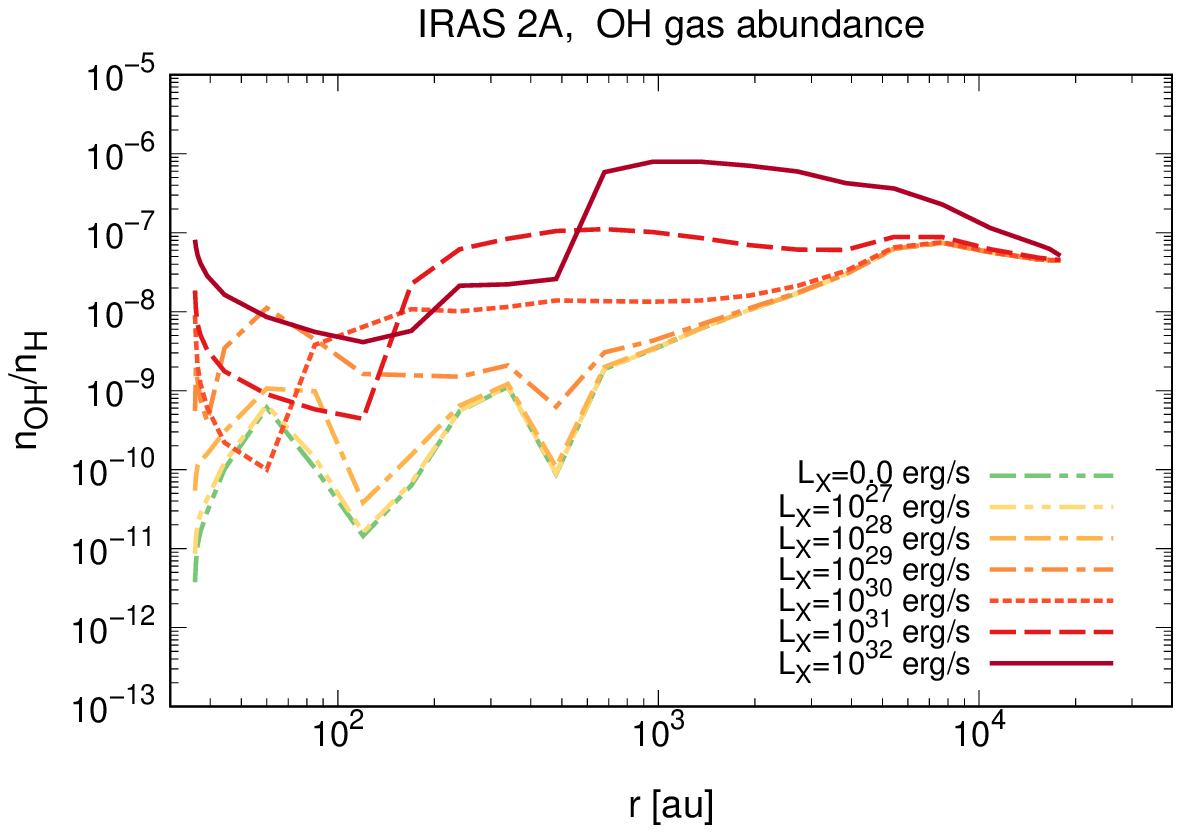}
\includegraphics[scale=0.67]{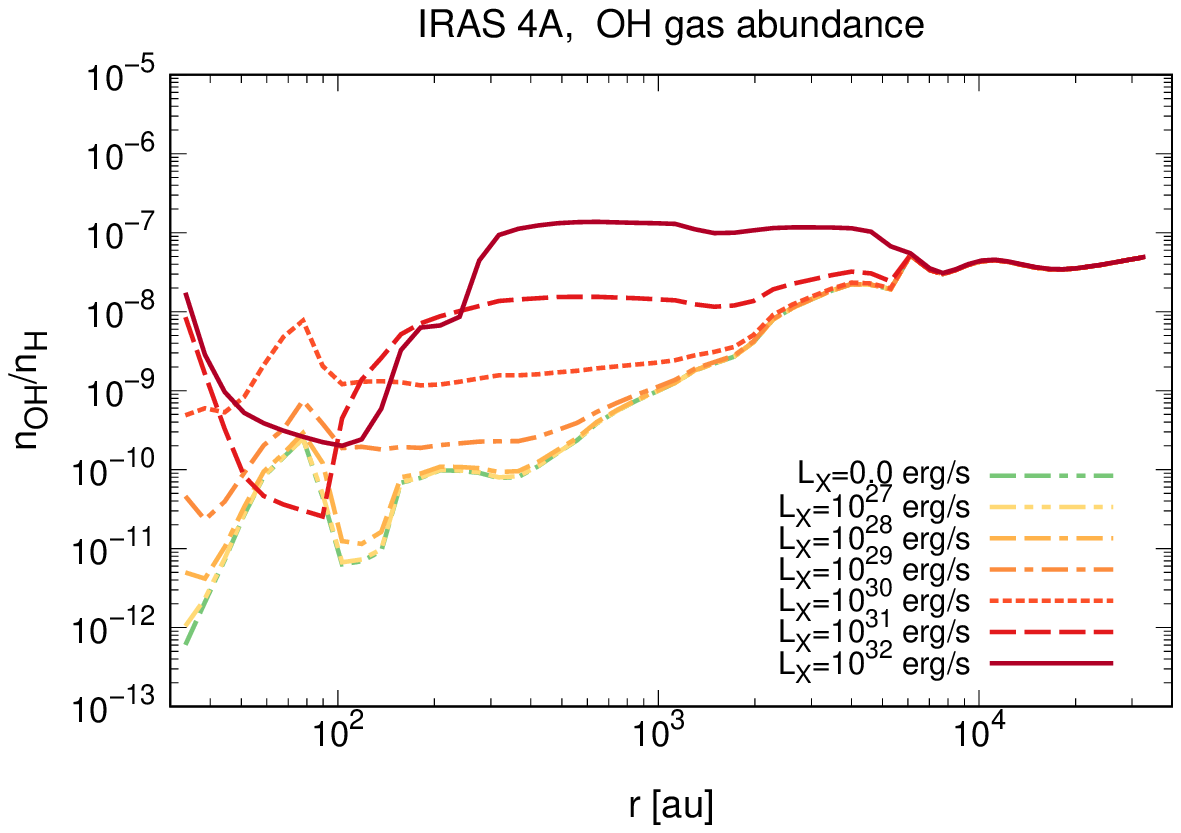}
\end{center}
\caption{
\noindent 
[Top panels]: The radial profiles of HCO$^{+}$ gas fractional abundances $n_{\mathrm{HCO}^{+}}$/$n_{\mathrm{H}}$ in NGC 1333-IRAS 2A (left panel) and NGC 1333-IRAS 4A (right panel) envelope models.
[Bottom panels]: 
The radial profiles of OH gas fractional abundances $n_{\mathrm{OH}}$/$n_{\mathrm{H}}$ in the IRAS 2A (left panel) and IRAS 4A (right panel) envelope models.
}\label{Figure6_HCO+&OHgas}
\end{figure*} 
The top panels of Figure \ref{Figure6_HCO+&OHgas} show the radial profiles of the HCO$^{+}$ fractional abundances $n_{\mathrm{HCO}^{+}}$/$n_{\mathrm{H}}$ in IRAS 2A (left panel) and IRAS 4A (right panel) envelope models, for the various X-ray luminosities.
According to our model, the HCO$^{+}$ abundances at $r\gtrsim10^{3}$ au ($\sim10^{-9}-10^{-8}$ for IRAS 2A, and $\sim10^{-10}-10^{-9}$ for IRAS 4A) do not change with different X-ray luminosities.
This is consistent with the input assumption that cosmic-ray ionization dominates at these radii (see Figure \ref{Figure2_FX} and Section 4.6).
\\ \\
The HCO$^{+}$ abundances at $r\lesssim10^{3}$ au are affected by strong X-ray fluxes.
For low X-ray luminosities,
HCO$^{+}$ abundances drop in the inner envelope, and reach $\lesssim10^{-13}$ within the water snowline, due to the efficient destruction by water (see below).
In contrast, for high X-ray luminosities, 
they become higher in the inner envelope, and reach more than 10$^{-9}$ (for IRAS 2A) and 10$^{-10}$ (for IRAS 4A) within water snowline.
The overall HCO$^{+}$ abundances are larger and X-ray effects are also stronger in the IRAS 2A model, since it has around $3-6$ times lower densities and thus higher X-ray fluxes than the IRAS 4A model has (see Figures \ref{Figure1_Tn} and \ref{Figure2_FX}, and Appendix A).
\\ \\
HCO$^{+}$ has been considered as a chemical tracer of the water snowline, since its most abundant destroyer in warm dense gas is water via the following reaction \citep{Jorgensen2013, Visser2015, vantHoff2018, Hsieh2019, Lee2020, Leemker2021}, 
\begin{equation}
\mathrm{H}_{2}\mathrm{O}+\mathrm{HCO}^{+}\rightarrow\mathrm{CO}+\mathrm{H}_{3}\mathrm{O}^{+}.
\end{equation}
Thus, a strong decline in HCO$^{+}$ (and its isotopologue H$^{13}$CO$^{+}$) is expected within the water snowline.
\citet{vantHoff2018} conducted spherically symmetric physical-chemical modeling using the same IRAS 2A temperature and number density model which we adopt (see Section 2.1.1 and Figure \ref{Figure1_Tn}). Their gas-grain chemical model included gas-phase cosmic-ray induced reactions, but did not include X-ray induced chemistry (see e.g., \citealt{Taquet2014}).
They reported an increase of H$^{13}$CO$^{+}$ emission just outside the water snowline and a spatial anti-correlation of H$^{13}$CO$^{+}$ and H$_{2}$$^{18}$O emission in the envelope around IRAS 2A.
The radial profiles of water and HCO$^{+}$ gas abundances in \citet{vantHoff2018} are similar to those in our model with $L_{\mathrm{X}}\lesssim10^{29}$ erg s$^{-1}$.
On the basis of our modeling, for high X-ray luminosities, the water gas abundance sharply decreases inside the water snowline and thus HCO$^{+}$ is not efficiently destroyed.
Formation of HCO$^{+}$ is dominated by the ion-molecule reaction between H$_{3}^{+}$ and CO \citep{Schwarz2018, vantHoff2018, Leemker2021}, and H$_{3}^{+}$ is mainly formed by the ionization of H$_{2}$. 
Therefore, the HCO$^{+}$ abundances increase as the X-ray ionisation rate increases, and they have relatively radially flat profiles for high X-ray luminosities (see Figure \ref{Figure6_HCO+&OHgas}).
Thus, our work suggested that HCO$^{+}$ and its isotopologue H$^{13}$CO$^{+}$ lines cannot be used as tracers of the water snowline position if X-ray fluxes are high and inner water gas is absent.
The X-ray ionisation rates $\xi_{\mathrm{X}}(r)$ where HCO$^{+}$ loses its efficacy as a water snowline tracer are $\gtrsim10^{-16}$ s$^{-1}$ (see Figure \ref{Figure2_FX}), which correspond $L_{\mathrm{X}}\gtrsim10^{30}-10^{31}$ erg s$^{-1}$, depending on density structures.
\\ \\
In Class II disks, HCO$^{+}$ and its isotopologues are considered to trace X-ray and high cosmic-ray ionisation rates with $\gtrsim10^{-17}$ s$^{-1}$ in the disk surface \citep{Cleeves2014}.
According to our calculations,
HCO$^{+}$ is the dominant cation in the outer envelopes where the cosmic-ray ionisation is dominant (see Figure 2), and also in the inner envelopes if $\xi_{\mathrm{X}}(r)$ is $\gtrsim10^{-16}$ s$^{-1}$.
Thus, in these cases HCO$^{+}$ line emission could be used to estimate the electron number densities and the ionization rates (see also \citealt{vantHoff2018} and Section 4.6 of this paper).
\subsection{OH fractional abundances}
The bottom panels of Figure \ref{Figure6_HCO+&OHgas} show the radial profiles of the OH gas fractional abundances $n_{\mathrm{OH}}$/$n_{\mathrm{H}}$ in IRAS 2A (left panel) and IRAS 4A (right panel) envelope models, for the various X-ray luminosities.
The OH abundances increase at $r\lesssim10^{4}$ au as values of X-ray luminosities become larger.
For low X-ray luminosities,
the OH abundances are around $10^{-9}-10^{-8}$ at $r\gtrsim10^{3}$ au, and become lower in the inner envelopes ($\sim10^{-10}-10^{-9}$ at $r\sim10^{2}$ au, and $\sim10^{-12}-10^{-11}$ at the inner edge).
For moderate X-ray luminosities,
the OH abundances become higher in the inner envelope, and reach more than 10$^{-9}$ within water snowline.
In addition, for high X-ray luminosities,
the OH abundances are much higher at $r\sim10^{2}-10^{4}$ au ($\sim10^{-8}-10^{-6}$), and become a bit lower ($\lesssim10^{-8}$) around and just inside the water snowline ($\lesssim10^{2}$ au).
\\ \\
OH is efficiently produced by X-ray induced photodissociation of H$_{2}$O gas and fragmental photodesorption of H$_{2}$O ice (see also Section 2.2.2), 
thus the OH abundances increase as the X-ray flux becomes larger (see Section 3.1 and Figure \ref{Figure3_H2Ogas}).
The former X-ray induced photodissociation reaction is dominant within the water snowline, 
whereas the fragmental photodesorption reaction is dominant outside the water snowline where a large amount of water ice is present on the dust-grain surface.
For example, at $r\sim480$ au and $L_{\mathrm{X}}=10^{32}$ erg s$^{-1}$ in the IRAS 2A model, the rate coefficient of the former X-ray induced photodissociation reaction, $k_{3}$, is 
$\sim5.6\times10^{-13}$ s$^{-1}$ \citep{Gredel1989, Heays2017}, and the reaction rate, $R(3)=$$k_{3}$$n_{\mathrm{H}_{2}\mathrm{O}}$, is $\sim1\times10^{-13}$ cm$^{-3}$ s$^{-1}$ at $t=10^{5}$ years.
In contrast, the rate coefficient of the latter photodesorption reaction, $k_{4}$, is $\sim1.9\times10^{-9}$ (\citealt{Oberg2009b, Walsh2015}, see also Section 2.2.2), and the reaction rate, $R(4)=$$k_{4}$$n_{\mathrm{H}_{2}\mathrm{O}, \mathrm{ice}}$, is $\sim1\times10^{-9}$ cm$^{-3}$ s$^{-1}$ at $t=10^{5}$ years.
As discussed in Sections 3.1 and 3.2, atomic oxygen is mainly produced by X-ray induced photodissociation of OH, and molecular oxygen is produced from OH in the gas phase (O$+$OH). 
Therefore, for high X-ray luminosities, the OH abundances decrease around and inside the water snowline where molecular and atomic oxygen abundances are high ($\sim10^{-4}$).
\subsection{CH$_{3}$OH fractional abundances}
\begin{figure*}
\begin{center}
\includegraphics[scale=0.67]{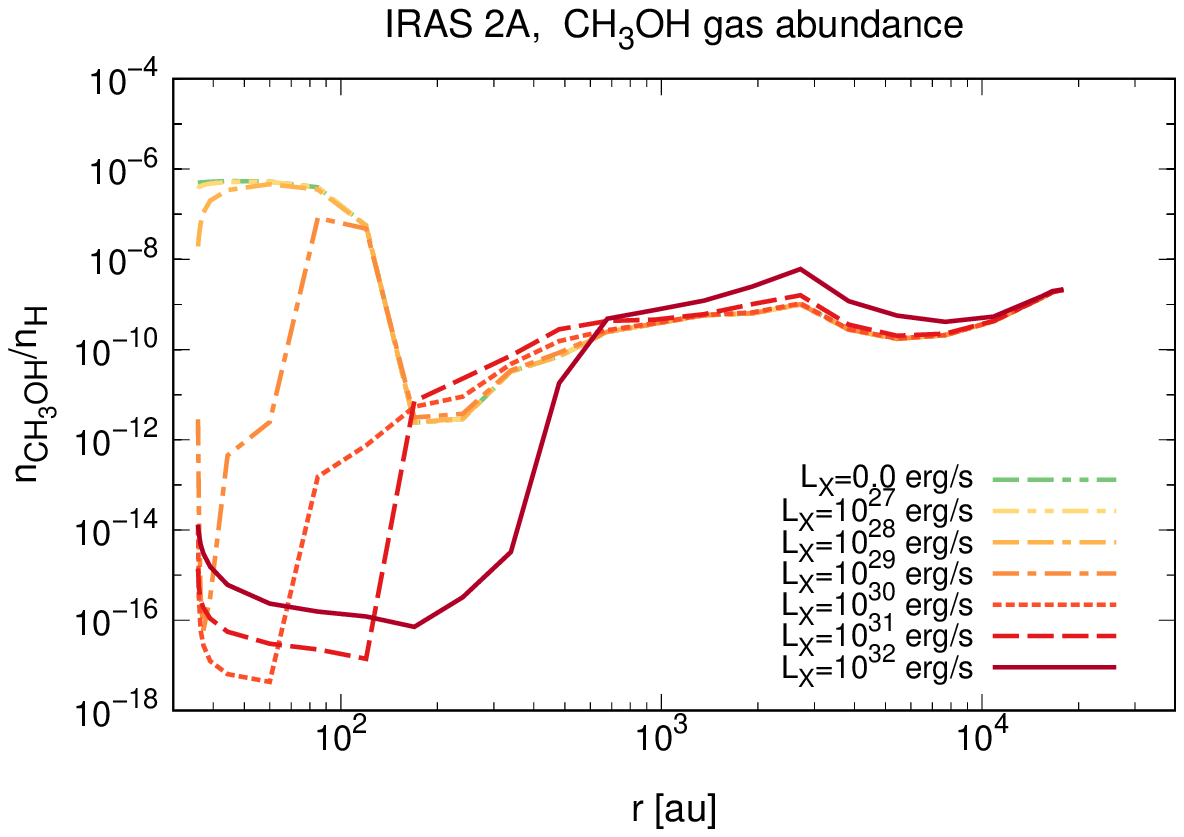}
\includegraphics[scale=0.67]{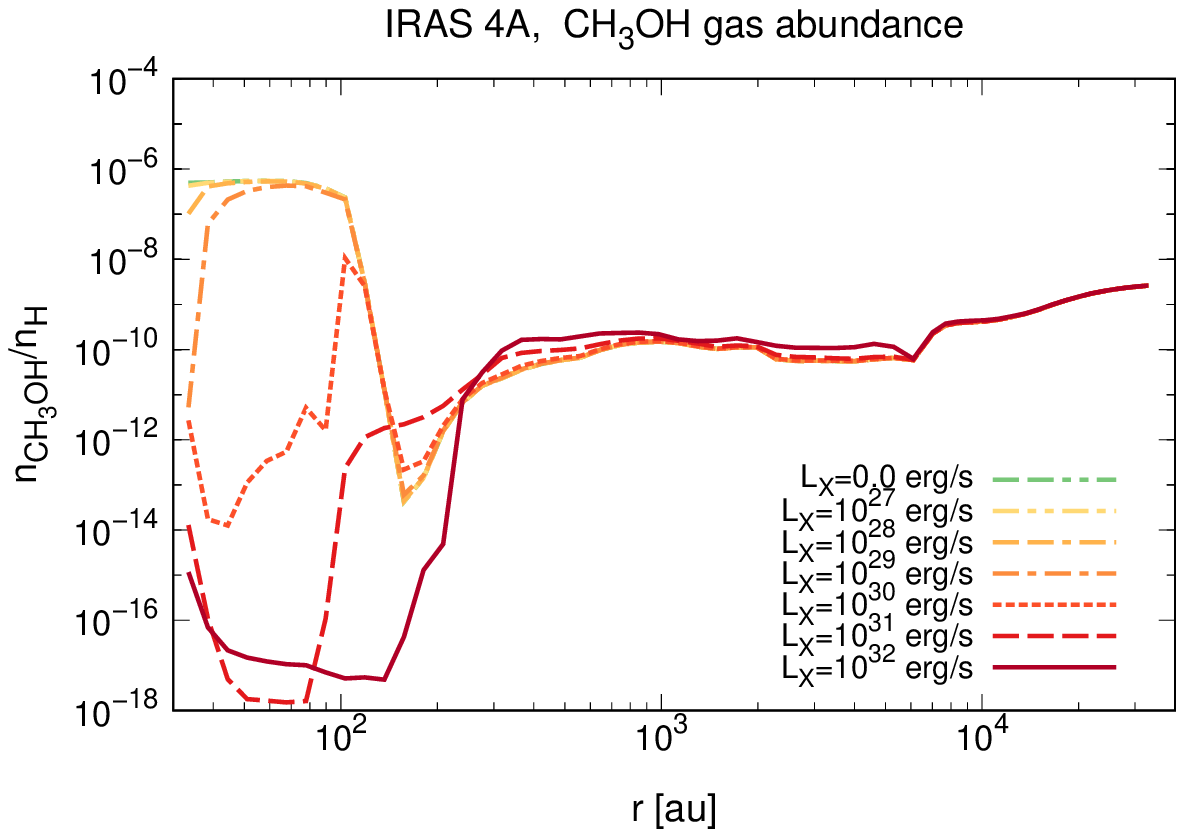}
\includegraphics[scale=0.67]{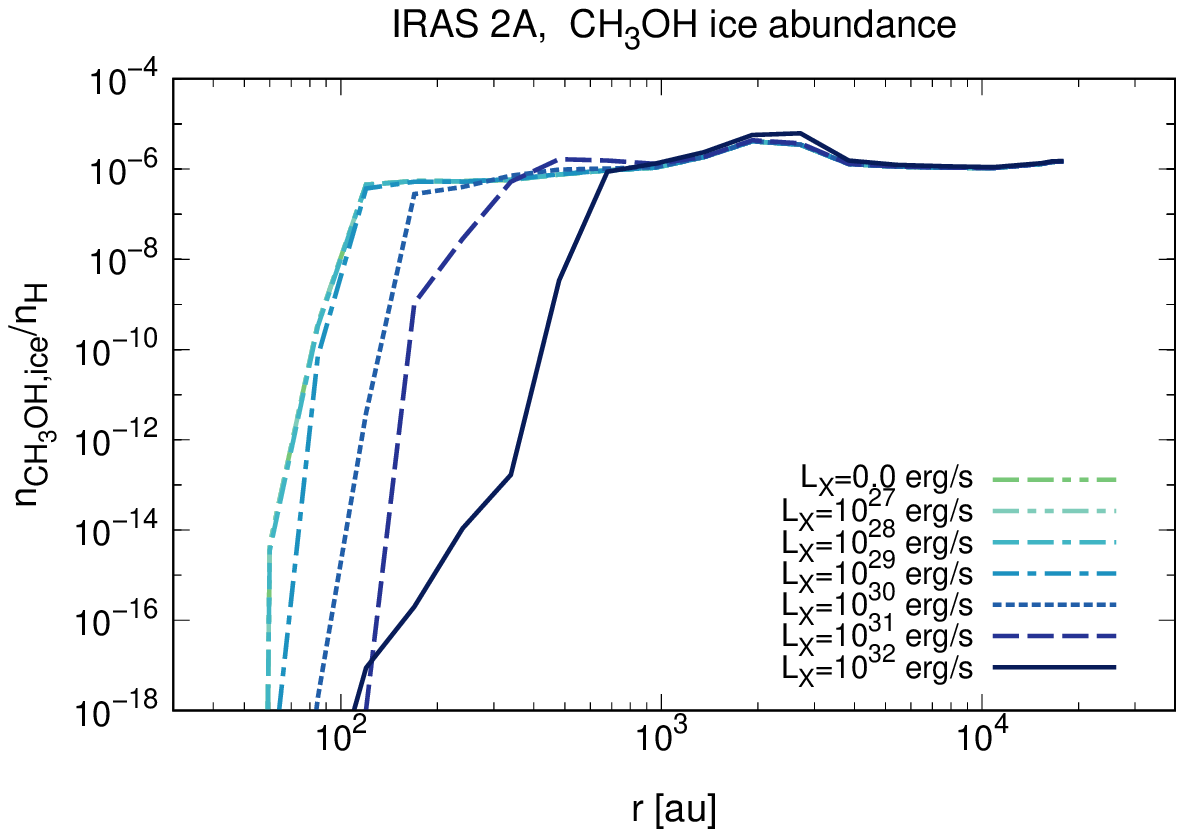}
\includegraphics[scale=0.67]{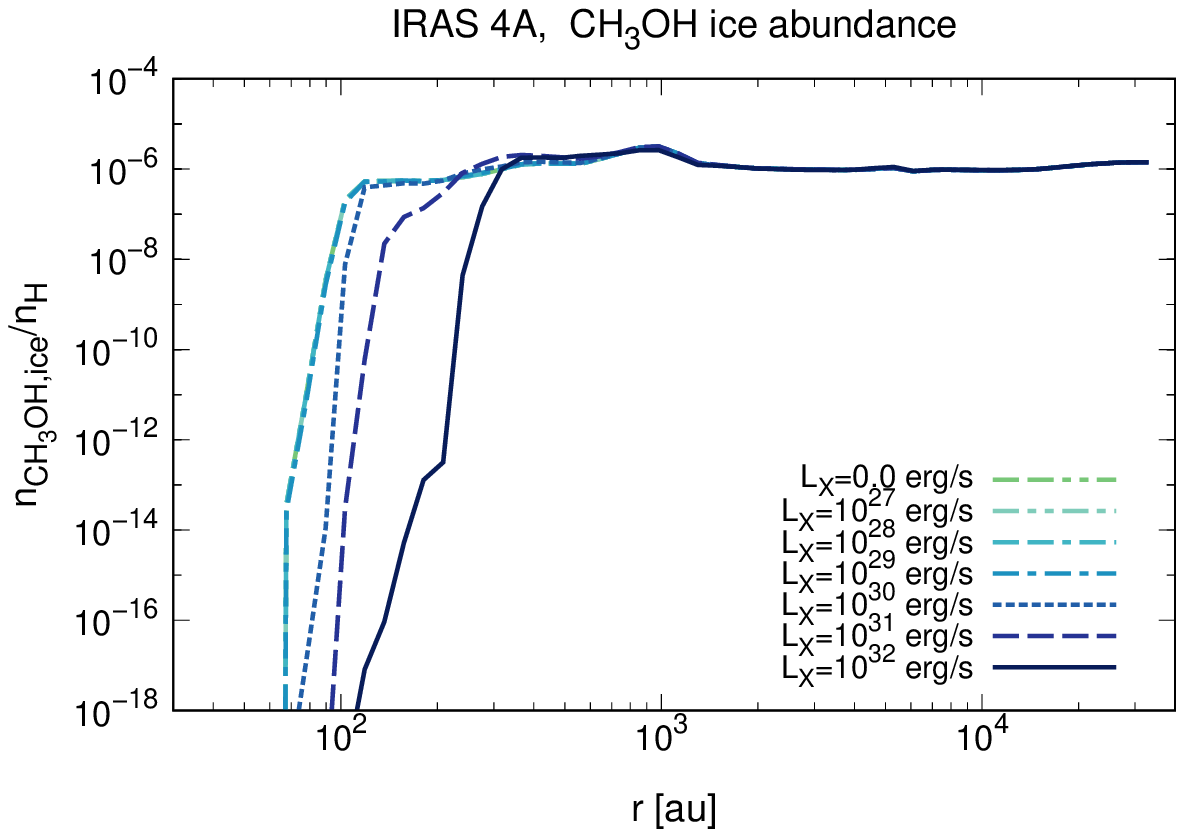}
\end{center}
\caption{
\noindent 
The radial profiles of methanol gas and ice fractional abundances $n_{\mathrm{CH}_{3}\mathrm{OH}}$/$n_{\mathrm{H}}$ (top panels) and $n_{\mathrm{CH}_{3}\mathrm{OH},\mathrm{ice}}$/$n_{\mathrm{H}}$ (bottom panels) in NGC 1333-IRAS 2A (left panels) and NGC 1333-IRAS 4A (right panels) envelope models.
%
}\label{Figure7_CH3OHgas&ice}
\end{figure*} 
Figure \ref{Figure7_CH3OHgas&ice} shows the radial profiles of the methanol gas fractional abundances $n_{\mathrm{CH}_{3}\mathrm{OH}}$/$n_{\mathrm{H}}$ and ice fractional abundances  $n_{\mathrm{CH}_{3}\mathrm{OH},\mathrm{ice}}$/$n_{\mathrm{H}}$ in IRAS 2A (left panel) and IRAS 4A (right panel) envelope models, for the various X-ray luminosities.
According to Table \ref{Table:2}, the binding energy of CH$_{3}$OH is somewhat smaller than that of H$_{2}$O ($E_{\mathrm{des}}$(CH$_{3}$OH)=3820 K and $E_{\mathrm{des}}$(H$_{2}$O)=4880 K), and the CH$_{3}$OH snowline position ($\sim2\times10^{2}$ au) is located outside the water snowline ($\sim10^{2}$ au).
Thus, CH$_{3}$OH has been considered to probe the $\gtrsim100$ K region in hot cores \citep{Nomura2004, Garrod2006, Herbst2009, Taquet2014}, and also provides an outer limit to the water snowline position in protostellar envelopes (e.g., \citealt{Jorgensen2013, vantHoff2018CH3OH, Lee2019, Lee2020}).
\\ \\
The CH$_{3}$OH abundances within $r\lesssim$ 200 au decrease as the values of X-ray luminosities become larger.
Outside the CH$_{3}$OH snowline, CH$_{3}$OH gas abundances are around $10^{-10}-10^{-9}$ with various X-ray fluxes.
Within the CH$_{3}$OH and H$_{2}$O snowlines, for low X-ray luminosities,
the CH$_{3}$OH gas abundances are around $10^{-7}-10^{-6}$.
In contrast, for high X-ray luminosities,
the CH$_{3}$OH gas abundances decrease and reach below $10^{-16}$ inside the water snowline.
\\ \\
According to previous studies (e.g., \citealt{Tielens1982, Watanabe2002, Cuppen2009, Fuchs2009, Drozdovskaya2014, Furuya2014, Walsh2016, Walsh2018, Bosman2018, Aikawa2020}), 
the main pathway to form methanol ice on or within icy mantles of dust grains is CO hydrogenation.
\citet{Drozdovskaya2014} discussed the methanol related chemistry both in gas and ice phases, and gas-phase methanol is supplied by the desorption of CH$_{3}$OH ice.
In our modeling, fragmental X-ray induced photodesorption reactions are included (see Section 2.2.2 of this paper and e.g., \citealt{Bertin2016}), and the photofragments of CH$_{3}$OH (e.g., CH$_{3}$, CH$_{2}$OH, CH$_{3}$O) will lead to larger and more complex molecules with grain-surface reactions (e.g., \citealt{Chuang2016, Drozdovskaya2016}).
The gas-phase production via ion-molecule reactions has been considered not to be efficient \citep{Charnley1992, Garrod2006, Geppert2006}.
In the presence of X-rays, gas-phase CH$_{3}$OH and other complex organic molecules (COMs) are mainly destroyed by X-ray induced photodissociation in the inner envelopes (e.g., \citealt{Garrod2006, Oberg2009b, Drozdovskaya2014, Taquet2016CH3OH}). 
Therefore, CH$_{3}$OH is predicted not to be an efficient tracer of the warm inner envelope and the water snowline position for moderate and high X-ray luminosities.
The X-ray ionisation rates $\xi_{\mathrm{X}}(r)$ where CH$_{3}$OH loses its efficacy as a water snowline tracer are $\gtrsim$ a few $\times10^{-17}$ s$^{-1}$ (see Figure \ref{Figure2_FX}).
%
\subsection{IRAS 4A sub-grid envelope models}
In Appendix B, Figure \ref{FigureB1_H2O&O2&O&OH&HCO+&CH3OHgas_reaction-rev} shows the radial profiles of H$_{2}$O, O$_{2}$, O, OH, HCO$^{+}$, and CH$_{3}$OH gas fractional abundances in the IRAS 4A envelope models, with X-ray luminosities between $L_{\mathrm{X}}=$$10^{30}$ and $10^{31}$ erg s$^{-1}$.
We plot these sub-grid model profiles since there is a large jump in abundances in this X-ray luminosity range (see Figures \ref{Figure3_H2Ogas}-\ref{Figure7_CH3OHgas&ice}).
For the abundance profiles of H$_{2}$O, HCO$^{+}$ and CH$_{3}$OH gas, between $10^{30}$ and $2\times10^{30}$ erg s$^{-1}$ seems to be the clear boundary.
Compared to them, the abundance profiles of O$_{2}$ and O gas gradually increase in the inner region as the values of $L_{\mathrm{X}}$ increase from $10^{30}$ to $\sim6\times10^{30}$ erg s$^{-1}$.
%
\subsection{Fractional abundances of other dominant oxygen, carbon, and nitrogen bearing molecules}
In Figures \ref{FigureC1_CO$_{2}$}, \ref{FigureC2_CO}, \ref{FigureD1_CH4&C2H&HCNgas}, and \ref{FigureE1_NH3&N2gas}, we show the radial fractional abundances of other dominant oxygen, carbon, and nitrogen bearing molecules (CO$_{2}$, CO, CH$_{4}$, C$_{2}$H, HCN, NH$_{3}$, and N$_{2}$) for the various X-ray luminosities.
According to these figures, as the X-ray flux becomes large, the fractional abundances of gas-phase CH$_{4}$, HCN, and NH$_{3}$ decrease within their own snowline positions.
The gas-phase \ce{CO2} abundances increase at $r\gtrsim3\times10^{2}$ au (outside the CO$_{2}$ snowline), as the X-ray fluxes become larger.
At $r\lesssim3\times10^{2}$ au and for low and moderate X-ray luminosities, the CO$_{2}$ gas abundance increases as the X-ray fluxes become larger, and they reach $\sim10^{-5}-10^{-4}$ for moderate X-ray luminosities.
In contrast, they decrease for high X-ray luminosities (below to $<10^{-6}$).
In addition, the radial CO and N$_{2}$ abundance profiles are constant for the various X-ray luminosities, and they are the dominant carbon and nitrogen carries under the strong X-ray fields.
The dependance of radial C$_{2}$H gas fractional abundances on X-ray fluxes are much smaller than other dominant molecules.
In Appendix C, D, and E, more details about their radial abundance profiles are described.
\section{Discussion}
\subsection{Dominant oxygen carriers}
\begin{figure*}
\begin{center}
\includegraphics[scale=1.4]{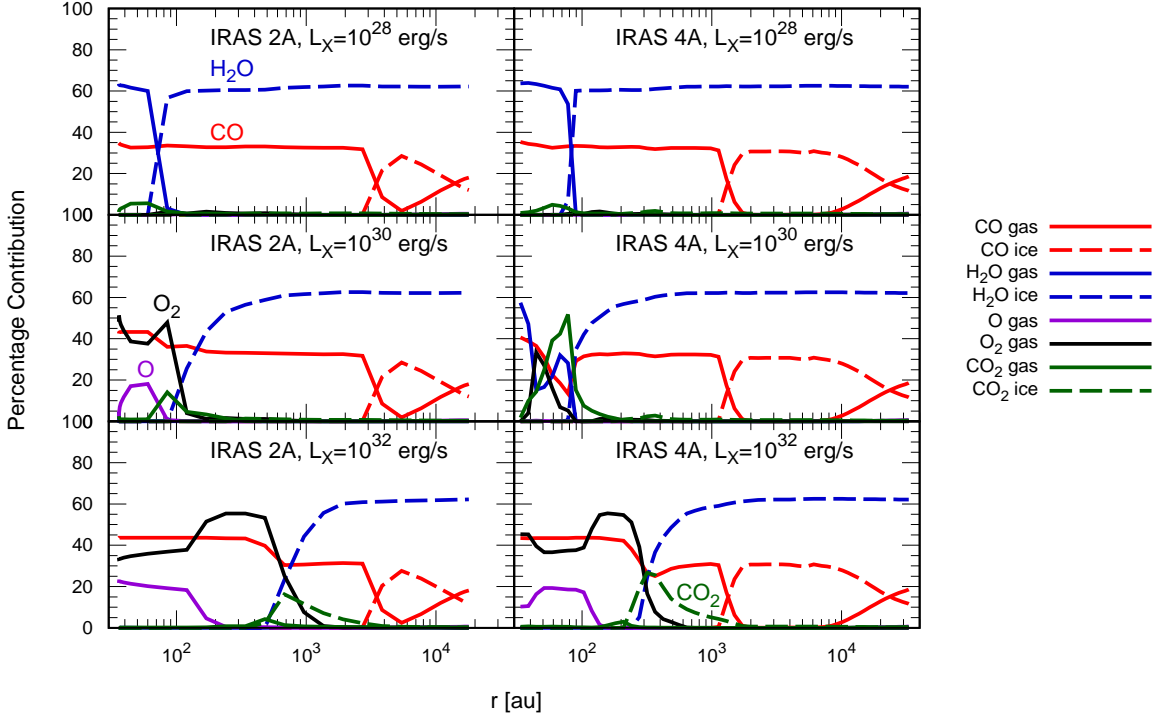}
\end{center}
\caption{
\noindent 
The radial profiles of percentage contributions of the dominant oxygen-bearing molecules to the total elemental oxygen abundance ($=3.2\times10^{-4}$) in the NGC 1333-IRAS 2A envelope model (left panels) and the NGC 1333-IRAS 4A envelope model (right panels).
Top, middle, and bottom panels show the radial profiles with $L_{\mathrm{X}}=10^{28}$, $10^{30}$, and $10^{32}$ erg s$^{-1}$, respectively.
Red, blue, purple, black, and green line profiles show the contribution of CO, H$_{2}$O, O, O$_{2}$, and CO$_{2}$ molecules, respectively.
Solid and dashed line profiles show the contribution of gaseous and icy molecules, respectively.
Since O$_{2}$ and CO$_{2}$ include two oxygen atom per molecule, percentage contributions are twice as much as those of CO, H$_{2}$O, and O when they have same fractional abundances with respect to hydrogen nuclei.
\vspace{0.3cm}
}\label{Figure8_Percentage}
\end{figure*} 
\begin{figure*}
\begin{center}
\includegraphics[scale=0.9]{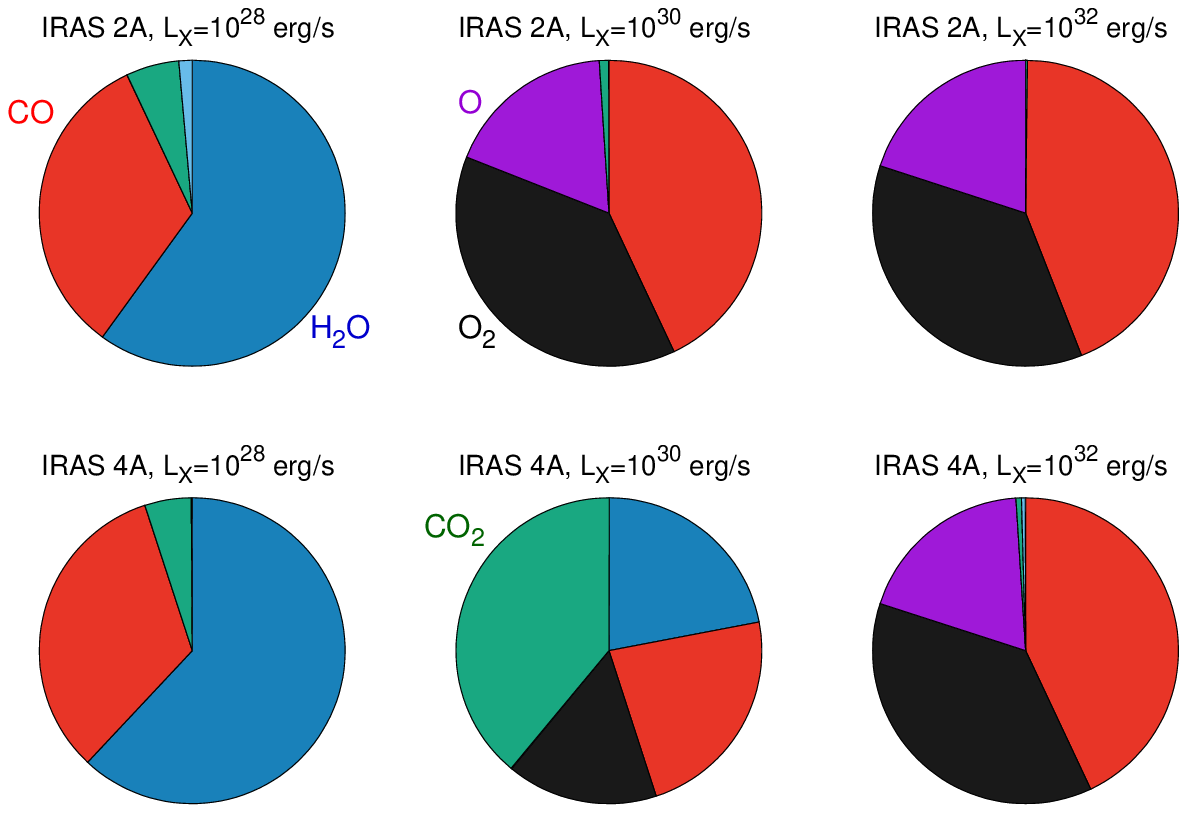}
\end{center}
\caption{
\noindent 
The pie charts of the percentage contributions of the dominant oxygen-bearing molecules to the total elemental oxygen abundance ($=3.2\times10^{-4}$) at $r\sim60$ au ($T_{\mathrm{gas}}\sim150$ K, inside the water snowline) in the NGC 1333-IRAS 2A envelope model (top three charts) and the NGC 1333-IRAS 4A envelope model (bottom three charts).
Left, middle, and right charts show the contributions with $L_{\mathrm{X}}=10^{28}$, $10^{30}$, and $10^{32}$ erg s$^{-1}$, respectively.
Red, dark blue, purple, black, green, and sky blue parts are the contributions of CO, H$_{2}$O, O, O$_{2}$, CO$_{2}$, and other molecules (such as CH$_{3}$OH), respectively.
\vspace{0.3cm}
}\label{Figure9_Percentage_PieChart}
\end{figure*} 
Figure \ref{Figure8_Percentage} shows the radial profiles of percentage contributions of the dominant oxygen-bearing molecules (CO, H$_{2}$O, O, O$_{2}$, and CO$_{2}$) to the total elemental oxygen abundance ($=3.2\times10^{-4}$) in the IRAS 2A envelope model and the IRAS 4A envelope model at the various assumed X-ray luminosities ($L_{\mathrm{X}}=10^{28}$, $10^{30}$, and $10^{32}$ erg s$^{-1}$).
Figure \ref{Figure9_Percentage_PieChart} shows the pie charts of the percentage contributions of the dominant oxygen-bearing molecules at $r\sim60$ au ($T_{\mathrm{gas}}\sim150$ K, inside the water snowline) in the IRAS 2A and IRAS 4A envelope models.
Table \ref{Table:3} in Appendix F shows the fractional abundances of major oxygen bearing molecules at $r=60$ au in the IRAS 2A and IRAS 4A envelope models for the various X-ray luminosities, and their percentage contributions. 
We note that O$_{2}$ and CO$_{2}$ include two oxygen atom per molecule, and thus percentage contributions are twice as much as those of CO, H$_{2}$O, and O when they have same abundances with respect to hydrogen nuclei.
On the basis of Figures \ref{Figure8_Percentage} and \ref{Figure9_Percentage_PieChart}, and Table \ref{Table:3}, for low X-ray luminosities,
H$_{2}$O and CO molecules are the dominant oxygen carriers ($>$90\%), both in the gas and ice. The percentage contributions of H$_{2}$O gas and ice are $\gtrsim$60\%, and that of CO gas and ice are $\gtrsim$30\% throughout the envelopes.
\\ \\
As the X-ray fluxes increase, the abundances of H$_{2}$O gas decrease at $r\lesssim10^{2}$ au, and those of H$_{2}$O ice also decrease just outside the water snowline ($r\gtrsim10^{2}$ au), where the X-ray induced photodesorption is considered to be efficient (see also Section 3.1).
Moreover, as the X-ray fluxes increase, the abundances of O$_{2}$ and O gas increase in the inner envelopes (inside and just outside the water snowline, see also Section 3.2).
For high X-ray luminosities, the water gas abundances at $r\lesssim10^{2}$ au 
become much smaller ($<<10^{-6}$), and O$_{2}$ and O gas are the dominant oxygen carriers along with CO at $r\lesssim$ a few $\times10^{2}$ au.
In these cases, the percentage contributions of O$_{2}$, O, and CO gas at these radii are $\approx$40\%, $\lesssim$20\%, and $\gtrsim$40\%, respectively.
In the outer envelopes where the X-ray induced photodesorption of water is not efficient, H$_{2}$O ice and CO gas and ice molecules are still dominant oxygen carriers.
In addition,
the percentage contributions of CO$_{2}$ gas or ice are around $10-40$\% for moderate X-ray luminosities and around $10-20$\% for high X-ray luminosities in the regions where the contributions of O$_{2}$ gas and H$_{2}$O are similar.
As discussed in Section 3.7 and Appendix C, CO$_{2}$ gas abundances at $r\lesssim3\times10^{2}$ au are highest (up to $\sim10^{-5}-10^{-4}$) for moderate X-ray luminosities.
The outer edge of the region where X-ray induced photodesorption of water is efficient spreads out from $r\sim10^{2}$ au to a few $\times10^{2}$ au as the values of $L_{\mathrm{X}}$ become larger.
\\ \\
On the basis of the our calculations and the discussion above, in order to estimate the total oxygen abundances in the inner envelopes of protostars under the various X-ray luminosities, not only CO and H$_{2}$O line observations, but also O$_{2}$ and O, and CO$_{2}$ line observations are important.
\\ \\
However, as discussed in Section 4.5, O$_{2}$ line observations are very difficult and only $^{16}$O$^{18}$O lines can be observed with ALMA.
The fine structure lines of O are available only at far-infrared wavelengths where dust opacity precludes probing the inner regions in the low-mass protostellar envelopes (see also Section 4.5).
\\ \\
In addition, because of the lack of a permanent dipole moment,
CO$_{2}$ can only be observed using ro-vibrational absorption or emission lines in the near-and mid-infrared wavelengths \citep{Boonman2003, Bosman2017}. 
These lines are included in the wavelengths coverage of James Webb Space Telescope (JWST), and one can probe the CO$_{2}$ abundances in the outer envelopes around low-mass protostars through these line observations with JWST, 
as done for high-mass protostellar envelopes using ISO \citep{vanDishoeck1996, Boonman2003}.
For low-mass protostellar envelopes, a hint of gas-phase CO$_{2}$ lines has been obtained using Spitzer (see e,g., \citealt{Poteet2013}).
We note that high dust opacities in these wavelengths make it difficult to probing the CO$_{2}$ gas abundances directly in the inner envelopes around low-mass protostars.
In Appendix C, the dependance of CO$_{2}$ abundances on X-rays in protostellar envelopes are discussed in detail.
\subsection{Comparison with observations for IRAS 4A}
\begin{figure*}
\begin{center}
\includegraphics[scale=0.67]{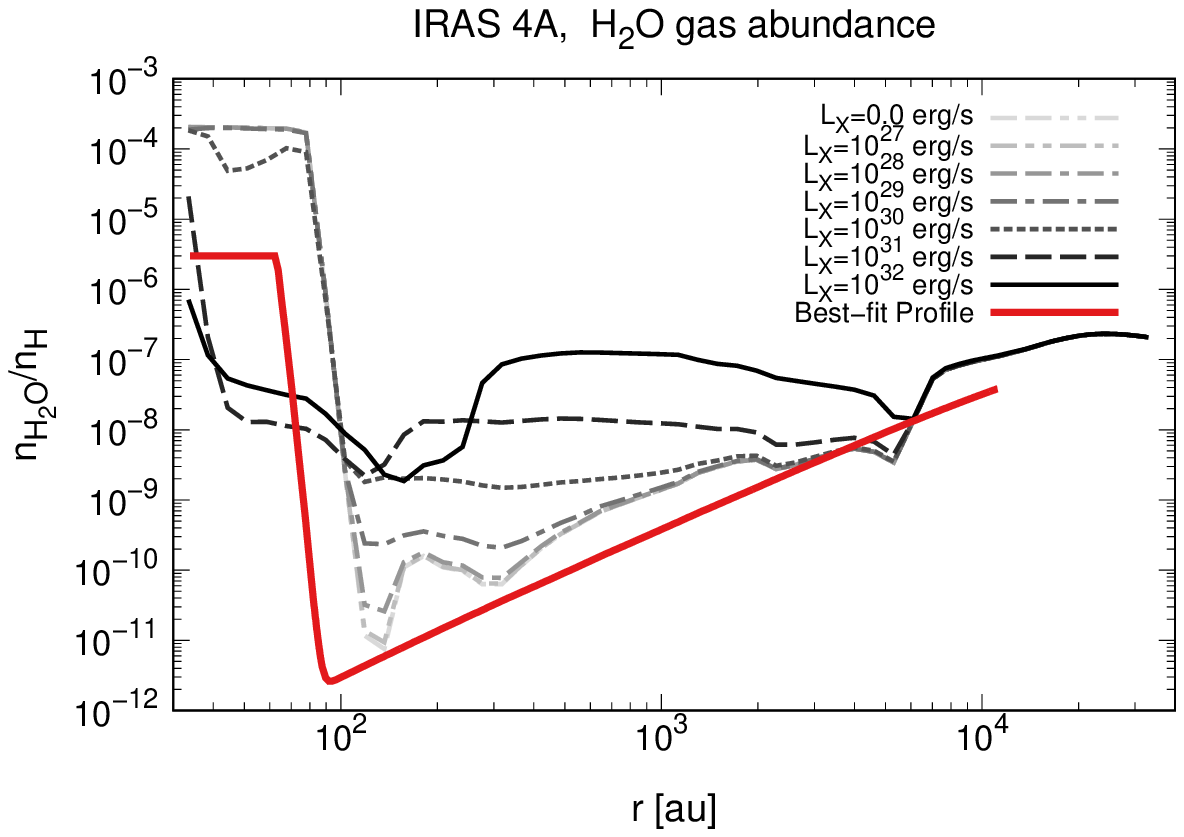}
\includegraphics[scale=0.67]{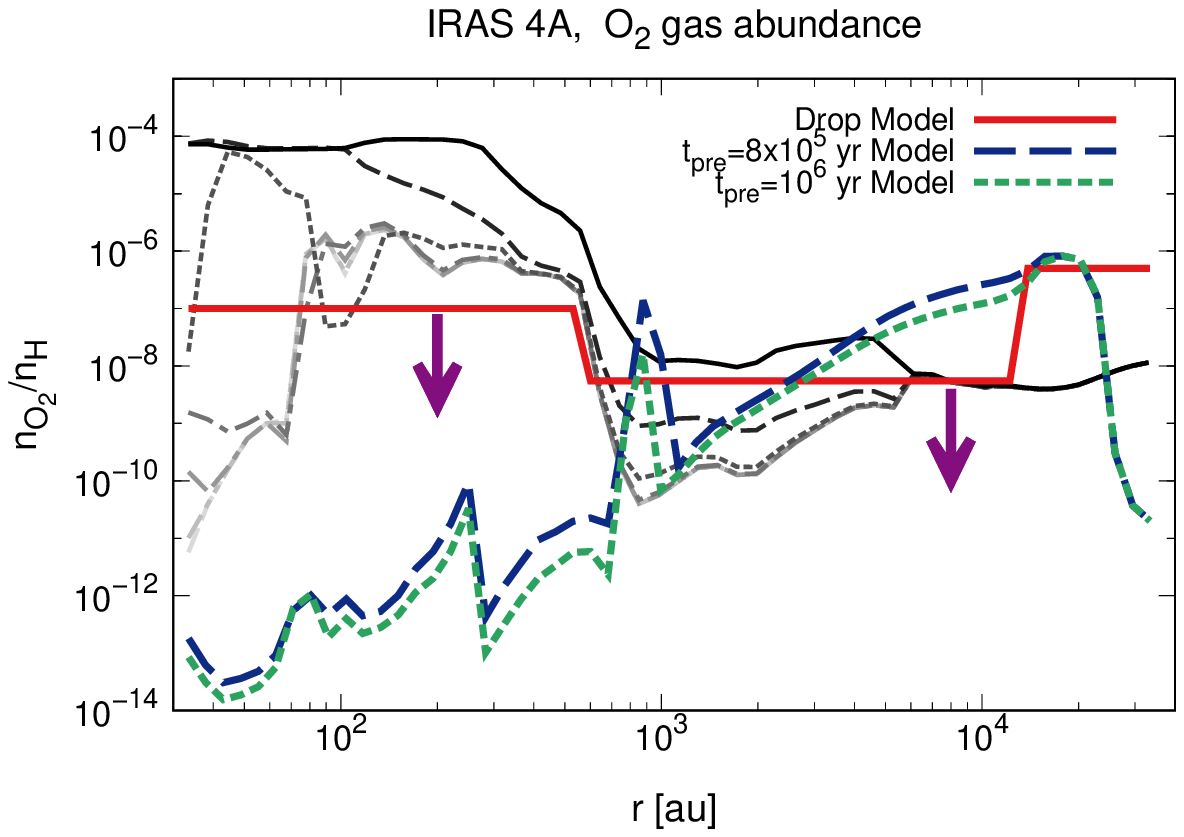}
\end{center}
\caption{
\noindent 
[Left panel]: 
The gray-scale plots (from white to black) are the radial profiles of water gas fractional abundances $n_{\mathrm{H}_{2}\mathrm{O}}$/$n_{\mathrm{H}}$ in the NGC 1333-IRAS 4A envelope model, which are the same as the color plots in the top right panel of Figure \ref{Figure3_H2Ogas}.
%
The red solid line shows the observational best-fit $n_{\mathrm{H}_{2}\mathrm{O}}$/$n_{\mathrm{H}}$ profile in the IRAS 4A envelope, obtained from \citet{vanDishoeck2021}.
This profile is based on analysis of $Herschel$/HIFI spectra which mainly trace the cold outer part \citep{Mottram2013, Schmalzl2014}, with the modification of the inner ($T_\mathrm{gas}>$100 K) water gas abundance from $>10^{-4}$ to 3$\times10^{-6}$ \citep{Persson2016}.
 [Right panel]:
The gray-scale plots (from white to black) are the radial profiles of molecular oxygen gas fractional abundances $n_{\mathrm{O}_{2}}$/$n_{\mathrm{H}}$ in the NGC 1333-IRAS 4A envelope model, which are the same as the color plots in the top right panel of Figure \ref{Figure5_O2&Ogas}.
The model abundance profiles obtained in \citet{Yildiz2013} are over-plotted.
The red solid line shows the drop gaseous O$_{2}$ abundance profile by using the C$^{18}$O modeling \citep{Yildiz2012} and assuming O$_{2}$ follows the same freeze-out and sublimation processes as C$^{18}$O. 
The blue dashed and the green dotted lines show the gaseous O$_{2}$ abundance profiles via their gas-grain modeling with $t_\mathrm{pre}=8\times10^{5}$ years and $t_\mathrm{pre}=10^{6}$ years, respectively.
The purple arrows are to indicate that \citet{Yildiz2013} only obtained the upper limit O$_{2}$ gas abundance for this object.
%
}\label{Figure10_H2O&O2gas_obs-plot}
\end{figure*} 
\begin{figure*}
\begin{center}
%
\includegraphics[scale=0.67]{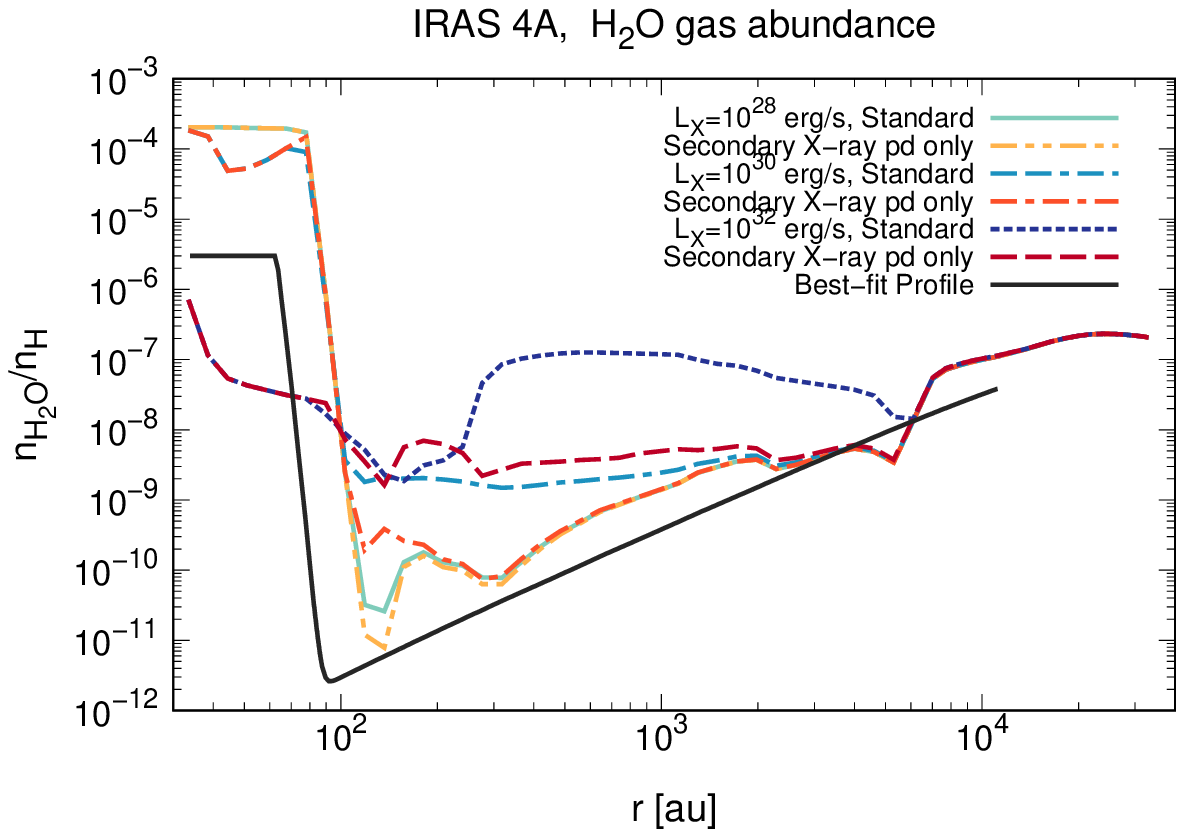}
\includegraphics[scale=0.67]{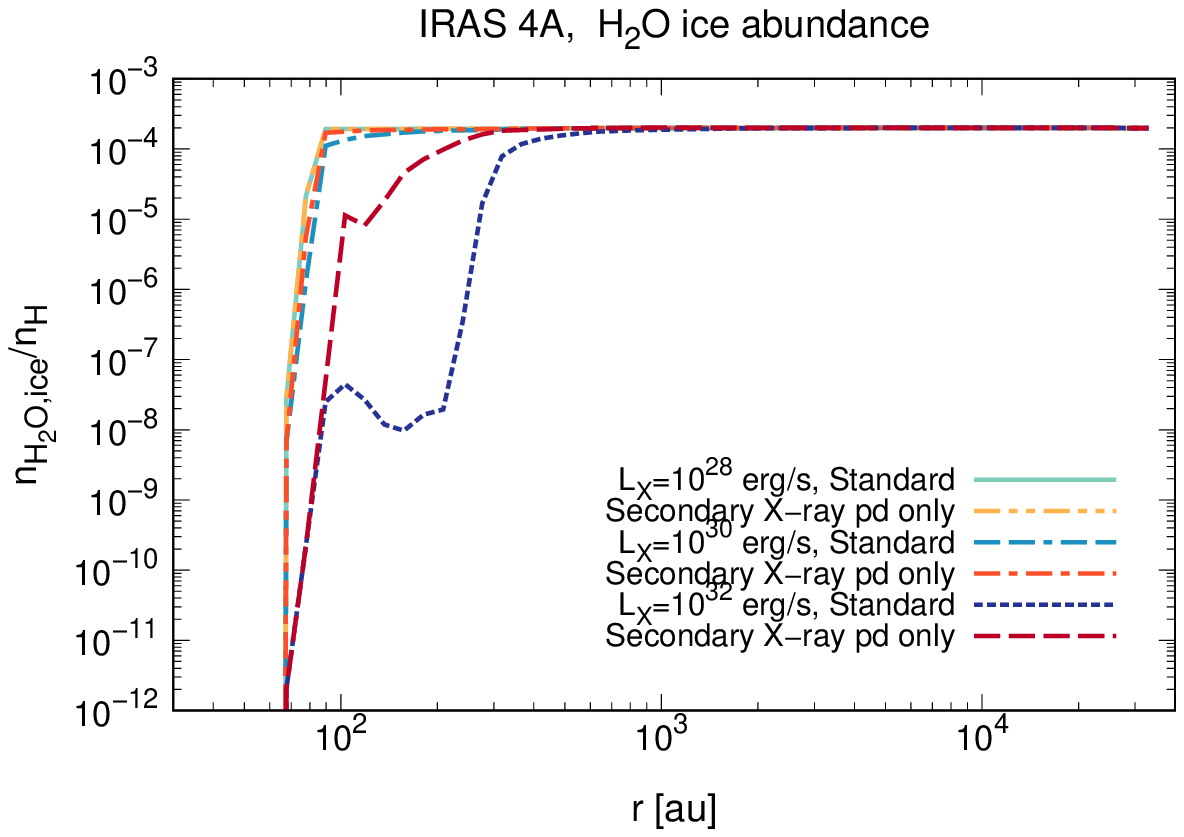}
\includegraphics[scale=0.67]{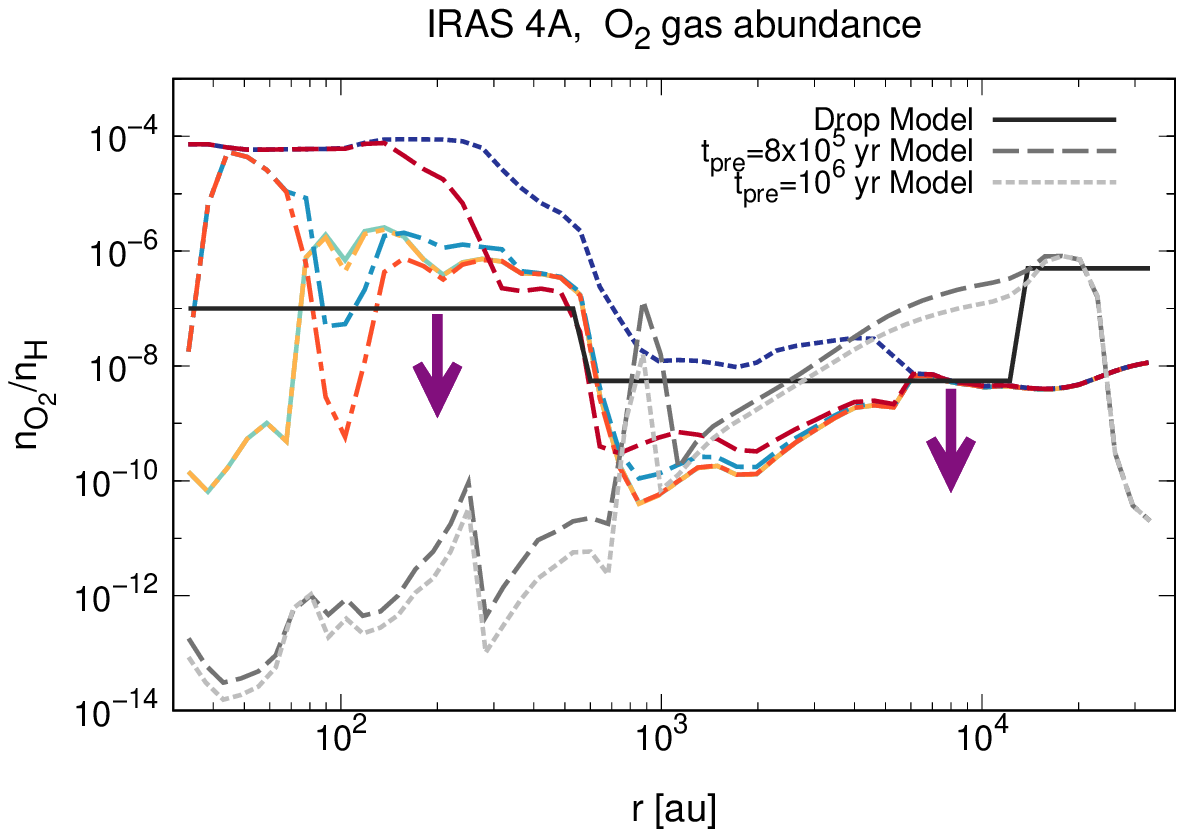}
\includegraphics[scale=0.67]{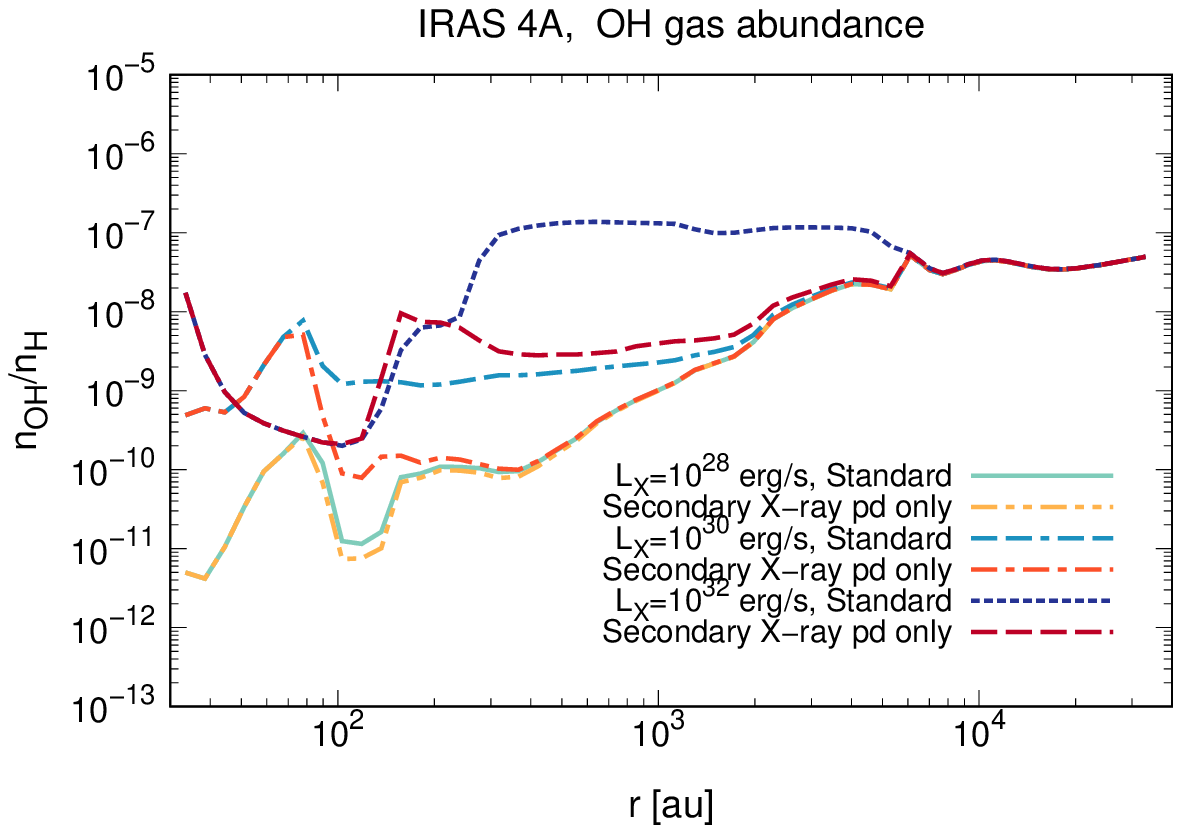}
%
\end{center}
\caption{
\noindent 
The radial profiles of gaseous fractional abundances 
of H$_{2}$O, O$_{2}$, and OH,
and icy fractional abundances of H$_{2}$O
%
in the NGC 1333-IRAS 4A envelope models.
The light-green solid lines, the cyan double-dashed dotted lines, and blue dotted lines show the radial profiles of our standard model calculations, for values of central star X-ray luminosities
$L_{\mathrm{X}}$=10$^{28}$, 10$^{30}$, and 10$^{32}$ erg s$^{-1}$, respectively (see also Figures \ref{Figure3_H2Ogas}, \ref{Figure4_H2Oice}, \ref{Figure5_O2&Ogas}, \ref{Figure6_HCO+&OHgas}).
The yellow dashed double-dotted lines, the scarlet dashed dotted lines, and the brown dashed lines show the radial profiles of our additional model calculations for $L_{\mathrm{X}}$=10$^{28}$, 10$^{30}$, and 10$^{32}$ erg s$^{-1}$, respectively.
In the additional model calculations, we include the photodesorption by UV photons generated internally via the interaction of secondary electrons produced by X-rays with H$_{2}$ molecules (the secondary (indirect) X-ray induced photodesorption, see Figures \ref{FigureG1_add_Xp-rev1_model-comparisons} and \ref{FigureG2_add_Xp-rev1_model-comparisons}), and we also switch off the direct X-ray induced photodesorption 
(see also Figure \ref{FigureG3_add_Xp-rev2_model-comparisons}). 
In the top left panel, the observational best-fit H$_{2}$O gas abundance profile obtained in \citet{vanDishoeck2021} is over-plotted with the black solid line (see also Figure \ref{Figure10_H2O&O2gas_obs-plot}).
In the bottom left panel, the three model O$_{2}$ gas abundance profiles obtained in \citet{Yildiz2013} are over-plotted (see also Figure \ref{Figure10_H2O&O2gas_obs-plot}).
}\label{Figure11_Xp-rev2_model-comparisons}
\end{figure*} 
%
In the left panel of Figure \ref{Figure10_H2O&O2gas_obs-plot}, the observational best-fit $n_{\mathrm{H}_{2}\mathrm{O}}$/$n_{\mathrm{H}}$ profile in the IRAS 4A envelope, obtained from \citet{vanDishoeck2021}, is overplotted on our model profiles (see also Section 3.1 and the top right panel of Figure \ref{Figure3_H2Ogas}). 
This profile is based on analysis of $Herschel$/HIFI spectra which mainly trace the cold outer part \citep{Mottram2013, Schmalzl2014}, with the modification of the inner ($T_\mathrm{gas}>$100 K) water gas abundance from $>10^{-4}$ to 3$\times10^{-6}$ \citep{Persson2012, Persson2014, Persson2016}.
In the cold outer part of the envelope ($T_{\mathrm{gas}}<10^{2}$ K, $r>10^{2}$ au), the best-fit profile is consistent with our model profiles for $L_{\mathrm{X}}\lesssim10^{28}$ erg s$^{-1}$.
In contrast, in the inner warm envelopes ($T_{\mathrm{gas}}\gtrsim150$ K, $r\lesssim60$ au), the gaseous water abundance in the best-fit profile is 3$\times10^{-6}$ \citep{Persson2016}, 
which suggests the possibility of efficient X-ray induced water destructions of gas-phase water molecules with $L_{\mathrm{X}}\gtrsim10^{30}$ erg s$^{-1}$ in these regions.
The reason for this discrepancy
between the inner and outer envelope is not clear.
\\ \\
In the cold outer part of the envelope, X-ray induced photodesorption of water molecules controls the water gas abundance.
Therefore, if the rates of X-ray induced photodesorption of water are much lower (e.g., $Y_{\mathrm{des}}$(H$_{2}$O)$\lesssim10^{-5}$ molecules photon$^{-1}$) than our adopted values, the water gas abundance profiles in the cold outer part of the envelope for $L_{\mathrm{X}}\gtrsim10^{30}$ erg s$^{-1}$ 
are expected to be more similar to
the observational profile. 
In Section 4.3, we discuss the rates of X-ray induced photodesorption in detail, with conducting additional model calculations.
\\ \\
%
The 3$\sigma$ upper limit O$_{2}$ gas abundance with respect to H$_{2}$ obtained by \citet{Yildiz2013} is $\leq6\times10^{-9}$ towards the entire envelope of IRAS 4A using $Herschel$/HIFI.
They estimated that the observed O$_{2}$ gas abundance cannot be more than $10^{-6}$ for the inner warm region ($r\lesssim10^{2}$ au). 
In the right hand panel of Figure \ref{Figure10_H2O&O2gas_obs-plot}, the three model abundance profiles calculated in \citet{Yildiz2013} are over-plotted on our model profiles (see also Section 3.2 and the top right panel of Figure \ref{Figure5_O2&Ogas}).
The black solid line shows the drop gaseous O$_{2}$ abundance profile 
obtained using the best-fit CO abundance profile produced from the observed C$^{18}$O line emission \citep{Yildiz2012}
%
and assuming O$_{2}$ has the same snowline position as CO ($=$ a constant O$_{2}$/CO abundance ratio).
The blue dashed and the green dotted lines show the gaseous O$_{2}$ abundance profiles via their gas-grain modeling with different pre-collapse lifetimes of $t_\mathrm{pre}=8\times10^{5}$ years and $t_\mathrm{pre}=10^{6}$ years, respectively.
As the basis for their gas-grain chemical network \citep{Yildiz2013}, the Ohio State University (OSU) gas-grain network \citet{Garrod2008} is used, which also included gas phase reactions, grain surface reactions, and thermal and non-thermal gas-grain interactions. Although the X-ray induced reactions were not contained, the cosmic-ray induced reactions were included in their calculations.
\\ \\
These three model profiles are consistent with the above observational upper limit in \citet{Yildiz2013} (the peak temperatures are similar between models and the observation), and within the values of our chemical modeling at $r\gtrsim10^{3}$ au. 
At $r\sim10^{2}-6\times10^{2}$ au, the gaseous O$_{2}$ abundance limit in the drop model is $10^{-7}$, which is similar to the values in our model profiles for $L_{\mathrm{X}}\lesssim10^{30}$ erg s$^{-1}$ within one order of magnitude. 
In contrast, at $r<6\times10^{2}$ au, the gaseous O$_{2}$ abundance in their gas-grain modeling are much smaller ($<<10^{-10}$) than those in our model profiles and that in the drop model.
\\ \\
We note that $L_{\mathrm{X}}$ of IRAS 4A is suggested to be $\gtrsim10^{30}$ erg s$^{-1}$ by comparing the results of our model calculations and the observationally estimated inner water gas abundances towards IRAS 4A (see discussions above).
Thus, on the basis of the discussions about H$_{2}$O and O$_{2}$ in this subsection, $L_{\mathrm{X}}$ of IRAS 4A is suggested to be around $10^{30}$ erg s$^{-1}$, although the discrepancy of suggested $L_{\mathrm{X}}$ between the inner and outer envelope discussed above is still remained (see also Section 4.3).
Since probing the O and CO$_{2}$ gas abundances in the inner envelopes are also difficult (see Sections 4.1 and 4.5, and Appendix C), observationally obtaining the abundance profiles of other tracers, especially HCO$^{+}$ and CH$_{3}$OH, is important to investigate the effects of X-ray induced chemistry and confine the values of $L_{\mathrm{X}}$ (see Section 4.6).
\subsection{The rates of X-ray induced photodesorption}
In our standard model calculations, we do not include the photodesorption by UV photons generated internally via the interaction of secondary electrons produced by X-rays with H$_{2}$ molecules, although we include X-ray induced photodissociation on dust-grains (see also Sections 2.2.2 and 2.2.3). 
Our adopted rates of X-ray induced photodesorption are an approximation based on the UV photodesorption rates (see also Section 2.2.2), since experimental constraints for X-ray induced photodesorption are limited (e.g., \citealt{Walsh2014}).
Recently, \citet{Dupuy2018} experimentally investigated X-ray induced photodesorption rates of H$_{2}$O, O$_{2}$, and other related molecules.
According to their experiments, photodesorption yields of H$_{2}$O and O$_{2}$ at 0.55 keV for a compact amorphous solid water ice at 15 K are $3.4\times10^{-3}$ and $4.0\times10^{-4}$ molecules photon$^{-1}$
The differences of these values and our adopted values are within a factor of a few (see Table \ref{Table:1}).
\\ \\
In addition, \citet{Dupuy2018} simply extrapolated X-ray photodesorption yields for higher X-ray photon energies using the absorption cross-sections of water gas.
They estimated that the yields of H$_{2}$O at 15 K would be $\lesssim10^{-4}$ at $>3$ keV, although further experimental studies will be needed to obtain accurate values.
If we estimate the local average X-ray induced photodesorption yields (by multiplying the energy-dependent photodesorption yields by the local X-ray spectrum) on the basis of their simply extrapolated results, the yields would become lower in the outer envelope with larger values of $N_{\mathrm{H}}$, where softer X-rays are more attenuated.  
\\ \\
Here we conduct the two types of additional calculations which focus on the rates of X-ray induced photodesorption and their effect on the chemistry.
In the first additional model (see Figures \ref{FigureG1_add_Xp-rev1_model-comparisons} and \ref{FigureG2_add_Xp-rev1_model-comparisons} in Appendix G), 
we include the photodesorption by UV photons generated internally via the interaction of secondary electrons produced by ``X-rays" with H$_{2}$ molecules.
We scale the cosmic-ray-induced photon flux (10$^{4}$ photons cm$^{-2}$ s$^{-1}$, \citealt{Walsh2014}) by the total ionisation rate (cosmic rays plus X-rays) relative to the cosmic-ray ionisation rate only, and use
the revised value in estimating the photodesorption rates.
According to our calculations (see Figures \ref{FigureG1_add_Xp-rev1_model-comparisons} and \ref{FigureG2_add_Xp-rev1_model-comparisons} in Appendix G), the effects of such additional secondary (indirect) X-ray induced photodesorption is marginal (the abundances are changed by <1\%). 
The direct photodesorption by X-ray photons is the dominant process in our calculations.
\\ \\
Next, in the second additional model (see Figures \ref{Figure11_Xp-rev2_model-comparisons} and \ref{FigureG3_add_Xp-rev2_model-comparisons}),
we switch off the direct X-ray induced photodesorption and include the secondary (indirect) X-ray induced photodesorption only. 
Through this calculation, we can also investigate the impact when the rates of the direct X-ray induced photodesorption are much smaller than our originally adopted values. 
As is seen in Figures \ref{Figure11_Xp-rev2_model-comparisons} and \ref{FigureG3_add_Xp-rev2_model-comparisons}, the effects of X-ray induced photodesorption are decreased relative to the previous case. 
In the case of H$_{2}$O, the gas-phase abundances outside the water snowline for $L_{\mathrm{X}}\gtrsim10^{30}$ erg s$^{-1}$ become around two orders of magnitude smaller than those in our standard model and the first additional model.
In addition, in the outer part of the envelope, the observational best-fit profile of H$_{2}$O gas is now roughly consistent with the models with $L_{\mathrm{X}}\lesssim10^{30}$ erg s$^{-1}$. 
We note that in cases of our standard model calculations (see Section 4.2),
the best-fit profile is consistent with model profiles for $L_{\mathrm{X}}\lesssim10^{28}$ erg s$^{-1}$ in the outer region, whereas with those for $L_{\mathrm{X}}\gtrsim10^{30}$ erg s$^{-1}$ in the inner region. 
Thus, lower rates (e.g., $Y_{\mathrm{des}}$(H$_{2}$O)$\lesssim10^{-5}$ molecules photon$^{-1}$) of direct X-ray induced photodesorption 
bring the models more in line with the observed abundance profiles, which calls into question whether the direct X-ray desorption rates are over-estimated.
We note that the OH gas abundances at $r\sim10^{2}-10^{4}$ au for $L_{\mathrm{X}}\gtrsim10^{30}$ erg s$^{-1}$ become $1-2$ orders of magnitude smaller than those in our standard model and the first additional model, since 
OH is efficiently produced by X-ray induced photodissociation of H$_{2}$O gas and fragmental photodesorption of H$_{2}$O ice (see also Section 3.4).
In addition, O$_{2}$ gas abundances at $r\sim10^{2}-10^{4}$ au for $L_{\mathrm{X}}\gtrsim10^{30}$ erg s$^{-1}$ become around several to ten times smaller,
since O$_{2}$ is formed in the gas-phase of O and OH (see Section 3.2).
For other molecules shown in Figure \ref{FigureG3_add_Xp-rev2_model-comparisons} in Appendix G, the differences in abundances between the standard model and the second additional model are much smaller than those in H$_{2}$O, OH, and O$_{2}$.
\\ \\
Future experimental and theoretical studies over a wider X-ray energy range are
needed to understand how X-ray induced photodesorption rates behave as a function of the X-ray energy spectrum.
X-ray induced photodesorption yields are also expected to vary for different ice composition.
\citet{Basalgete2021a, Basalgete2021b} recently investigated the X-ray induced photodesorption yields of CH$_{3}$OH experimentally, and they estimated that the intact yields for mixed methanol-water ices would be more than around two orders of magnitude smaller than those for pure methanol ices.
\subsection{Model assumptions for chemistry}
\begin{figure*}
\begin{center}
%
\includegraphics[scale=0.67]{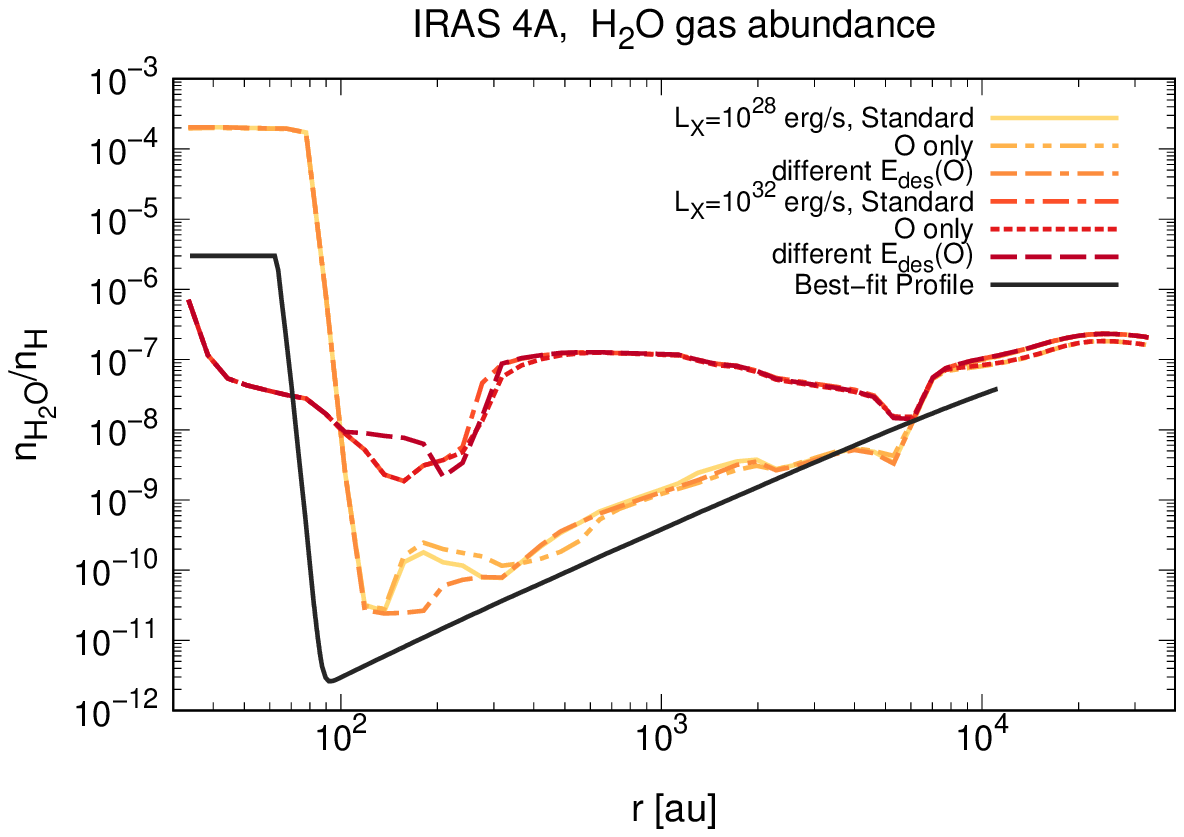}
\includegraphics[scale=0.67]{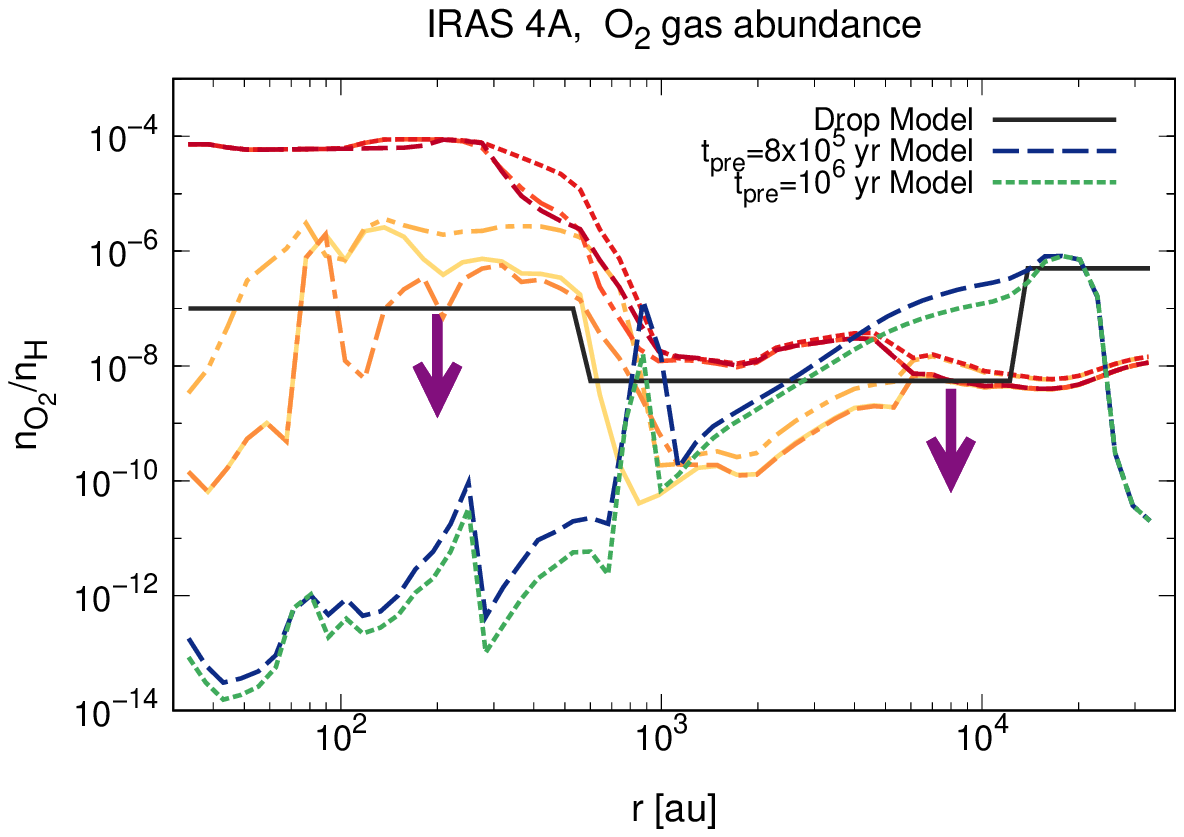}
\includegraphics[scale=0.67]{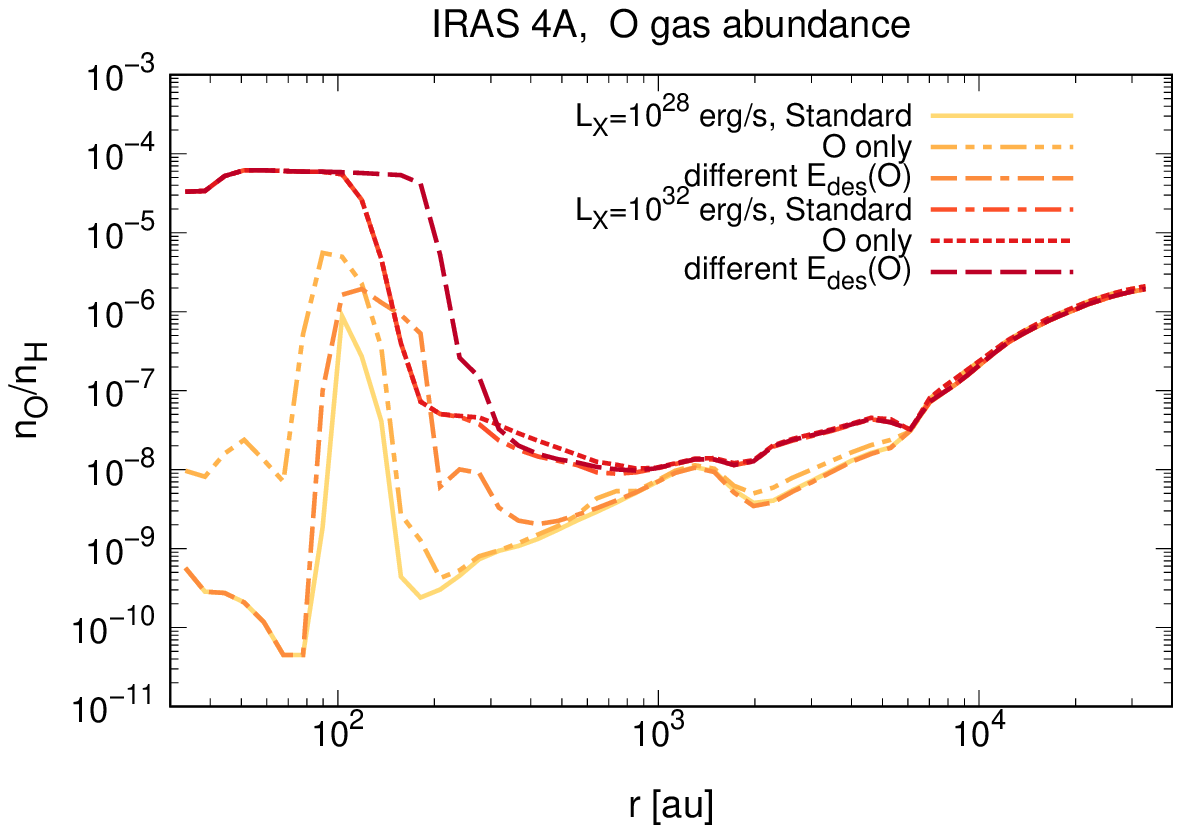}
\includegraphics[scale=0.67]{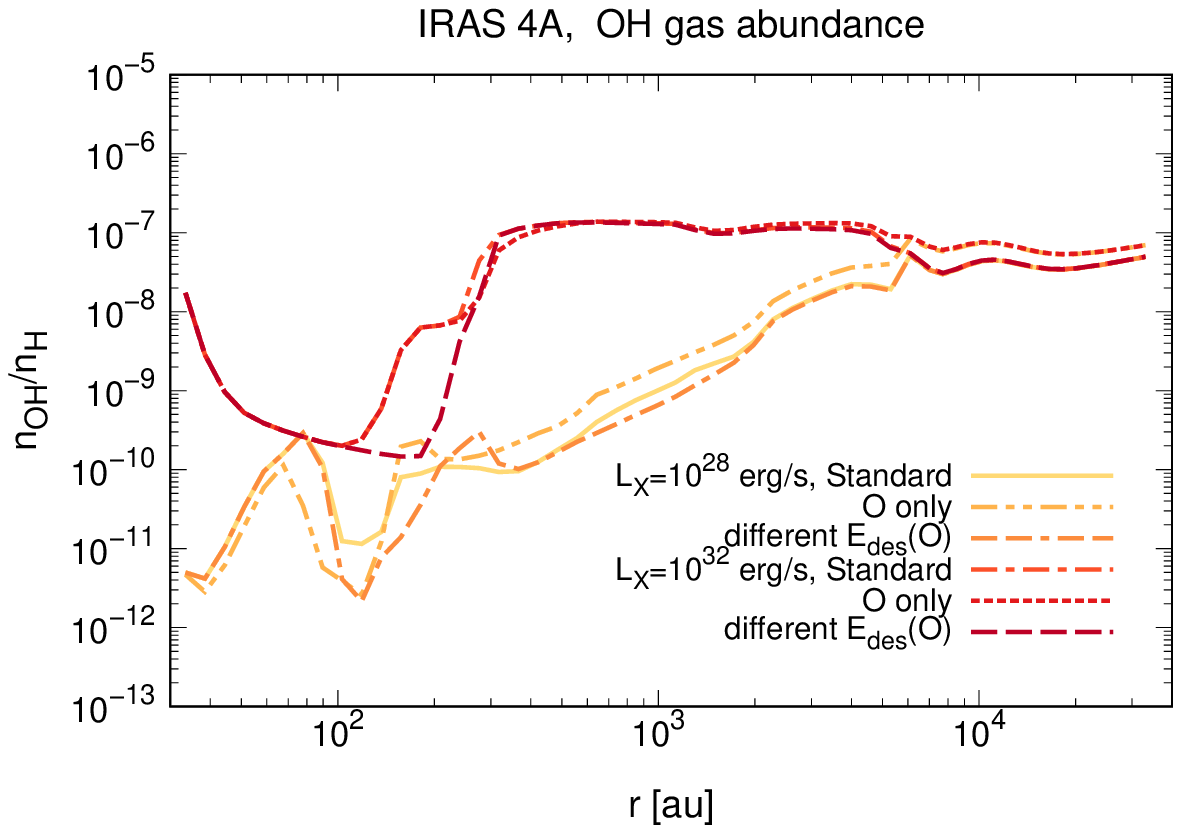}
\end{center}
\caption{
\noindent 
The radial profiles of gaseous fractional abundances of H$_{2}$O, O$_{2}$, O, and OH
in NGC 1333-IRAS 4A envelope models.
The yellow solid lines and scarlet dashed dotted lines show the radial profiles of our standard model calculations, for values of central star X-ray luminosities $L_{\mathrm{X}}$=10$^{28}$ and 10$^{32}$ erg s$^{-1}$, respectively (see also Figures \ref{Figure3_H2Ogas}, \ref{Figure5_O2&Ogas}, and \ref{Figure6_HCO+&OHgas}).
The orange dashed double-dotted lines and red dotted lines show those of our test calculations assuming that the product of H$_{2}$O photodissociation is 100\% atomic oxygen, unlike our standard model (100\% OH) (see also Figure \ref{FigureH1_H2O&O2&O&OHgas_reaction-rev}).
The orange double-dashed dotted lines and brown dashed lines show those of our test calculations assuming the smaller $E_{\mathrm{des}}(j)$ for atomic oxygen ($=800$ K) than that in our fiducial model (=1660 K) (see also Figure \ref{FigureH2_H2O&O2&O&OHgas_Edes-O-rev}).
In the top left panel, the observational best-fit H$_{2}$O gas abundance profile obtained in \citet{vanDishoeck2021} is over-plotted with the black solid line (see also Figure \ref{Figure10_H2O&O2gas_obs-plot}).
In the top right panel, the three model O$_{2}$ gas abundance profiles obtained in \citet{Yildiz2013} are over-plotted (see also Figure \ref{Figure10_H2O&O2gas_obs-plot}).
}\label{Figure12_H2O&O2&O&OHgas_model-comparisons}
\end{figure*} 
In our adopted chemical reaction network (see Section 2), the X-ray induced and cosmic-ray-induced photodissociation of H$_{2}$O in both gas and ice is the following route \citep{McElroy2013};
\begin{equation}
\label{reaction14}
\mathrm{H}_{2}\mathrm{O}+h\nu\rightarrow\mathrm{OH}+\mathrm{H}.
\end{equation}
According to \citet{vanHarrevelt2008} and \citet{Heays2017}, however, there is another route of photodissociation of H$_{2}$O;
\begin{equation}
\label{reaction15}
\mathrm{H}_{2}\mathrm{O}+h\nu\rightarrow\mathrm{O}+\mathrm{H}_{2}/2\mathrm{H}.
\end{equation}
\citet{Heays2017} calculated the photodissociation branching ratio of H$_{2}$O using various radiation fields as OH/O$\sim3$, although it depends on wavelength.
\\ \\
In order to investigate the impacts of this photodissociation branching ratio on the abundances of H$_{2}$O and related molecules, in Figures \ref{Figure12_H2O&O2&O&OHgas_model-comparisons} and \ref{FigureH1_H2O&O2&O&OHgas_reaction-rev} (see Appendix H) we show the gas-phase abundance profiles of H$_{2}$O, O$_{2}$, O, and OH, which are calculated assuming the extreme case that the product of H$_{2}$O photodissociation is 100\% atomic oxygen (Reaction \ref{reaction15} only both in the gas and ice), and compare them with the profiles in our standard models (see also Figures \ref{Figure3_H2Ogas}, \ref{Figure5_O2&Ogas}, and \ref{Figure6_HCO+&OHgas}).
Since the photodissociation of H$_{2}$O is important just inside the snowline (see Section 3.1), the abundances around $r\approx10^{2}$ au are mainly affected.
The OH gas abundances become smaller by a factor of a few at $r\lesssim10^{2}$ au than those in our standard model.
For $L_{\mathrm{X}}\lesssim10^{30}$ erg s$^{-1}$, the O and O$_{2}$ gas abundances increase by around $1-2$ orders of magnitude at $r\lesssim10^{2}$ au compared with those in our standard model.
For $L_{\mathrm{X}}\gtrsim10^{31}$ erg s$^{-1}$ where O and O$_{2}$ gas become dominant oxygen carries, those abundances are similar between these different models.
Outside the water snowline $r>10^{2}$ au, these abundances are almost unchanged.
\\ \\
Except atomic oxygen, the differences of binding energies $E_{\mathrm{des}}(j)$ for dominant molecules between our adopted values and those in recent other chemical modeling are up to around several tens of percent (see Table \ref{Table:2} and e.g., \citealt{Cuppen2017, Penteado2017}).
In our adopted chemical model, $E_{\mathrm{des}}$(O) is assumed to be 1660 K, on the basis of recent experimental measurements \citep{He2014, He&Vidali2014}.
\citet{Taquet2016} and \citet{Eistrup2019} adopted similar value of $E_{\mathrm{des}}$(O).
However, some of other recent chemical models (e.g., \citealt{Eistrup2016, Bosman2018}) adopted the older estimated value of 800 K, which is around two times smaller than our adopted values.
This value of 800 K has been widely used in many chemical models (e.g., \citealt{Tielens1982, Hasegawa1992, Garrod2011}), but \citet{He2014} pointed out that it has no strong theoretical derivation and no experimental confirmation. 
\\ \\
In order to investigate the impacts of different binding energies of O on molecular abundances, in Figures \ref{Figure12_H2O&O2&O&OHgas_model-comparisons} and \ref{FigureH2_H2O&O2&O&OHgas_Edes-O-rev} (see Appendix H)
we show the gas-phase abundance profiles of H$_{2}$O, O$_{2}$, O, and OH, which are obtained from our test calculations assuming the smaller $E_{\mathrm{des}}$(O) ($=800$ K) than that in our fiducial model ($=1660$ K),  and compare them with the profiles in our standard models (see also Figures \ref{Figure3_H2Ogas}, \ref{Figure5_O2&Ogas}, and \ref{Figure6_HCO+&OHgas}).
The O gas abundances become larger by up to one order of magnitude at $r\sim(1-4)\times10^{2}$ au than those in our standard model.
In contrast, only for $L_{\mathrm{X}}\lesssim10^{28}$ erg s$^{-1}$ the O$_{2}$ gas abundances become smaller by factor of a few at $r\sim(1-4)\times10^{2}$ au than those in our standard model.
The OH and H$_{2}$O gas abundances are similar to those in our standard model.
\subsection{H$_{2}$O, O$_{2}$, and O line observations for other protostars}
%
On the basis of previous PdBI (NOEMA) and $Herschel$/HIFI observations \citep{Persson2012, Persson2014, Persson2016, Visser2013}, the water gas abundance is around 10$^{-4}$ in the inner warm envelope and disk of IRAS 2A.
In contrast, the water gas abundances in the inner envelopes and disks of IRAS 4A and 4B are lower by $1-3$ orders of magnitude.
%
Through recent ALMA observations, \citet{Jensen2019} estimated that 
the H$_{2}$$^{18}$O column densities in the warm inner envelopes around three isolated low-mass Class 0 protostars (L483, B335 and BHR71-IRS1) are around a few $\times10^{15}$ cm$^{-2}$, which is similar to that of IRAS 4B, and around 10 times smaller than that of IRAS 2A \citep{Persson2014}.
Compared with the results of our chemical modeling, the observational results of IRAS 2A are consistent with the profiles with $L_{\mathrm{X}}\lesssim10^{29}$ erg s$^{-1}$.
In contrast, the reported inner water gas abundances for IRAS 4A, IRAS 4B, and other objects ($\sim 10^{1}-10^{3}$ times lower than that for IRAS 2A) are close to the values of our chemical modeling with $L_{\mathrm{X}}\gtrsim10^{30}-10^{31}$ erg s$^{-1}$.
Thus, X-ray induced destruction processes can explain the lower water abundances in the inner envelopes of these objects. 
\\ \\
\citet{Harsono2020} show that water vapor is not abundant in the warm envelopes and disks around Class I protostars, and upper limit values of the water gas abundance averaged over the inner warm disk with $T_{\mathrm{gas}}$$>100$ K are $\sim10^{-7}-10^{-5}$.
These lower water gas abundances might also be caused by efficient water gas destruction through X-ray induced chemistry, in addition to locking up water in icy dust grains.
Future detailed water line observations using e.g., ALMA for more Class 0 and I objects with various X-ray luminosities, and chemical modeling with detailed physical structure models (e.g., disk$+$envelope, see also \citealt{vanDishoeck2014, vanDishoeck2021, Furuya2017}) will clarify the effects of X-ray induced chemistry on water and other related molecules and the water trail from protostellar envelopes to planet-forming regions in disks. 
\\ \\
In low-mass protostar observations, only an upper limit and a tentative detection are reported for O$_{2}$ lines.
This is partly because O$_{2}$ does not possess electric dipole-allowed rotational transition lines, and it possess only magnetic-allowed lines \citep{Crownover1990}.
In addition to the report of an upper limit toward IRAS 4A using $Herschel$/HIFI \citep{Yildiz2013} (see Sections 1 and 4.2 of this paper),
\citet{Taquet2018} recently performed a deep search for the $^{16}$O$^{18}$O 234 GHz $2_{1}-0_{1}$ line  ($E_{\mathrm{up}}=11.2$ K) toward the inner envelope around a low-mass protostar IRAS 16293-2422 B with ALMA, and reported a residual emission at a 3$\sigma$ level after subtraction of the contaminated two brighter transitions at $\pm1$ km s$^{-1}$. However, they considered the detection as ``tentative", since there is a velocity offset of $0.3-0.5$ km s$^{-1}$ relative to the source velocity.
Assuming that the $^{16}$O$^{18}$O was not detected and using CH$_{3}$OH as a reference species, \citet{Taquet2018} obtained a [O$_{2}$]/[CH$_{3}$OH] abundance ratio $<2-5$, which is $3-4$ times lower abundance ratio than that in comet 67P/Churyumov-Gerasimenko.
\\ \\
The strong far-infrared [O I] lines (such as the $^{3}\mathrm{P}_{1}$-$^{3}\mathrm{P}_{2}$ 63.2 $\mu$m line) have been used as an outflow and jet tracers in low-mass protostars (e.g., \citealt{Karska2013, Nisini2015, Kristensen2017}), but the inner regions in the protostellar envelopes cannot be probed using these lines because of high dust opacities at far-infrared wavelengths.
Spectrally resolved [O I] lines profiles (which can be obtained using such as SOFIA/GREAT) can probe the atomic oxygen abundances in the outer envelopes, if the outflow can be properly disentangled (see also discussions in \citet{vanDishoeck2021}).
\subsection{HCO$^{+}$, CH$_{3}$OH, and other molecular line observations}
On the basis of our chemical modeling,
for $L_{\mathrm{X}}$$\gtrsim10^{30}-10^{31}$ erg s$^{-1}$, 
HCO$^{+}$ is not efficiently destroyed within the water snowline, and its abundances remain $\gtrsim10^{-9}$ both inside and just outside the water snowline (see Section 3.3).
In addition, CH$_{3}$OH is considered not to be an efficient tracer of the warm inner envelope and the water snowline position for $L_{\mathrm{X}}$$\gtrsim10^{30}-10^{31}$ erg s$^{-1}$ (see Section 3.5).
In these cases, the CH$_{3}$OH gas abundances are expected to be much smaller (below to $10^{-16}$) than those with $L_{\mathrm{X}}\lesssim10^{29}-10^{30}$ erg s$^{-1}$ ($\sim10^{-7}-10^{-6}$).
Thus, observationally obtaining the abundance profiles of HCO$^{+}$ (including its isotopologue H$^{13}$CO$^{+}$) and CH$_{3}$OH is important to investigate the effects of X-ray induced chemistry in protostellar envelopes.
\\ \\
\citet{vantHoff2018} observationally reported an increase of H$^{13}$CO$^{+}$ emission just outside the water snowline in the IRAS 2A envelope. It is consistent with the profiles of our modeling 
with $L_{\mathrm{X}}\lesssim10^{29}$ erg s$^{-1}$ (see also Section 3.3).
We note that observationally obtained inner water gas abundances are also consistent with our model profiles for $L_{\mathrm{X}}\lesssim10^{29}$ erg s$^{-1}$ (see Section 4.5).
\\ \\
%
\citet{vantHoff2018CH3OH} and \citet{Lee2019} reported spatially resolved line images of CH$_{3}$OH and other COMs (complex organic molecules) with ALMA toward the disk around the embedded protostar V883 Ori, which is a well-known FU Orionis star (e.g., \citealt{Cieza2016}). They discussed that the radial extent of CH$_{3}$OH gas is around 100 au.
In the FU Orionis type stars,
sudden increases in the luminosity of the central star will quickly expand the sublimation front (the so-called snowline) to larger radii, which provide good opportunities to study the abundances of COMs in the planet forming materials \citep{Lee2019}.
\\ \\
\citet{Lee2020} obtained spatially resolved images of CH$_{3}$OH and H$^{13}$CO$^{+}$ emission lines with ALMA towards the embedded protostar EC53, in which quasi-periodic emission was reported 
by the near-infrared monitoring observations \citep{Hodapp2012} and the submillimeter (JCMT\footnote[7]{James Clerk Maxwell Telescope}) monitoring survey \citep{Herczeg2017,Yoo2017}, which strongly suggests variable accretion rates.
Its luminosities $L_{\mathrm{bol}}$ of 1.7-4.8$L_{\odot}$ (e.g., \citealt{Evans2009}) and envelope mass $M_{\mathrm{env}}$ of 0.86-1.25 \citep{Lee2020} are similar to those in IRAS 2A and IRAS 4A (the differences are within a factor of 4-5 times).
This source is classified as Class I \citep{Giardino2007}, and $L_{\mathrm{X}}$ is around $1-3\times10^{30}$ erg s$^{-1}$, according to XMM-Newton X-ray observations \citep{Preibisch2003} and Chandra X-ray observations \citep{Giardino2007}.
In this observation, H$^{13}$CO$^{+}$ line emission is depleted near the continuum peak, where the CH$_{3}$OH line emission is present.
The CH$_{3}$OH emission is more extended than the expected water snowline from the current luminosity of the central star, indicating previous outburst events.
However, the derived CH$_{3}$OH gas abundance for EC 53 is two orders of magnitude lower than the CH$_{3}$OH abundance of $\sim10^{-8}$ for V883 Ori \citep{Lee2019}, despite the similar size of methanol emitting region.
Comparing with the profiles of our modeling, those observed profiles of H$^{13}$CO$^{+}$ and CH$_{3}$OH abundances for EC 53 will be explained if $L_{\mathrm{X}}$ is around $\times10^{29}-10^{30}$ erg s$^{-1}$, which is roughly consistent with the observed values of $L_{\mathrm{X}}$.
\\ \\
\citet{Hsieh2019} reported the detections of the HCO$^{+}$ (3-2) line with ALMA toward the envelopes around 18 Class 0 and 11 Class I protostars in the Perseus molecular cloud, and also the detections of the CH$_{3}$OH 254.015 GHz ($2_{0,2}-1_{-1,1}$) line toward six of the above sources in which the HCO$^{+}$ line was detected.
They discussed that in four sources where the CH$_{3}$OH line was detected, the measured HCO$^{+}$ peak radii broadly agreed with the CH$_{3}$OH emission extents, except two Class 0 sources with very weak CH$_{3}$OH emission.
In these two Class 0 objects (L1455-IRS4 and L1448-IRS2), both HCO$^{+}$ and CH$_{3}$OH emissions have a similar peak position at the center.
Our modeling would suggest that these two sources have strong X-ray emission, destroying H$_{2}$O and CH$_{3}$OH in the inner regions, leading to weak CH$_{3}$OH and strong HCO$^{+}$ emission, while in the other sources, X-ray luminosities would be lower.
This can be tested by independent determination of the X-ray luminosities for these sources, although observations of X-ray luminosities toward embedded Class 0 protostars are difficult (see Section 2.1.2).  
\\ \\
CH$_{3}$OH and C$_{2}$H
are the representative products of hot corino chemistry and warm carbon chain chemistry (WCCC) in star-forming cores, respectively (e.g., \citealt{Sakai2013, Imai2016, Oya2016, Higuchi2018, Aikawa2020, Yang2021}).
As shown in Figures \ref{Figure7_CH3OHgas&ice} and \ref{FigureD1_CH4&C2H&HCNgas}, the dependance of radial profiles of C$_{2}$H gas fractional abundances on X-ray fluxes are much smaller than those of CH$_{3}$OH gas fractional abundances, in the inner envelopes ($r\lesssim 300$ au).
As the X-ray flux increases, the CH$_{3}$OH gas fractional abundances significantly decrease (from $10^{-7}-10^{-6}$ to $<10^{-15}$) at $r\lesssim 300$ au, whereas the C$_{2}$H gas fractional abundances change within two orders of magnitude. 
\citet{Aikawa2020} investigated the physical conditions which affect the hot corino chemistry and WCCC, and deficiency of COMs (including CH$_{3}$OH) in prototypical WCCC sources is hard to reproduce within their models. They discussed that gas-phase destruction processes of CH$_{3}$OH and other COMs within several 10$^{4}$ years after sublimation from dust grains would be important \citep{Charnley1992, Nomura2009, Taquet2016CH3OH}. X-ray induced destruction reactions discussed in this paper would help to destroy these molecules within the above timescale.
\\ \\
Recently, the line observations of CH$_{3}$OH and other molecules (including COMs and C$_{2}$H) with much higher spatial resolutions ($\Delta r\lesssim$ several tens of au) have been conducted with ALMA and VLA towards disks and inner envelopes around Class 0 and I protostars (e.g., \citealt{Sahu2019, Bianchi2020, DeSimone2020}, and FAUST\footnote[8]{\url{http://faust-alma.riken.jp}}). 
These detailed observations can also be used to constrain the effects of X-ray induced chemistry on the abundance profiles of HCO$^{+}$, CH$_{3}$OH, and other molecules. 
In addition, future molecular line observations with e.g., ngVLA\footnote[9]{the next generation Very Large Array} will also be helpful.
They are expected to constrain the inner gas abundances of CH$_{3}$OH, other COMs \citep{Oberg2018}, HCO$^{+}$, and NH$_{3}$, which are also affected by X-ray induced chemistry (see Appendix E and \citealt{Zhang2018}), creating a more complete picture of the oxygen chemistry and opening a window into the independent nitrogen chemistry. 
Moreover, since the dust opacities in the frequencies of ngVLA are smaller than those of ALMA (see also e.g., \citealt{DeSimone2020}), these observations will be useful to trace the inner gas abundances more precisely.
\\ \\
Particles are accelerated in shocks along the protostellar jets and on the protostellar surfaces, and they can enhance the cosmic-ray ionization rates in protostellar envelopes (e.g., \citealt{Padovani2016}). 
However, because of the differences in energies, they are much more transparent compared with X-rays, and they affect the ionization rates and thus abundances of ion molecules (such as HCO$^{+}$ and N$_{2}$H$^{+}$) not only in the inner regions, but also in the outermost regions, such as $r>10^{3}$ au \citep{Ceccarelli2014, Favre2017}.
\citet{Ceccarelli2014} reported a HCO$^{+}$/N$_{2}$H$^{+}$ abundance ratio of around 3-4 in the outer envelopes ($r>$ a few $\times 10^{3}$ au) around the protostar OMC-2 FIR4 from $Herschel$ observations, which was very low compared to that in other protostellar envelopes ($>>10$).
They suggested that the cosmic-ray ionization rate is around $10^{-14}$ s$^{-1}$, which is much higher than the average value in dense clouds ($\sim10^{-17}$ s$^{-1}$, \citealt{Umebayashi2009}).
\\ \\
In addition, \citet{vantHoff2018} 
conducted a first order approximation of the effect of cosmic-ray ionization rates on the HCO$^{+}$ gas abundances in the outer envelopes, by considering the main formation (CO+H$_{3}^{+}$) and destruction (HCO$^{+}$ + e$^{-}$) reactions of HCO$^{+}$ outside the water snowline.
Assuming steady state and the similar abundances for HCO$^{+}$ and electron ($n_{\mathrm{HCO}^{+}}/n_{\mathrm{H}}\approx$$n_{\mathrm{e}^{-}}/n_{\mathrm{H}}$),
they obtained the following expression for the HCO$^{+}$ gas densities $n_{\mathrm{HCO}^{+}}$ in the outer envelopes,
\begin{equation}
n_{\mathrm{HCO}^{+}}=\sqrt{\frac{\xi_{\mathrm{CR}}n_{\mathrm{H}_{2}}}{k_{5}}},
\end{equation}
where $k_{5}$ is the rate coefficient of the main destruction reaction of HCO$^{+}$ + e$^{-}$ (for more details, see Appendix B of \citet{vantHoff2018}).
Thus, with investigating the abundances of ion molecules such as HCO$^{+}$ both the inner and outer regions, we could distinguish the effects of X-ray induced ionization and cosmic-ray induced ionization accelerated by e.g., protostellar jets.
According to our model calculations ($\xi_{\mathrm{CR}}(r)=$$10^{-17}$ s$^{-1}$), the HCO$^{+}$ abundances at $r\gtrsim10^{3}$ au are $\sim10^{-9}-10^{-8}$ for IRAS 2A, and $\sim10^{-10}-10^{-9}$ for IRAS 4A.
Thus, if the HCO$^{+}$ abundances at these radii are more than an order of magnitude higher, $\xi_{\mathrm{CR}}(r)$ is estimated to be $\gtrsim10^{-15}$ s$^{-1}$.
\\ \\
We note that \citet{Bruderer2009} and \citet{Benz2016} offered an alternative explanation for enhanced HCO$^{+}$, originating in the UV irradiated warm outflow cavity walls. 
Spatially resolved observations for HCO$^{+}$ lines are needed to distinguish the scenarios.
\subsection{Chemical evolution from envelopes to disks}
In our chemical modeling, we assume that the physical structures (especially the radial $n_{\mathrm{H}_{2}}$ and $T_{\mathrm{gas}}$ profiles) are constant throughout $10^{5}$ years.
However, the timescale of the main accretion phase of protostars (Class 0-I) is around a few $10^{5}$ years \citep{Dunham2014, Kristensen2018}, and material in the envelopes moves inward (e.g., \citealt{Visser2009, Visser2011, Harsono2015, Furuya2017}).
According to our calculations, for the highest X-ray luminosities of $L_{\mathrm{X}}\gtrsim10^{31}$ erg $s^{-1}$, the X-ray induced reactions in the inner envelopes proceed with shorter time scales of $t<10^{3}$ years than the timescale of the main accretion phase, although the timescale of grain surface chemistry are longer ($t>>10^{5}$ years, see also \citealt{Yoneda2016}).
\citet{Aikawa2008, Aikawa2020} suggested that the infalling material passes through the region with $T_{\mathrm{gas}}\sim10^{1}-10^{2}$ K in several $10^4$ years,
and fall into the central star and the disk $\sim10^{2}$ years after they enter the region with $T_{\mathrm{gas}}>10^{2}$ K.
Therefore, the molecular abundances in the inner envelopes, especially in the innermost region, would be affected because of such inward accretion, unless they enter a rotating disk-like structure \citep{Schoier2002}.
\\ \\
In our chemical modeling of this paper, we assume that X-ray luminosities are constant throughout $10^{5}$ years.
However, X-ray luminosities of central protostars and X-ray fluxes in surrounding envelopes are expected to be changed with time, 
because protostar X-ray flares are the dominant X-ray source and they occur repeatedly once in around 10 days \citep{Takasao2019}.
In the presence of strong X-ray fluxes, water is considered to be efficiently destroyed with very short timescale of $t<10^{3}$ years \citep{Stauber2006} in the disks and the inner envelopes around protostars. 
Thus repetition of sudden increase (and decrease) of X-ray luminosities would also affect molecular abundances in the disks and inner envelopes around Class 0 and I protostars.
\\ \\
As future studies, the detailed chemical modeling with time-dependent physical structures of disks and envelopes around Class 0 and I protostars will be important to understand the effects of X-ray induced chemistry on the abundance profiles of water and related molecules in detail.
We note that \citet{Cleeves2017} reported the time variation of the H$^{13}$CO$^{+}$ J$=3-2$ line intensities in a Class II disk between three observational epochs, and they discussed that the 
enhancement of HCO$^{+}$ abundance in the upper layer of the disk would be explained by X-ray driven chemistry during large X-ray flare events.
\citet{Waggoner2019} discussed that the day-scale impulsive increase and decrease of the H$_{2}$O gas abundances in the surfaces of Class II disks could be caused by the time dependent chemistry driven by X-ray flares.
\\ \\
According to our chemical modeling, X-ray induced chemistry affects the abundances profiles of H$_{2}$O and other dominant molecules, such as O, O$_{2}$, HCO$^{+}$, CH$_{3}$OH, OH, CO$_{2}$, \ce{CH4}, \ce{HCN}, and \ce{NH3}.
In the presence of strong X-ray fields (with $L_{\mathrm{X}}$$\gtrsim10^{30}-10^{31}$ erg s$^{-1}$), the abundances of H$_{2}$O, CH$_{3}$OH, CH$_{4}$, HCN, and NH$_{3}$ significantly decrease in the inner envelopes around protostars, 
and CO, O$_{2}$, O become the dominant oxygen carriers.
In addition, on the basis of Figures \ref{FigureC1_CO$_{2}$}-\ref{FigureE1_NH3&N2gas}, CO and N$_{2}$ become the dominant carbon and nitrogen carries under such strong X-ray fields.
Note that the material in the protostellar envelopes accretes into disks, thus the molecular abundances in protostellar envelopes determine the initial abundances of chemical evolution in disks, where planet formation occurs (see also the recent review by \citealt{Oberg2021}).
In many studies of chemical modeling in disks (e.g., \citealt{Walsh2015, Bosman2018}), 
initial chemical abundances were assumed to be inherited from dark clouds, pre-stellar cores, and protostellar envelopes, and they are water-rich, 
on the basis of previous observations (e.g., \citealt{Visser2009, Visser2011, Boogert2015}).
However, whether the disk chemical evolution is started from initial abundance conditions of the chemical reset (by e.g., irradiation, accretion shocks) or the inheritance from the dark clouds and 
protostellar envelopes is an important question (e.g., \citealt{Yoneda2016, Coutens2020, Jorgensen2020, vantHoff2020, Oberg2021}).
\citet{Eistrup2016, Eistrup2018} and \citet{Notsu2020} discussed that the chemical abundances in Class II disks are strongly affected by ionisation rates in disks and the adopted initial molecular abundances (inheritance or reset).
In the presence of strong X-ray fields (with $L_{\mathrm{X}}$$\gtrsim10^{30}-10^{31}$ erg s$^{-1}$), the molecular abundances in protostellar envelopes are also altered from inheritance initial molecular abundances.
\\ \\
In future studies, the chemical modeling in disks with initial abundances which consider the effects of X-ray induced chemistry discussed in this paper will be important to understand the chemical evolution history in disks and the chemical compositions of exoplanets (e.g., \citealt{Notsu2020, Turrini2021}).
%
\renewcommand{\labelitemi}{-}
\section{Conclusions}
   We investigated the radial dependence of the abundances of water and related molecules on X-rays in Class 0 low-mass protostellar envelopes, and identify potential oxygen carriers other than water.
   We used a detailed gas-grain chemical reaction network including X-ray-induced chemical processes.
    Gas-phase reactions, thermal and non-thermal gas-grain interactions, and grain-surface reactions are included in our adopted chemical reaction network.
  For the physical structures of the Class 0 protostellar envelopes, we adopted two type of spherically symmetric radial gas temperature $T_{\mathrm{gas}}$ and molecular hydrogen number density $n_{\mathrm{H}_{2}}$ profiles for IRAS 2A and IRAS 4A, in order to examine the effect of density differences on X-ray induced chemistry. 
  Our findings can be summarized as follows:
 \\
\begin{itemize}
\item Outside the water snowline ($T_{\mathrm{gas}}<10^{2}$~K, $r>10^{2}$ au), if X-ray luminosities of the central protostars $L_{\mathrm{X}}$ are larger than $10^{30}$ erg s$^{-1}$, water gas fractional abundances are increased (up to $n_{\mathrm{H}_{2}\mathrm{O}}$/$n_{\mathrm{H}}$$\sim10^{-8}-10^{-7}$), compared with the values ($n_{\mathrm{H}_{2}\mathrm{O}}$/$n_{\mathrm{H}}$$\sim10^{-10}$)
for $L_{\mathrm{X}}<10^{30}$ erg s$^{-1}$. 
In addition, water ice abundances are around $2\times10^{-4}$ outside the water snowline for $L_{\mathrm{X}}\lesssim10^{30}$ erg s$^{-1}$, and they become much smaller
(below to $n_{\mathrm{H}_{2}\mathrm{O}, \mathrm{ice}}$/$n_{\mathrm{H}}$$\sim10^{-8}$ at a few $\times10^{2}$ au) for $L_{\mathrm{X}}\gtrsim10^{31}$ erg s$^{-1}$.  
X-ray induced photodesorption of water ice affects in this region.
Since there are limited experimental constraints for X-ray induced photodesorption rates, future theoretical and experimental studies for the X-ray induced photodesorption over a wider X-ray energy ranges are important.
 \\
 \item
Inside the water snowline ($T_{\mathrm{gas}}>10^{2}$~K, $r<10^{2}$ au), for $L_{\mathrm{X}}\lesssim10^{29}-10^{30}$ erg s$^{-1}$, water maintains a high abundance of $\sim 10^{-4}$, and water and CO are the dominant oxygen carriers.
For $L_{\mathrm{X}}\gtrsim10^{30}-10^{31}$ erg s$^{-1}$, the water gas abundances significantly decrease just inside the water snowline ($T_{\mathrm{gas}}\sim100-250$~K, down to $n_{\mathrm{H}_{2}\mathrm{O}}$/$n_{\mathrm{H}}$$\sim10^{-8}-10^{-7}$) and in the innermost regions ($T_{\mathrm{gas}}\sim250$~K, $n_{\mathrm{H}_{2}\mathrm{O}}$/$n_{\mathrm{H}}$$\sim10^{-6}$).
In the presence of strong X-ray fields, gas-phase water is mainly destroyed with the ion-molecule reactions and the X-ray induced photodissociation. 
In our chemical modeling, the former ion-molecule reactions are dominant processes for the water gas destruction inside the water snowline.
For $L_{\mathrm{X}}\gtrsim10^{29}-10^{30}$ erg s$^{-1}$, the X-ray ionization rates $\xi_{\mathrm{X}}(r)$ are larger than 
our adopted cosmic ray ionization rate $\xi_{\mathrm{CR}}(r)$($=$$1.0\times10^{-17}$ s$^{-1}$) within the water snowline.
In the innermost hot region, water abundances become relatively large since the two-body water formation reaction (OH+H$_{2}$) becomes efficient.
\\
\item
As the X-ray fluxes become larger, the O$_{2}$ and O gas abundances become larger both inside and outside the water snowline.
Inside the water snowline, both O$_{2}$ and O gas abundances are much smaller ($<10^{-8}$) for $L_{\mathrm{X}}\lesssim10^{28}-10^{29}$ erg s$^{-1}$.
In contrast, for $L_{\mathrm{X}}\gtrsim10^{29}-10^{30}$ erg s$^{-1}$, their abundances become larger, and reach about 10$^{-4}$ with $L_{\mathrm{X}}\gtrsim10^{31}$ erg s$^{-1}$.
Compared with the water gas abundances, both O$_{2}$ and O gas abundances have opposite dependence on X-ray fluxes. For $L_{\mathrm{X}}\gtrsim10^{30}-10^{31}$ erg s$^{-1}$, O$_{2}$, O, and CO become the dominant oxygen carriers in the inner envelopes ($r\lesssim$ a few $\times10^{2}$ au).
\\
\item
According to previous studies, the most abundant destroyer of \ce{HCO+} in warm gas is water, and the radius of the \ce{CH3OH} snowline ($\sim2\times10^{2}$ au) is around two times larger than that of the water snowline ($\sim10^{2}$ au). 
Thus, \ce{CH3OH} and \ce{HCO+} (and also H$^{13}$CO$^{+}$) gas lines have been used as good tracers of the water snowline.
In our modeling, the \ce{HCO+} and \ce{CH3OH} gas abundances are increased/decreased within the water snowline, respectively, as the X-ray fluxes become larger.
For $L_{\mathrm{X}}\gtrsim10^{30}-10^{31}$ erg s$^{-1}$,
the \ce{HCO+} abundances within the water snowline increase by four orders of magnitude, and reach more than $10^{-9}-10^{-10}$, which are similar to those outside the water snowline.
In contrast, \ce{CH3OH} gas abundance in these radii decrease from $\sim10^{-7}-10^{-6}$ to $<10^{-16}$.
Therefore, both \ce{HCO+} and \ce{CH3OH} cannot be used as tracers of the water snowline position for $L_{\mathrm{X}}\gtrsim10^{30}-10^{31}$ erg s$^{-1}$.
Observationally obtaining the abundance profiles of HCO$^{+}$, H$^{13}$CO$^{+}$, and CH$_{3}$OH is important to investigate the effects of X-ray induced chemistry in protostellar envelopes.
\\
%
\item
The gas-phase fractional abundances of OH and \ce{CO2} increase in the outer disk (> a few hundred au), as the X-ray fluxes become larger.
At $r<$ a few hundred au, for $L_{\mathrm{X}}\lesssim10^{29}-10^{30}$ erg s$^{-1}$, OH and CO$_{2}$ gas abundances increase as the X-ray fluxes become larger. 
CO$_{2}$ gas abundances are $\sim10^{-5}-10^{-4}$ at $L_{\mathrm{X}}\sim10^{29}-10^{30}$ erg s$^{-1}$.
In these cases, CO$_{2}$ also becomes one of dominant oxygen bearing molecules, especially in the regions where the abundances of O$_{2}$ gas and H$_{2}$O are similar. 
However, OH and CO$_{2}$ gas abundances decrease for $L_{\mathrm{X}}\gtrsim10^{30}-10^{31}$ erg s$^{-1}$, and CO$_{2}$ gas abundances are 
$\lesssim10^{-6}$ at $L_{\mathrm{X}}\gtrsim10^{31}-10^{32}$ erg s$^{-1}$.
%
\\
\item
As X-ray fluxes become large, the fractional abundances of gas-phase CH$_{4}$, HCN, and NH$_{3}$ are decreasing within their own snowline positions.
The radial CO and N$_{2}$ abundance profiles are constant for the various X-ray luminosities, and they are the dominant carbon and nitrogen carries under the strong X-ray fields.
\\
\item
The effects of X-ray induced chemistry are larger in the IRAS 2A model than those in the IRAS 4A model, which has $3-6$ times larger in densities.
%
\\
\item
Comparing the results of our modeling with the observationally obtained inner gas abundances of H$_{2}$O and H$^{13}$CO$^{+}$, $L_{\mathrm{X}}$ of IRAS 2A is estimated to be $\lesssim10^{29}$ erg s$^{-1}$.
In addition, our models with $L_{\mathrm{X}}\sim10^{30}$ erg s$^{-1}$ would explain both the low inner water gas abundances and the upper limit values of O$_{2}$ gas abundances obtained by previous observations towards IRAS 4A.
However, in the cold outer part of the envelope, the best-fit profile obtained from observations is consistent with our model profiles for $L_{\mathrm{X}}\lesssim10^{28}$ erg s$^{-1}$.
The discrepancy of suggested $L_{\mathrm{X}}$ between the inner and outer envelope is remained, unless the rates of direct X-ray induced photodesorption of water are around two orders of magnitude lower than our adopted values.
Since probing the O and CO$_{2}$ gas abundance in the inner envelopes are difficult, observationally obtaining the abundance profiles of other tracers, especially HCO$^{+}$ and CH$_{3}$OH, is important to investigate the effects of X-ray induced chemistry and confine the values of $L_{\mathrm{X}}$.
\\
\end{itemize}
On the basis of our chemical modeling, X-ray induced chemistry strongly affects the abundances of water and other related molecules (such as O$_{2}$, O, HCO$^{+}$, CH$_{3}$OH, CO$_{2}$, OH, CH$_{4}$, HCN, and NH$_{3}$) especially in the inner regions, and can explain the observed low water abundances in the inner protostellar envelopes.
We find that gas-phase destruction of molecules by X-rays as well as X-ray-induced photodesorption processes are important. 
Future molecular line observations towards the disks and envelopes around low-mass protostars, using e.g., ALMA and ngVLA, will constrain the effects of X-ray induced chemistry.
In addition, it will be important to discuss how the X-ray induced chemistry at protostar phases affect the initial abundances and chemical evolution in planet forming disks.
\\ \\
\begin{acknowledgements}
\\
We are grateful to Daniel Harsono, Umut A. Y{\i}ld{\i}z, Joseph C. Mottram, Merel L. R. van't Hoff, and Lars E. Kristensen for giving us the data of 
temperature, number density, and water abundance profiles estimated from previous observations of Class 0 protostellar envelopes.
We thank Shinsuke Takasao and Masanobu Kunitomo for their important comments about X-ray fields in protostars.
We are also grateful to Yuri Aikawa for her useful comments on chemical evolutions and to Nami Sakai for her comments on the possibility of future ngVLA observations.
We thank the referee for important suggestions and comments.
Our numerical studies were carried out on PC cluster at Center for Computational Astrophysics (CfCA), National Astronomical Observatory of Japan (NAOJ), and on computer systems at Leiden Observatory, Leiden University. 
S.N. is grateful for support from JSPS (Japan Society for the Promotion of Science) Overseas Research Fellowships, RIKEN Special Postdoctoral Researcher Program (Fellowships), and 
MEXT/JSPS Grants-in-Aid for Scientific Research (KAKENHI) 20K22376, 20H05845, and 20H05847.
C.W.~acknowledges financial support from the University of Leeds and from the Science and Technology Facilities Council (grant numbers ST/R000549/1 and ST/T000287/1).
H.N. is supported by MEXT/JSPS Grants-in-Aid for Scientific Research (KAKENHI) 18H05441, 19K03910 and 20H00182, NAOJ ALMA Scientific Research grant No. 2018-10B, and FY2019 Leadership Program at NAOJ.
\end{acknowledgements}
\begin{appendix}
\section{The dependance of X-ray induced chemistry on gas number density}
In this section, we investigate the dependance of X-ray induced chemistry on the gas number density.
The rates of formation or destruction for gas-phase species Y due to the X-ray induced photoionisation and photodissociation reactions are scaled with $\zeta_{\mathrm{X}}(r)\times$$n_{\mathrm{Y}}$, and these are first-order kinetic processes with regard to gas densities.
In contrast, ion-molecule reactions, neutral-neutral reactions, are second-order kinetic processes with regard to gas densities.
Thus, given the same X-ray ionisation rates $\zeta_{\mathrm{X}}(r)$ and gas temperatures $T_{\mathrm{gas}}$, the more important the latter ``second-order'' processes are, and the less important the former ``first-order'' processes are, as the gas density increases.
\\ \\
At $r\sim60$ au (inside the water snowline), the gas density in IRAS 4A ($n_{\mathrm{H}_{2}}$$=1.1\times10^{9}$ cm$^{-3}$) is around 5.5 times larger than that in IRAS 2A ($n_{\mathrm{H}_{2}}$$=2.0\times10^{8}$ cm$^{-3}$), whereas the gas temperatures are similar ($T_{\mathrm{gas}}\sim140$ K for IRAS 2A and $\sim150$ K for IRAS 4A).
For $L_{\mathrm{X}}=10^{32}$ erg s$^{-1}$, $\zeta_{\mathrm{X}}(r)$ in IRAS 4A ($2.0\times10^{-14}$ s$^{-1}$) is around 4.4 times lower than that in IRAS 2A ($8.8\times10^{-14}$ s$^{-1}$).
We conduct a test chemical calculation, in which we adopt the IRAS 4A physical structure ($n_{\mathrm{H}_{2}}$ and $T_{\mathrm{gas}}$) at $r\sim60$ au and $L_{\mathrm{X}}=10^{32}$ erg s$^{-1}$, and rescale $\zeta_{\mathrm{X}}(r)$ to the value at a similar radius in the IRAS 2A model. 
We compare the rates of reactions of this test calculation with those of the standard IRAS 2A model.
\\ \\
In this test calculation, at $t=10^{5}$ years, 
the fractional abundance and absolute number density of water are
are $6.9\times10^{-8}$ and $1.5\times10^{2}$ cm$^{-3}$, respectively, and the same for HCO$^{+}$
are $4.2\times10^{-9}$ and $9.4$ cm$^{-3}$, respectively.
On the basis of these values, the rate coefficient of the ion-molecule reaction with H$_{2}$O+HCO$^{+}$$\rightarrow$CO+H$_{3}$O$^{+}$, $k_{6}$, is
$\sim3.5\times10^{-9}$ cm$^{3}$ s$^{-1}$ \citep{Adams1978}, and the reaction rate,
$R(6)=$$k_{6}$$n_{\mathrm{H}_{2}\mathrm{O}}$$n_{\mathrm{HCO}^{+}}$, is $\sim2.2\times10^{-5}$ cm$^{-3}$ s$^{-1}$ at $t=10^{5}$ years.
In contrast, the rate coefficient of X-ray-induced photodissociation leading to H+OH, $k_{7}$, is $\sim8.6\times10^{-11}$ s$^{-1}$ \citep{Gredel1989}, 
and the reaction rate, $R(7)=$$k_{7}$$n_{\mathrm{H}_{2}\mathrm{O}}$, is $\sim1.3\times10^{-8}$ cm$^{-3}$ s$^{-1}$ at $t=10^{5}$ years.
\\ \\
Comparing these reaction rates with those of the standard IRAS 2A model (see Section 3.1),
the differences of the reaction rates are larger in the former ion-molecule reaction ($R(6)/R(1)$$\sim25$) than the latter X-ray-induced photodissociation ($R(7)/R(2)$$\sim2.5$). 
It is because the former reaction is the``second-order'' process and the latter reaction is the ``first-order'' process.
Thus, as the gas densities become larger, the``second-order'' processes including ion-molecule reactions become much more dominant compared with the``first-order'' processes including X-ray-induced photodissociation.
\section{Sub-grid calculations for the IRAS 4A envelope models}
\begin{figure*}
\begin{center}
\includegraphics[scale=0.67]{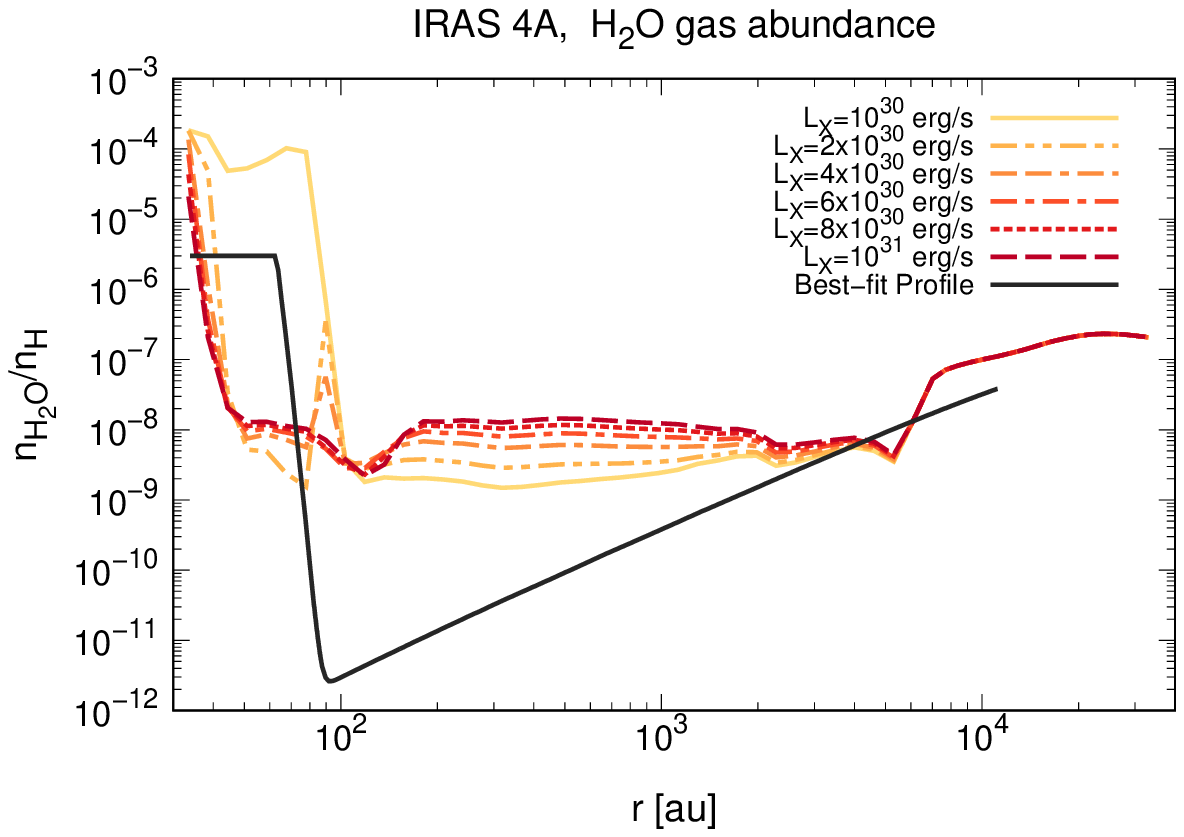}
\includegraphics[scale=0.67]{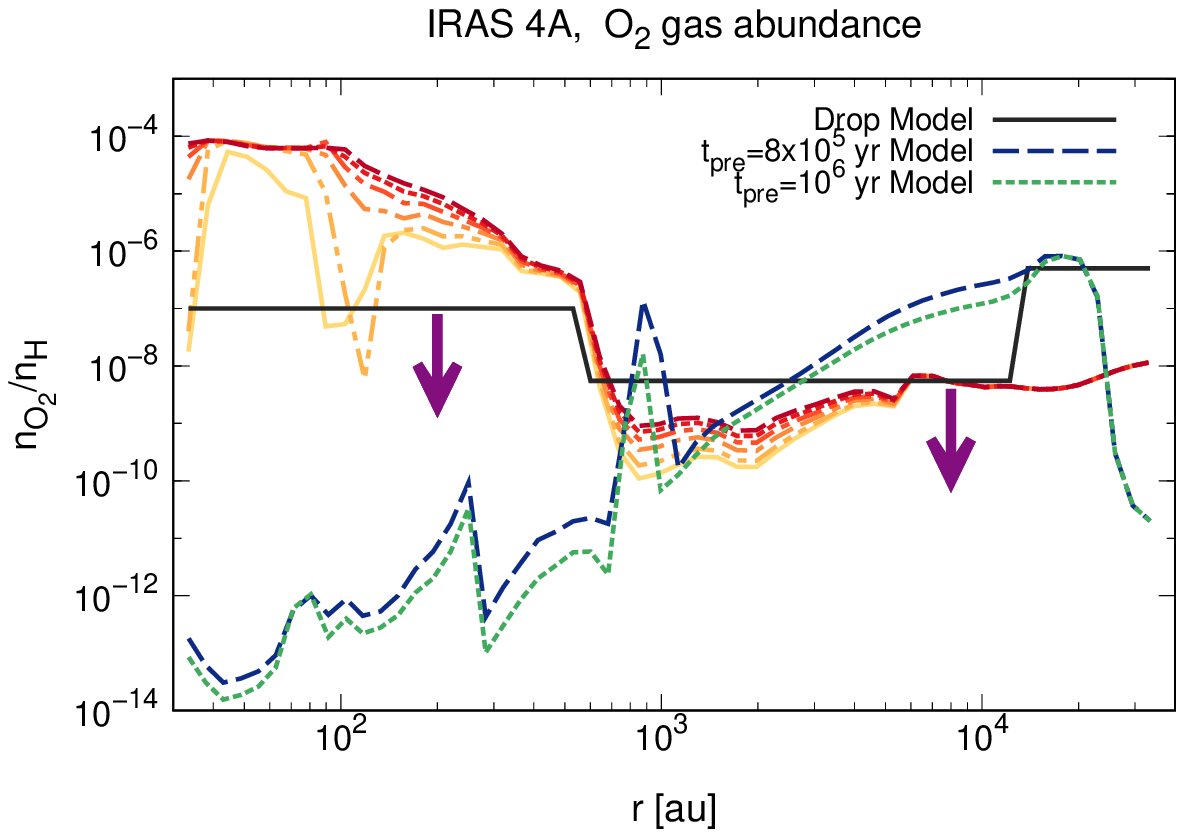}
\includegraphics[scale=0.67]{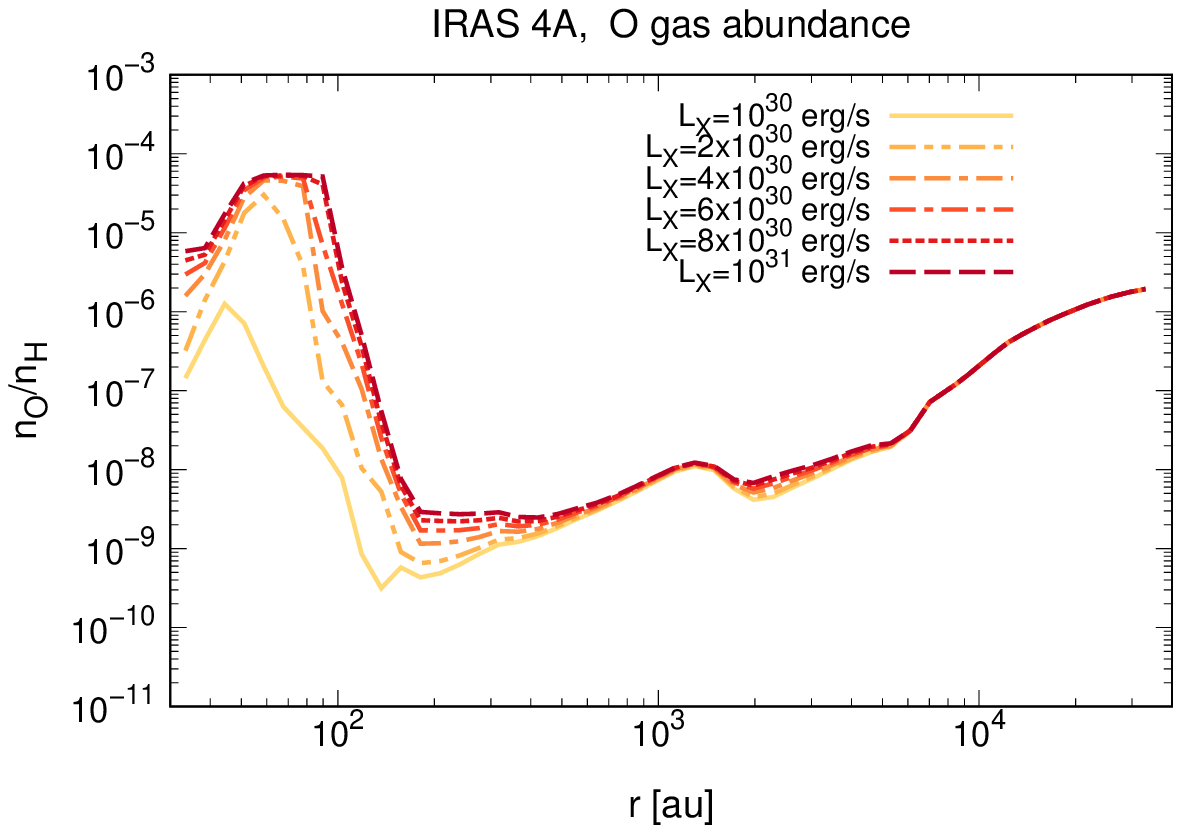}
\includegraphics[scale=0.67]{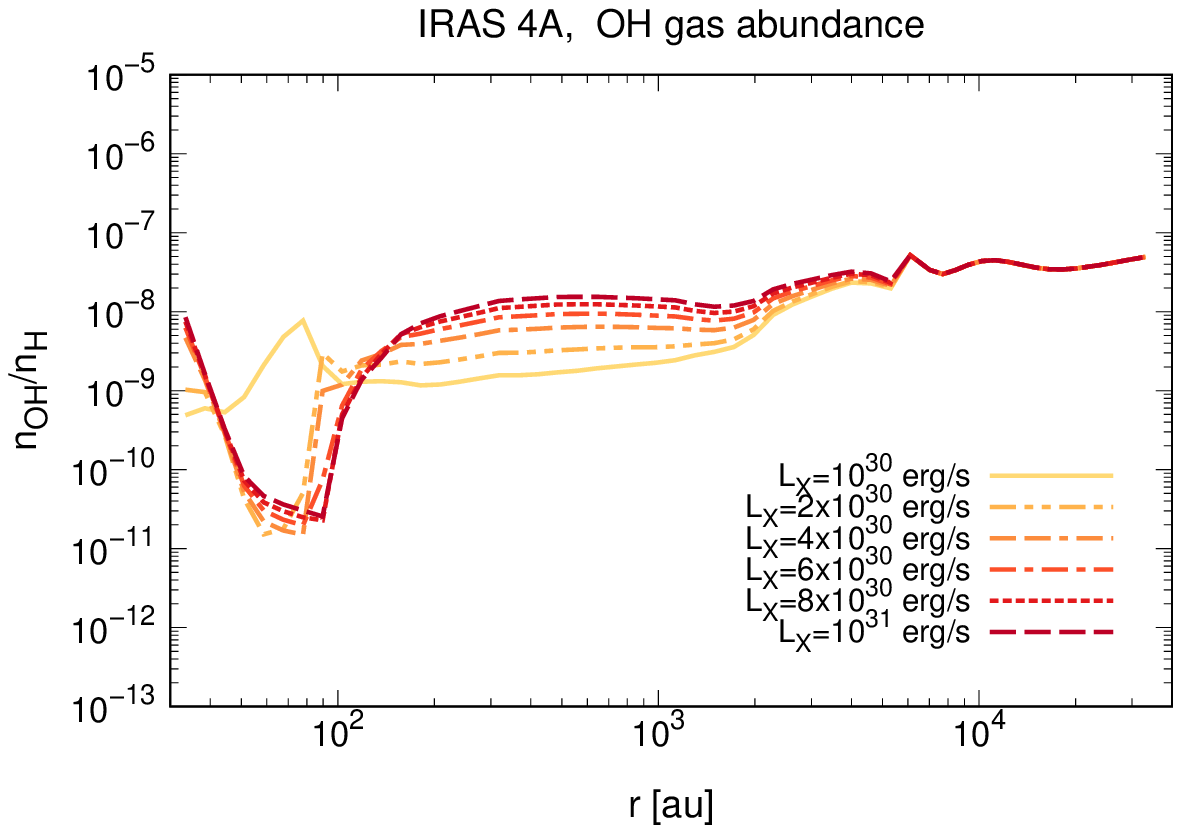}
\includegraphics[scale=0.67]{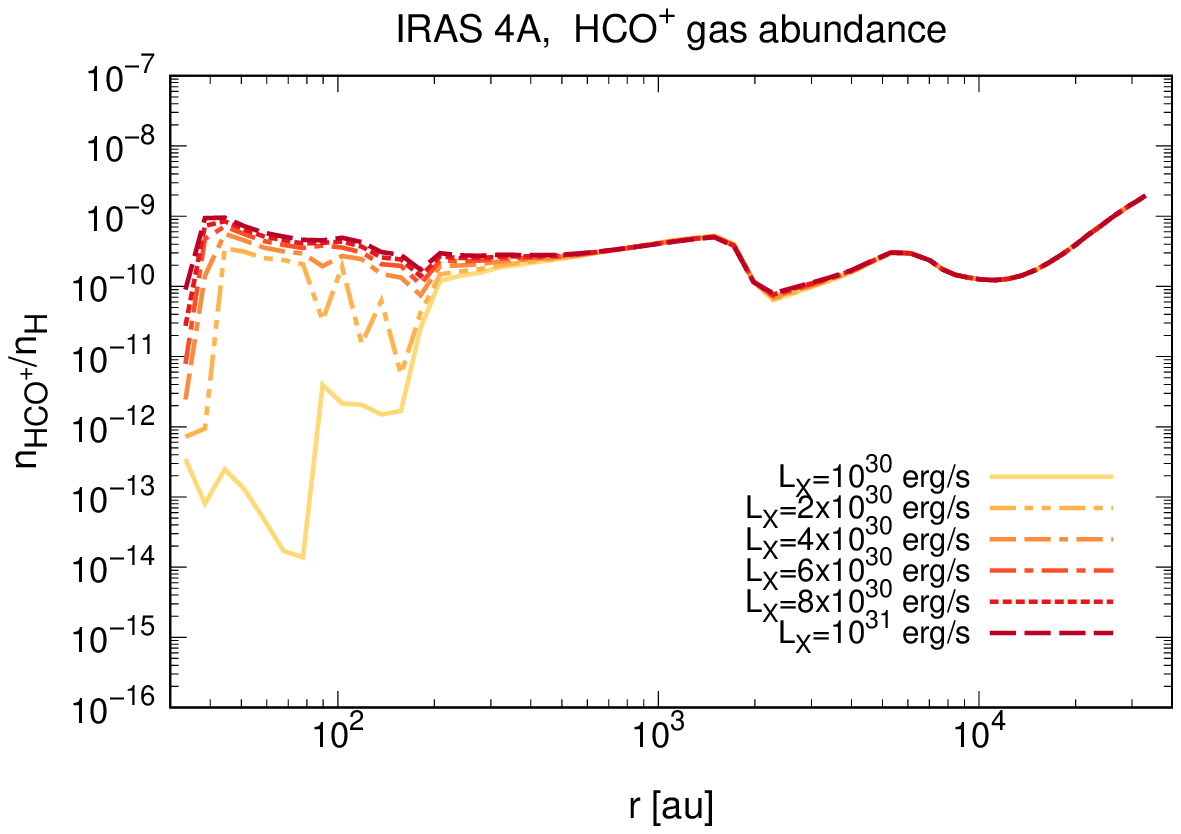}
\includegraphics[scale=0.67]{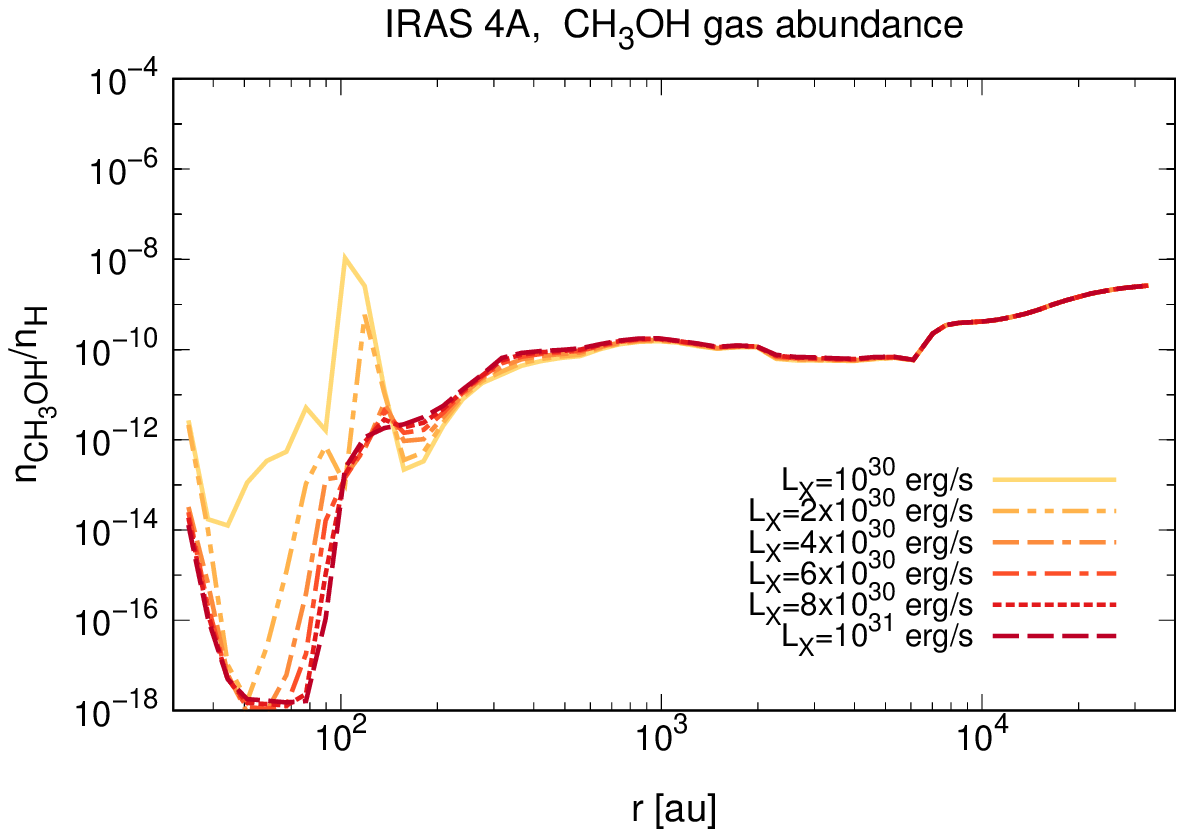}
\end{center}
\caption{
\noindent 
The radial profiles of gaseous fractional abundances of of H$_{2}$O, O$_{2}$, O, OH, HCO$^{+}$, and CH$_{3}$OH
in the NGC 1333-IRAS 4A envelope models, with X-ray luminosities between $L_{\mathrm{X}}=$$10^{30}$ and $10^{31}$ erg s$^{-1}$. The line colors are gradually change from yellow to orange, red, and brown as the values of $L_{\mathrm{X}}$ increase.
In the top left panel, the observational best-fit H$_{2}$O gas abundance profile obtained in \citet{vanDishoeck2021} is over-plotted, as we plot in Figure \ref{Figure10_H2O&O2gas_obs-plot}.
In the top right panel, the three model O$_{2}$ gas abundance profiles obtained in \citet{Yildiz2013} are over-plotted, as we plot in Figure \ref{Figure10_H2O&O2gas_obs-plot}.
}\label{FigureB1_H2O&O2&O&OH&HCO+&CH3OHgas_reaction-rev}
\end{figure*} 
Figure \ref{FigureB1_H2O&O2&O&OH&HCO+&CH3OHgas_reaction-rev} shows the radial profiles of H$_{2}$O, O$_{2}$, O, OH, HCO$^{+}$, and CH$_{3}$OH gas fractional abundances in the IRAS 4A envelope models, with X-ray luminosities between $L_{\mathrm{X}}=$$10^{30}$ and $10^{31}$ erg s$^{-1}$.
We plot these sub-grid model profiles since there is a large jump in abundances in this X-ray luminosity range (see Figures \ref{Figure3_H2Ogas}-\ref{Figure7_CH3OHgas&ice}).
For the abundance profiles of H$_{2}$O gas, between $10^{30}$ and $2\times10^{30}$ erg s$^{-1}$ seems to be the clear boundary, and they decrease from $\sim10^{-4}$ to $<10^{-7}$ at $r\lesssim10^{2}$ au.
For the abundance profiles of HCO$^{+}$ and CH$_{3}$OH gas, between $10^{30}$ and $2\times10^{30}$ erg s$^{-1}$ also seems to be the clear boundary.
Compared to them, the abundance profiles of O$_{2}$ and O gas gradually increase in the inner region as the values of $L_{\mathrm{X}}$ increase from $10^{30}$ to $\sim6\times10^{30}$ erg s$^{-1}$.
\section{CO$_{2}$ and CO fractional abundances}
\begin{figure*}
\begin{center}
\includegraphics[scale=0.67]{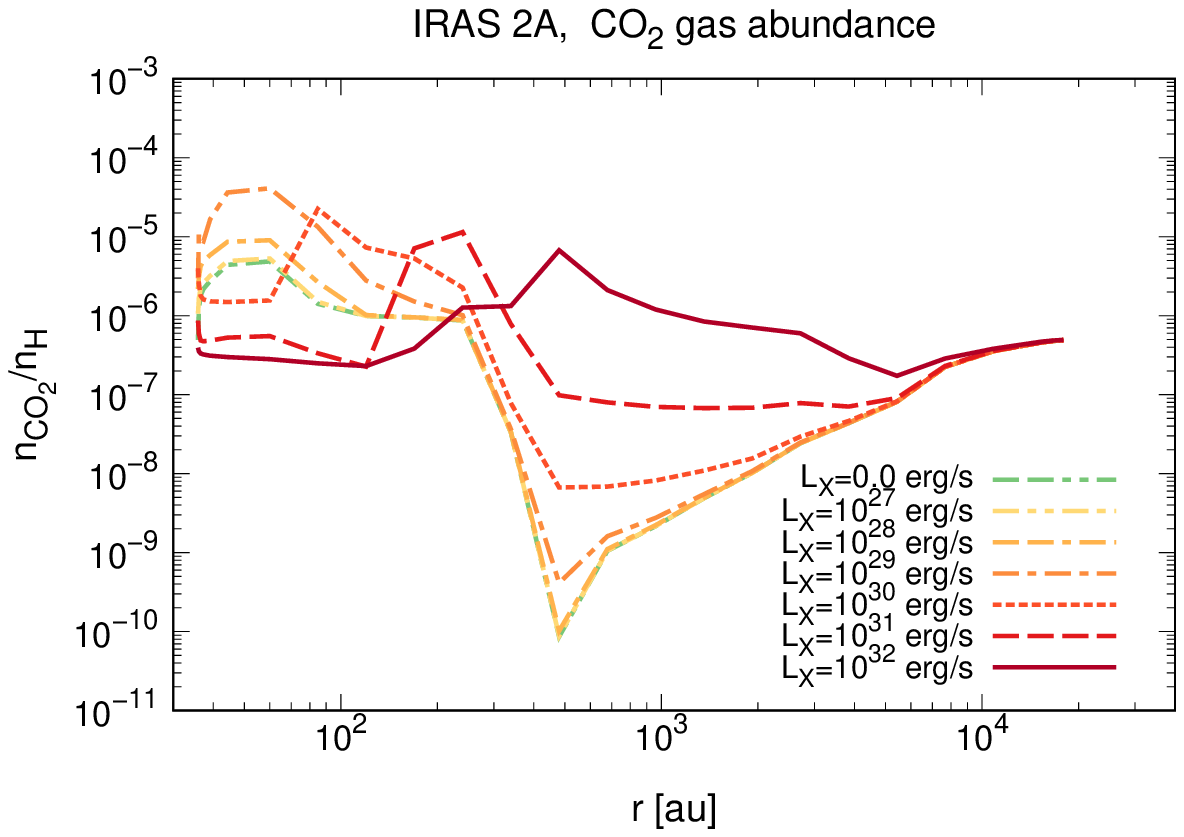}
\includegraphics[scale=0.67]{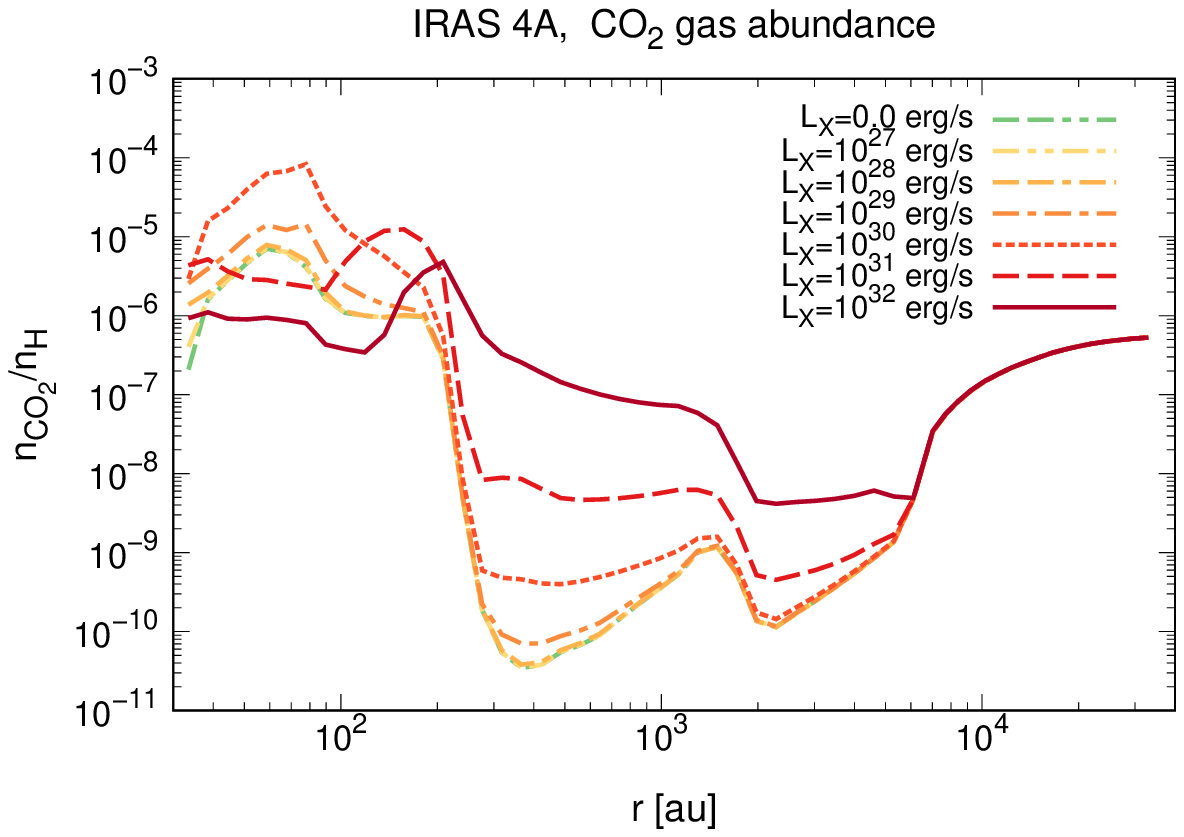}
\includegraphics[scale=0.67]{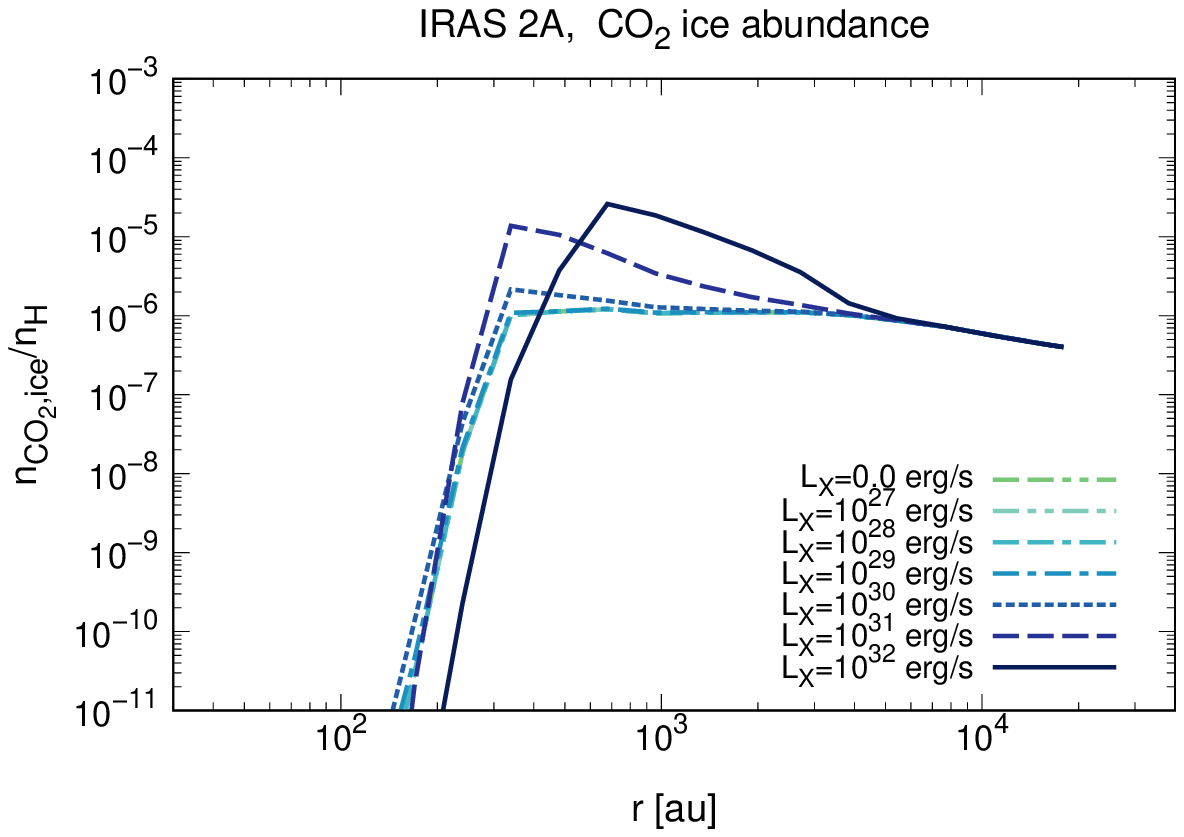}
\includegraphics[scale=0.67]{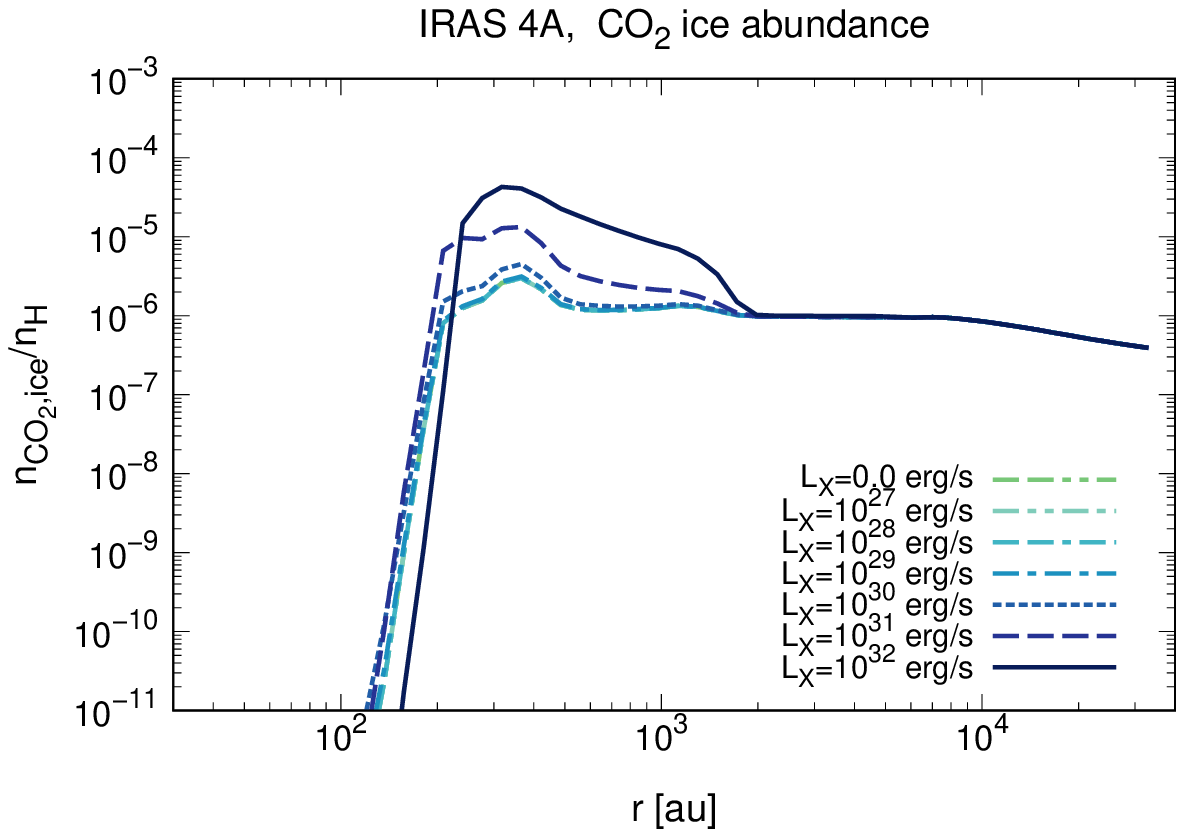}
\end{center}
\caption{
\noindent 
The radial profiles of CO$_{2}$ gas and ice fractional abundances $n_{\mathrm{CO}_{2}}$/$n_{\mathrm{H}}$ (top panels) and $n_{\mathrm{CO}_{2},\mathrm{ice}}$/$n_{\mathrm{H}}$ (bottom panels) in NGC 1333-IRAS 2A (left panels) and NGC 1333-IRAS 4A (right panels) envelope models.
%
}\label{FigureC1_CO$_{2}$}
\end{figure*}
\begin{figure*}
\begin{center}
\includegraphics[scale=0.67]{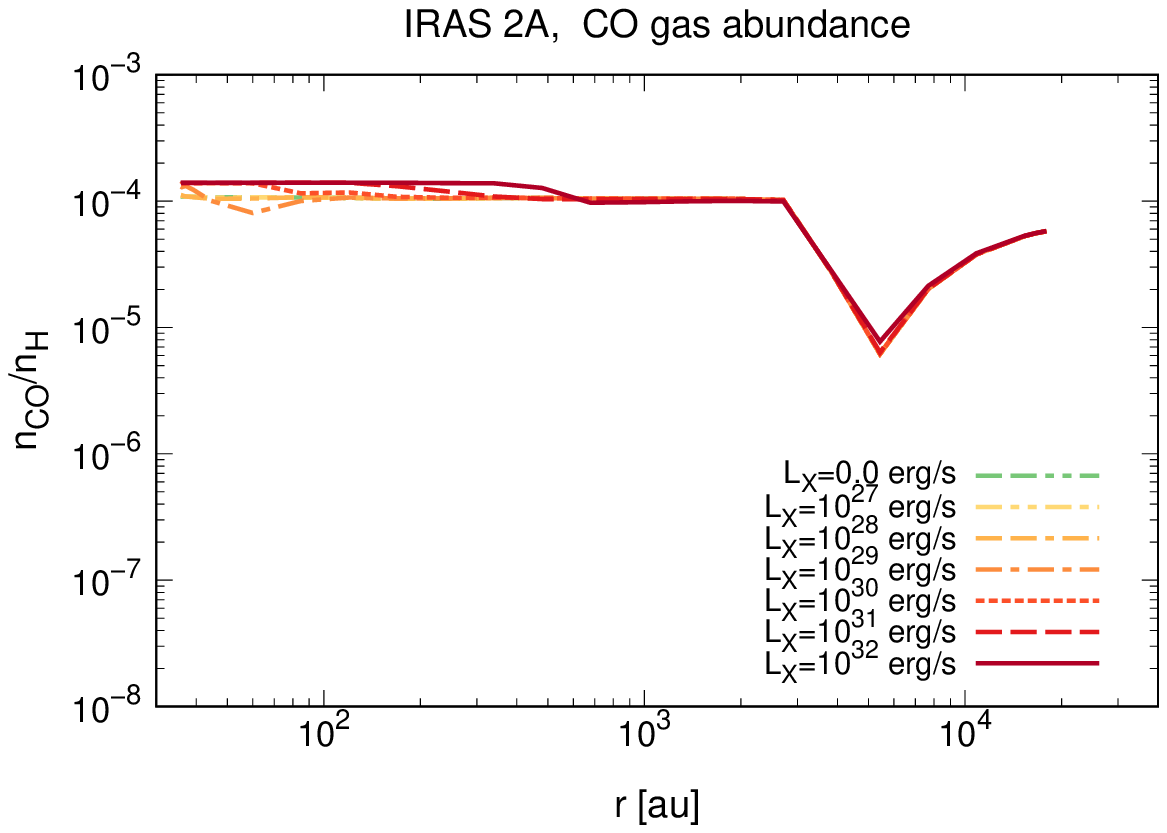}
\includegraphics[scale=0.67]{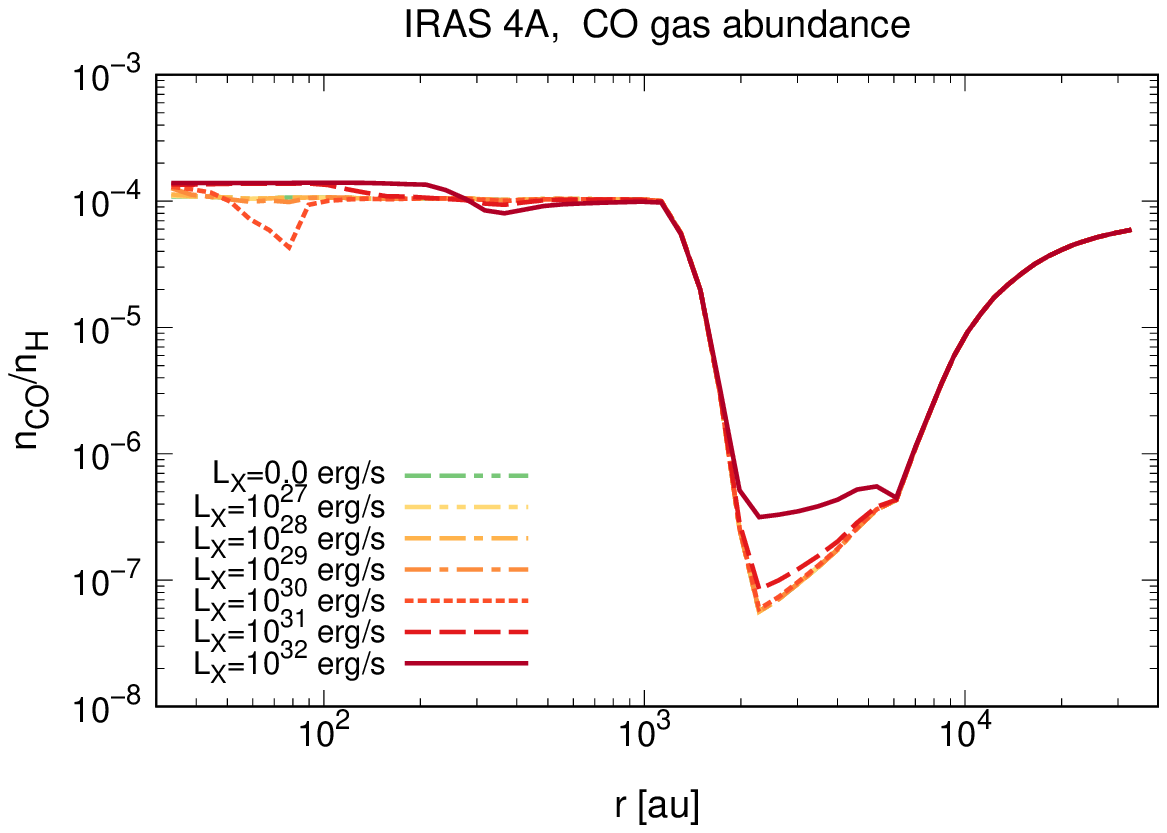}
\includegraphics[scale=0.67]{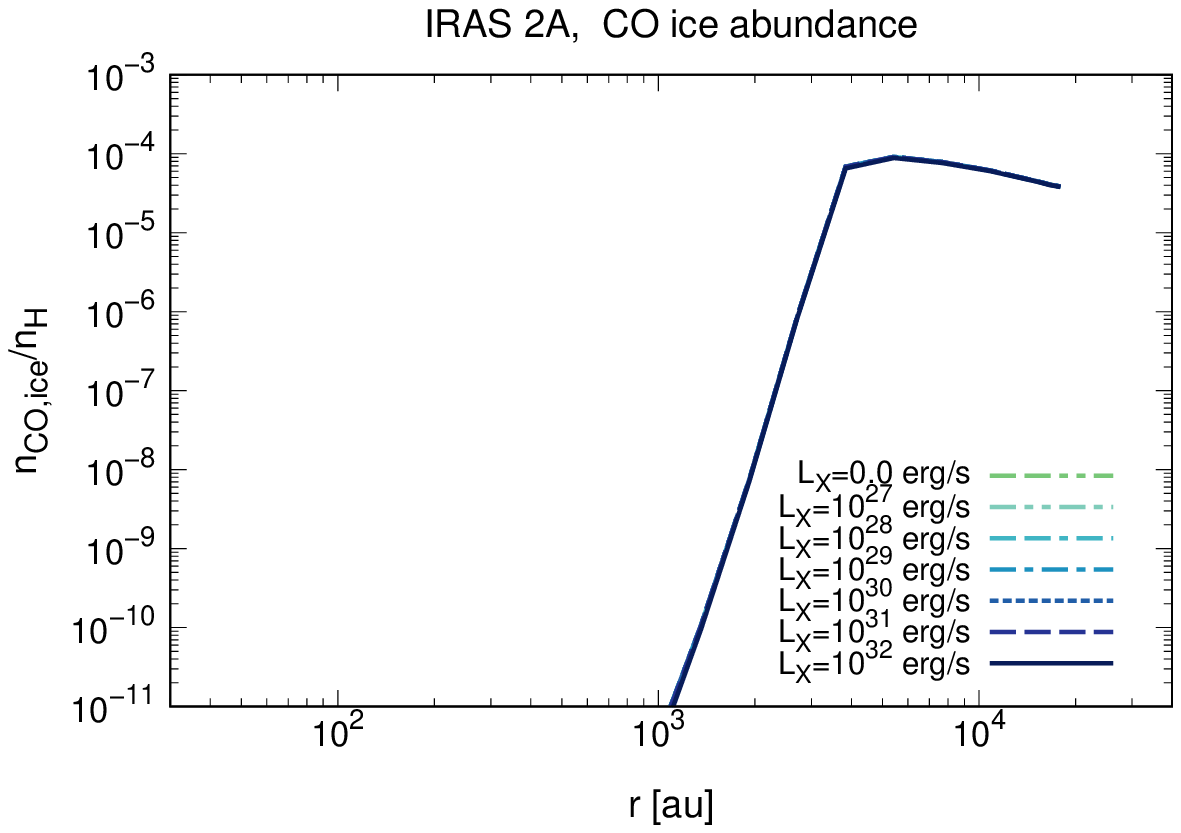}
\includegraphics[scale=0.67]{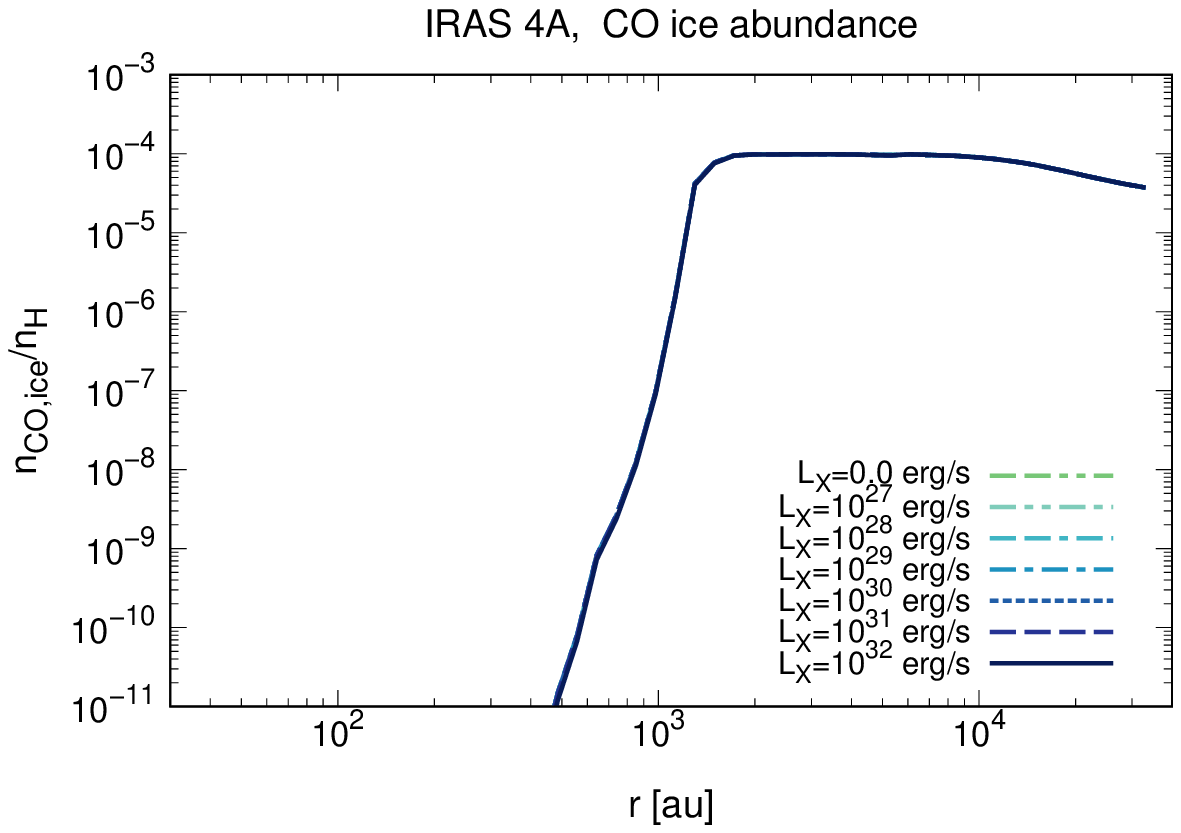}
\end{center}
\caption{
\noindent 
The radial profiles of CO gas and ice fractional abundances $n_{\mathrm{CO}}$/$n_{\mathrm{H}}$ (top panels) and $n_{\mathrm{CO},\mathrm{ice}}$/$n_{\mathrm{H}}$ (bottom panels) in NGC 1333-IRAS 2A (left panels) and NGC 1333-IRAS 4A (right panels) envelope models.
}\label{FigureC2_CO}
\end{figure*}
Figure \ref{FigureC1_CO$_{2}$} shows the radial profiles of CO$_{2}$ gas and ice fractional abundances $n_{\mathrm{CO}_{2}}$/$n_{\mathrm{H}}$ in IRAS 2A (left panels) and IRAS 4A (right panels) envelope models, for the various X-ray luminosities.
At $r\sim3\times10^{2}-4\times10^{3}$ au (in the IRAS 2A model) and $r\sim3\times10^{2}-2\times10^{3}$ au (in the IRAS 4A model), which are the regions between CO$_{2}$ and CO snowlines, CO$_{2}$ ice abundances are around 10$^{-6}$ for $L_{\mathrm{X}}\lesssim10^{30}$ erg s$^{-1}$.
In contrast, for $L_{\mathrm{X}}\gtrsim10^{31}$ erg s$^{-1}$, CO$_{2}$ ice abundances increase (up to 10$^{-5}$) in these regions.
On the basis of \citet{Drozdovskaya2016}, \citet{Eistrup2016, Eistrup2018}, and \citet{Bosman2018}, in the presence of X-ray fluxes,
the X-ray induced photodissociation of H$_{2}$O ice forms OH radicals within the ice mantle, which subsequently react with CO on grain surfaces to form CO$_{2}$ ice inside the CO snowline. 
\\ \\
CO$_{2}$ gas abundances 
at $r\lesssim10^{4}$ au in the IRAS 2A model and at $r\lesssim6\times10^{3}$ au in the IRAS 4A model are affected by strong X-ray fluxes.
For low and moderate X-ray luminosities\footnote[10]{For the definition of the values of low, moderate, and high X-ray luminosities, please see Section 3.2.},
CO$_{2}$ gas abundances become smaller, and reach $\lesssim10^{-9}$ at $r\sim3\times10^{2}-10^{3}$ au (outside the CO$_{2}$ snowline).
In contrast, for 
high X-ray luminosities,
CO$_{2}$ gas abundances become larger, and they reach around 10$^{-6}$ (for IRAS 2A) and 10$^{-7}$ (for IRAS 4A) at these radii for $L_{\mathrm{X}}\sim10^{32}$ erg s$^{-1}$.
At $r\lesssim3\times10^{2}$ au (inside CO$_{2}$ snowline), CO$_{2}$ abundances are around $10^{-6}-10^{-5}$ for low X-ray luminosities.
In addition, like OH, the CO$_{2}$ gas abundances increase as the X-ray fluxes become larger, and they reach $\sim10^{-5}-10^{-4}$ for moderate X-ray luminosities,
However, they decrease for high X-ray luminosities.
At $L_{\mathrm{X}}\gtrsim10^{31}$ erg s$^{-1}$ in the IRAS 2A model and $L_{\mathrm{X}}\gtrsim10^{32}$ erg s$^{-1}$ in the IRAS 4A model, the CO$_{2}$ gas abundances are around $3\times10^{-7}-10^{-6}$.
\\ \\
Outside the CO$_{2}$ snowline, CO$_{2}$ gas is supplied by X-ray induced photodesorption reaction of CO$_{2}$ ice. 
In the inner envelope, gas-phase CO$_{2}$ is mainly formed by two body reaction of CO$+$OH \citep{Bosman2017}, and destroyed with X-ray induced photodissociation.
Therefore, CO$_{2}$ abundance profiles strongly depend on the radial profiles of X-ray fluxes and OH abundances.
\\ \\
The CO$_{2}$ abundances in the outer envelopes around low-mass protostars can be probed through the observations
CO$_{2}$ ro-vibrational lines with JWST (see also Section 4.1), as done for high-mass protostellar envelopes using ISO \citep{vanDishoeck1996, Boonman2003}.
For low-mass protostellar envelopes, a hint of gas-phase CO$_{2}$ lines has been obtained using Spitzer (see e,g., \citealt{Poteet2013}).
We note that high dust opacities in the near and mid-infrared wavelengths make it difficult to probing the CO$_{2}$ gas abundances directly in the inner envelopes around low-mass protostars.
\\ \\
\citet{Boonman2003} and \citet{Bosman2018CO2} noted the disagreements of CO$_{2}$ gas abundances within the CO$_{2}$ snowline between models ($\sim10^{-5}$) and observations ($\sim10^{-7}$), both for high-mass protostar envelopes and Class II disks.
\citet{Bosman2018CO2} discussed that the CO$_{2}$ should be destroyed within $10^{4}$ years after the sublimation of CO$_{2}$ ice.
X-ray induced destruction reactions discussed in this paper would help to destroy CO$_{2}$ molecules within the above timescale, assuming that CO$_{2}$ chemistry is similar among these type of sources.
\\ \\
Figure \ref{FigureC2_CO} shows the radial profiles of CO gas and ice fractional abundances $n_{\mathrm{CO}}$/$n_{\mathrm{H}}$ in IRAS 2A (left panels) and IRAS 4A (right panels) envelope models, for the various X-ray luminosities.
Unlike other dominant oxygen-bearing molecules (e.g., H$_{2}$O, O$_{2}$, O, CO$_{2}$), 
both of CO gas and ice abundances do not depend on X-ray fluxes.
CO gas fractional abundances are around 10$^{-4}$ at $r\lesssim3\times10^{3}$ au at IRAS 2A and $r\lesssim10^{3}$ au at IRAS 4A (within the CO snowline), and CO ice fractional abundances are around 10$^{-4}$ at $r\gtrsim4\times10^{3}$ au at IRAS 2A and $r\gtrsim10^{3}$ au at IRAS 4A (outside the CO snowline).
\\ \\
With an ISM level cosmic-ray ionisation rate of $\xi_{\mathrm{CR}}(r)=$$1.0\times10^{-17}$ s$^{-1}$, longer timescale ($>10^{6}$ years, which is a typical age of Class II disks) than that in our modeling ($10^{5}$ years, which is a typical age of Class 0 protostars) is required to achieve a CO depletion by a factor of 10 and more \citep{Bosman2018, Eistrup2018, Schwarz2018}.
Moreover, in the inner envelopes where $\xi_{\mathrm{X}}(r)$$>\xi_{\mathrm{CR}}(r)$, efficient X-ray induced destruction reactions of other dominant molecules, especially H$_{2}$O, consider to supply more atomic oxygen, and it reacts with C$^{+}$ to returning to CO.
According to previous studies of chemical modeling 
with a timescale of $>10^{6}$ years, CO is chemically processed and the carbon is sequestered into less volatile species such as CH$_{3}$OH, CH$_{4}$, and CO$_{2}$ (e.g., \citealt{Furuya2014, Yu2016, Bosman2018, Schwarz2018, Schwarz2019, Krijt2020}). 
\\ \\ 
According to recent observations with e.g., ALMA, CO gas abundances in the Class 0-I disks with $<10^{6}$ years are consistent with the ISM abundance ($\sim10^{-4}$) within a factor of 2 \citep{Harsono2014, vantHoff2020, Zhang2020}, nearly one order of magnitude higher than the average value in $10^{6}-10^{7}$ years Class II disks (e.g., \citealt{Ansdell2016, Long2017, Zhang2019, Bergner2020}). 
\section{CH$_{4}$, HCN, and C$_{2}$H fractional abundances}
\begin{figure*}
\begin{center}
\includegraphics[scale=0.67]{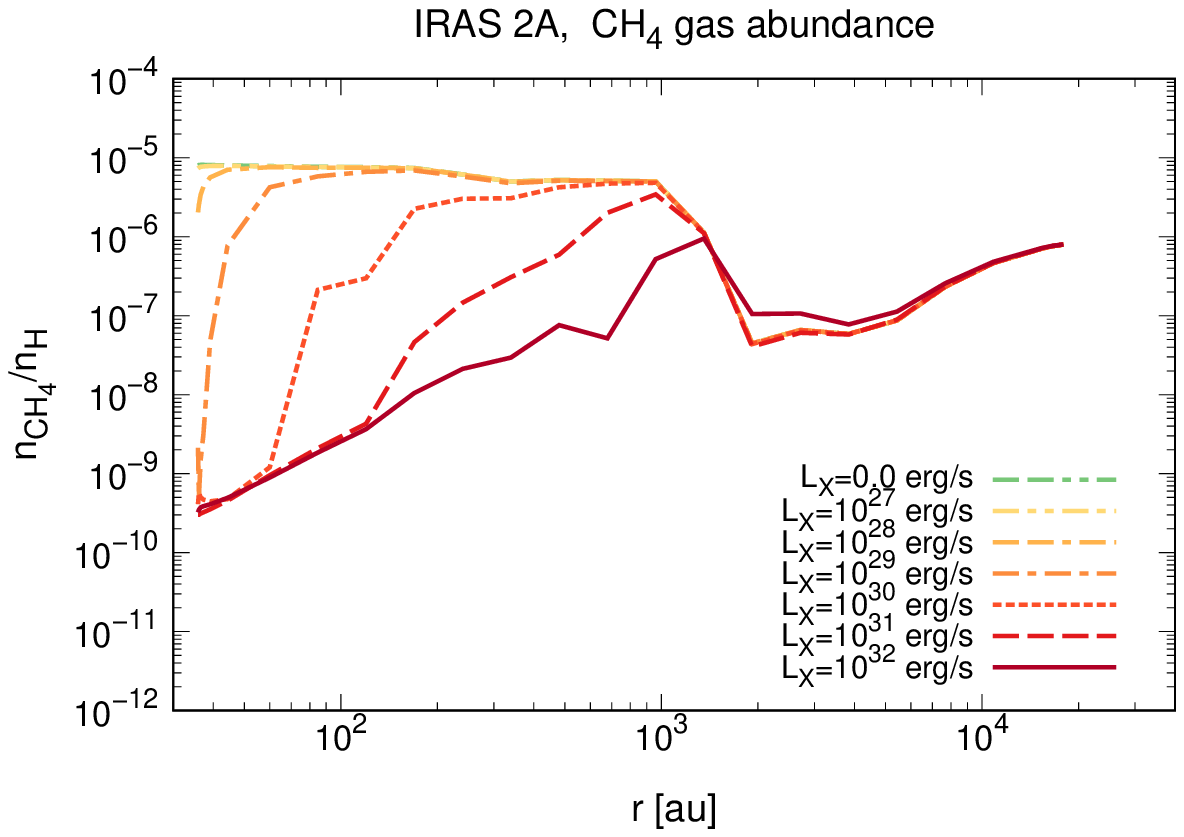}
\includegraphics[scale=0.67]{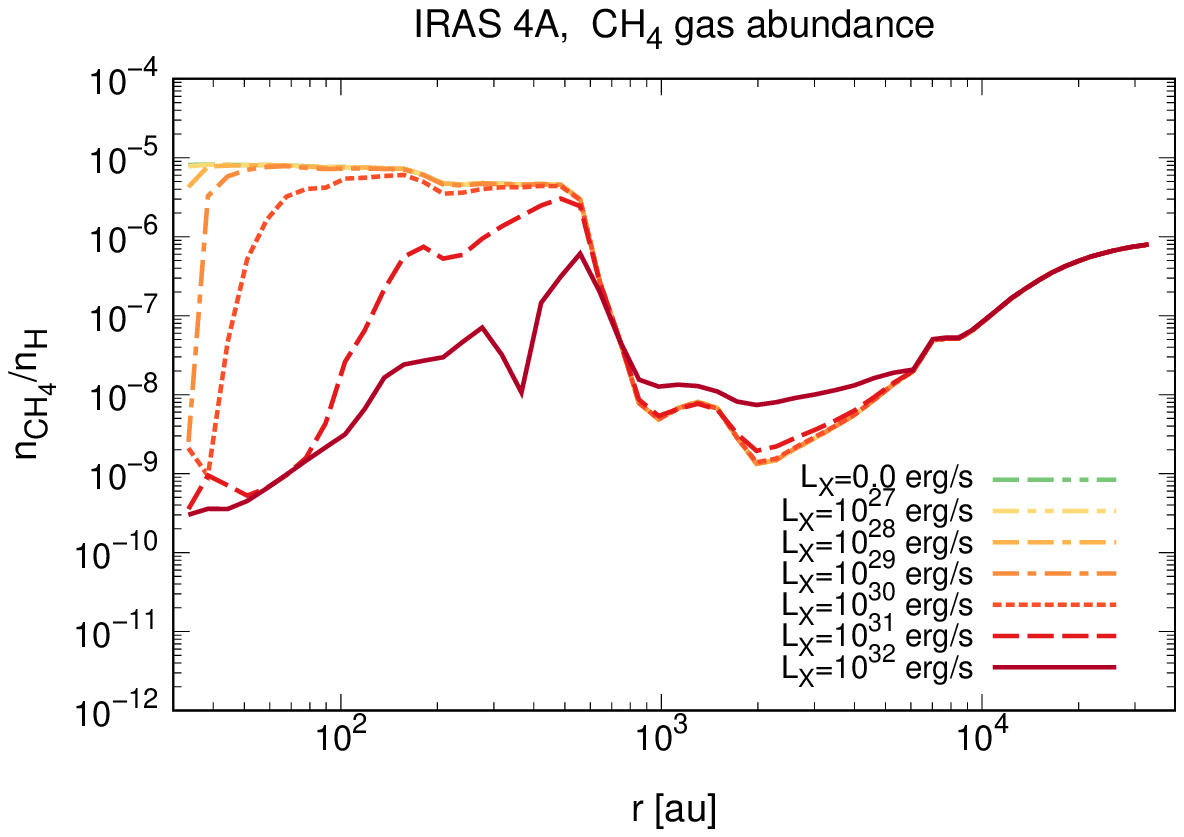}
\includegraphics[scale=0.67]{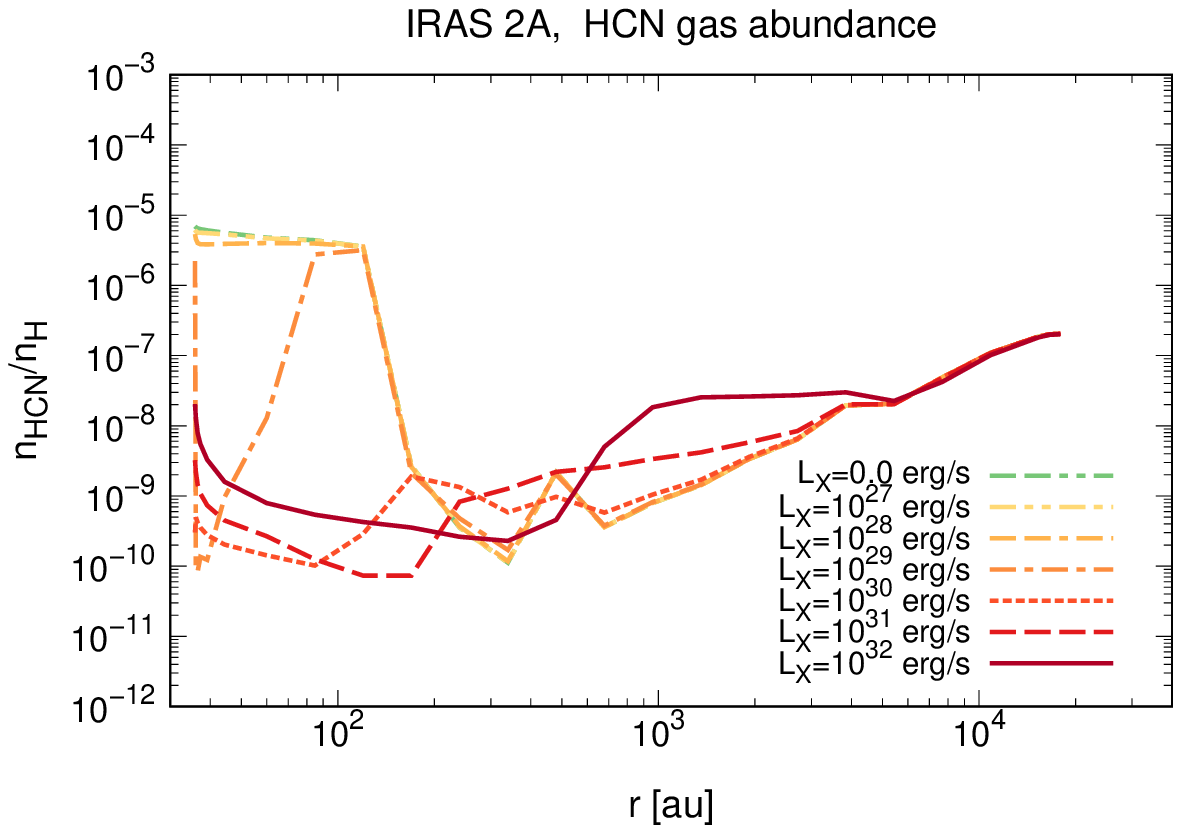}
\includegraphics[scale=0.67]{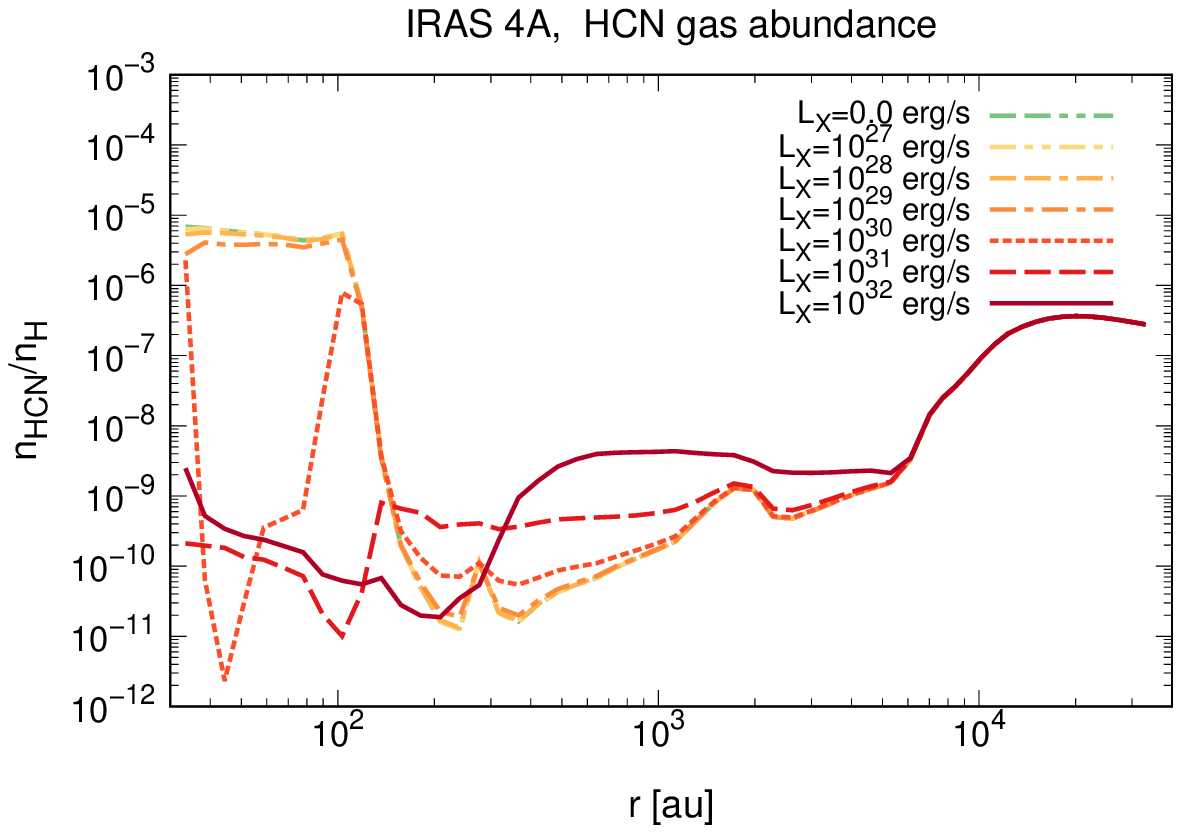}
\includegraphics[scale=0.67]{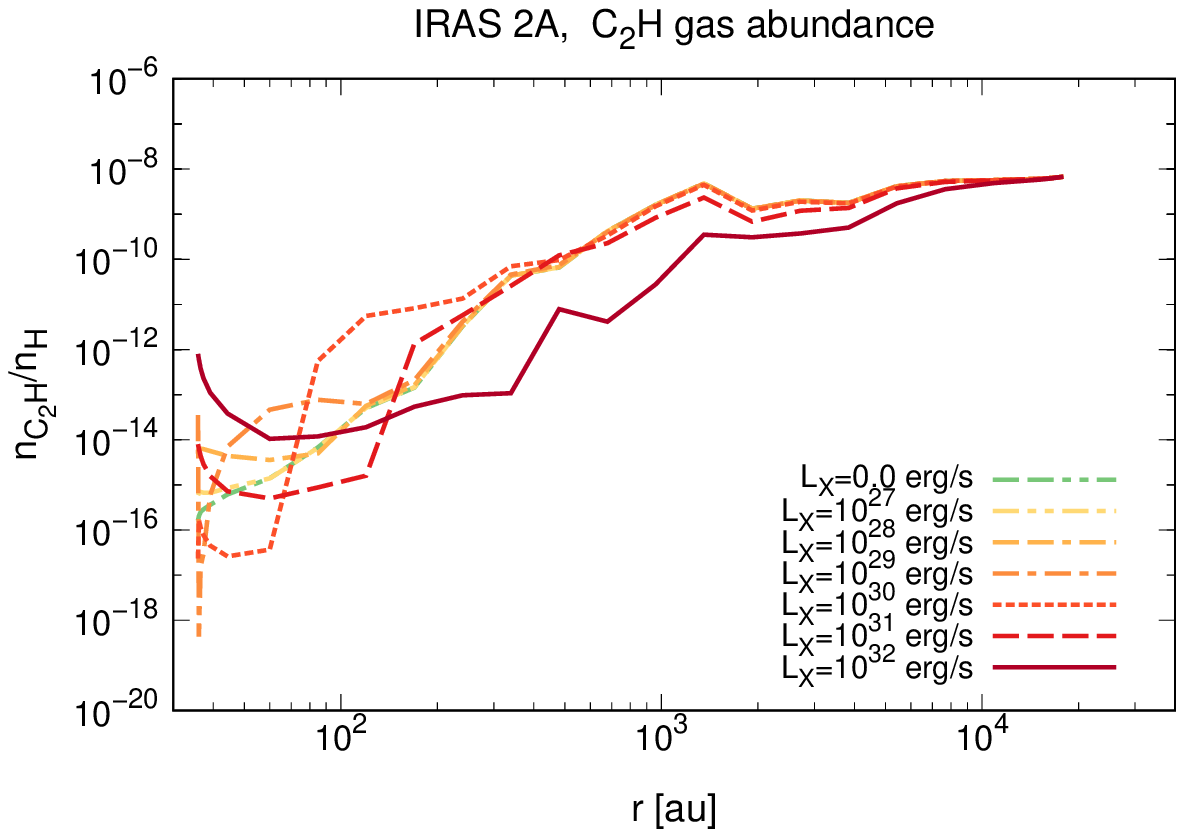}
\includegraphics[scale=0.67]{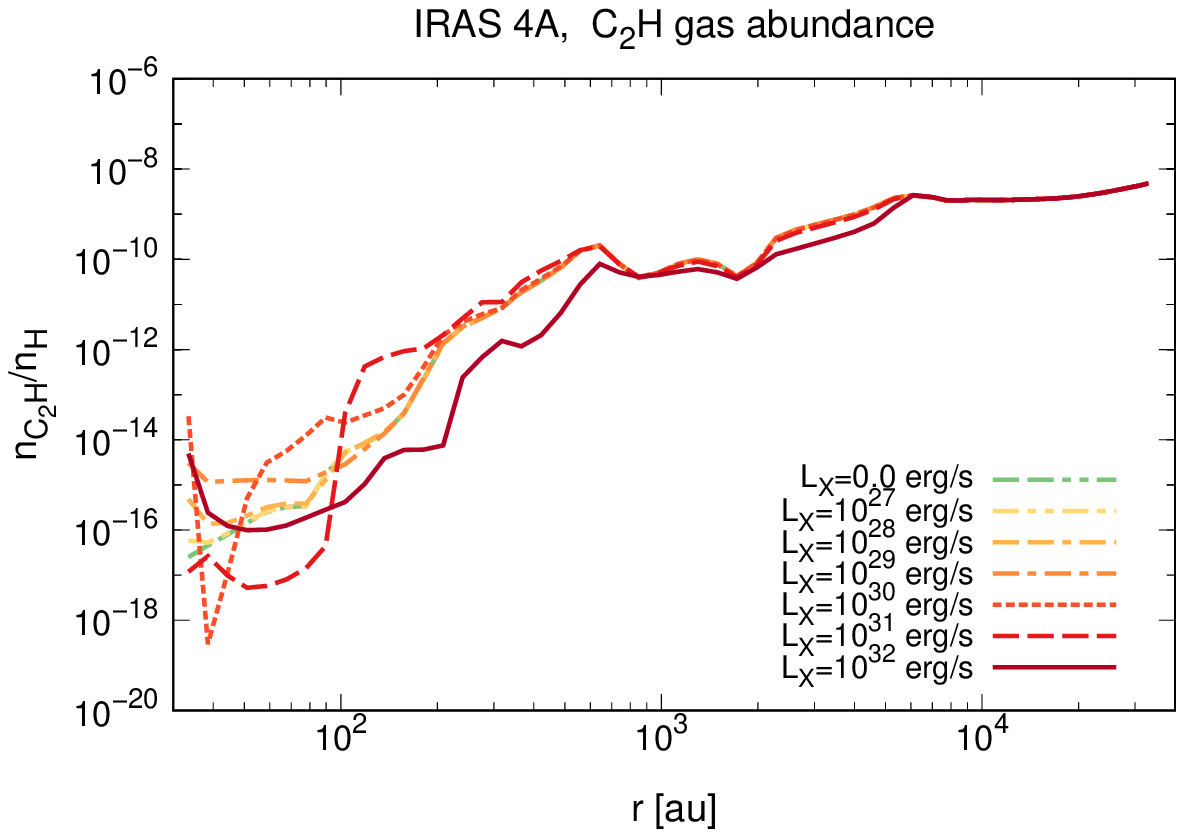}
\end{center}
\caption{
\noindent 
[Top panels]: The radial profiles of CH$_{4}$ fractional abundances $n_{\mathrm{CH}_{4}}$/$n_{\mathrm{H}}$ in NGC 1333-IRAS 2A (left panel) and NGC 1333-IRAS 4A (right panel) envelope models.
[Middle panels]: 
The radial profiles of HCN gas fractional abundances $n_{\mathrm{HCN}}$/$n_{\mathrm{H}}$ in the IRAS 2A (left panel) and IRAS 4A (right panel) envelope models.
[Bottom panels]: 
The radial profiles of C$_{2}$H gas fractional abundances $n_{\mathrm{C}_{2}\mathrm{H}}$/$n_{\mathrm{H}}$ in the IRAS 2A (left panel) and IRAS 4A (right panel) envelope models.
}\label{FigureD1_CH4&C2H&HCNgas}
\end{figure*}
Top panels of Figure \ref{FigureD1_CH4&C2H&HCNgas} show the radial profiles of CH$_{4}$ gas fractional abundances $n_{\mathrm{CH}_{4}}$/$n_{\mathrm{H}}$ in IRAS 2A (left panel) and IRAS 4A (right panel) envelope models, for the various X-ray luminosities.
For $L_{\mathrm{X}}\lesssim10^{28}$ erg s$^{-1}$, the CH$_{4}$ gas abundances are around $(3-8)\times10^{-6}$ at $r\lesssim10^{3}$ au in the IRAS 2A model and at $r\lesssim6\times10^{2}$ au in the IRAS 4A model (within the CH$_{4}$ snowline).
As the X-ray flux becomes large, the CH$_{4}$ gas abundances in these radii decrease.
For $L_{\mathrm{X}}\gtrsim10^{31}$ erg s$^{-1}$, the CH$_{4}$ gas abundances are around $10^{-9}-10^{-8}$ at $r\lesssim10^{2}$ au.
\\ \\
The X-ray induced photodissociation and ion-molecule reactions (with He$^{+}$, H$_{3}^{+}$ etc.) are considered to be dominant destruction processes of CH$_{4}$ in the inner envelopes (e.g., \citealt{Aikawa1999, Eistrup2016}).
We note that ion-molecule reactions of gas-phase CH$_{4}$ with e.g., C$^{+}$ within the CH$_{4}$ snowline are important to form unsaturated carbon chain molecules, such as C$_{2}$H, C$_{3}$H$_{2}$, and C$_{4}$H (WCCC, e.g., \citealt{Hassel2008, Sakai2008, Sakai2013, Aikawa2012, Aikawa2020}).
\\ \\
Middle panels of Figure \ref{FigureD1_CH4&C2H&HCNgas} show the radial profiles of HCN gas fractional abundances $n_{\mathrm{HCN}}$/$n_{\mathrm{H}}$ in IRAS 2A (left panel) and IRAS 4A (right panel) envelope models, for the various X-ray luminosities.
The binding energy of HCN is somewhat smaller than that of H$_{2}$O and similar to that of CH$_{3}$OH ($E_{\mathrm{des}}$(HCN)=3610 K, $E_{\mathrm{des}}$(H$_{2}$O)=4880 K, $E_{\mathrm{des}}$(CH$_{3}$OH)=3820 K, see Table \ref{Table:2}), and the HCN snowline position ($\sim2\times10^{2}$ au) exists outside the water snowline  ($\sim10^{2}$ au).
For low X-ray luminosities, the HCN gas abundances are around $\sim10^{-6}-10^{-5}$ within the HCN snowline.
As the X-ray flux becomes large, the HCN gas abundances in these radii decrease.
For high X-ray luminosities,
the HCN gas abundances are $\lesssim10^{-9}$ within the HCN snowline.
The X-ray induced photodissociation and ion-molecule reactions (with He$^{+}$, H$_{3}^{+}$ etc.) are considered to be dominant destruction processes of HCN in the inner envelopes (e.g., \citealt{Huntress1977, vanDishoeck2006, Walsh2015}).
\\ \\
Bottom panels of Figure \ref{FigureD1_CH4&C2H&HCNgas} show the radial profiles of C$_{2}$H gas fractional abundances $n_{\mathrm{C}_{2}\mathrm{H}}$/$n_{\mathrm{H}}$ in IRAS 2A (left panel) and IRAS 4A (right panel) envelope models, for the various X-ray luminosities.
The C$_{2}$H gas fractional abundances are around $10^{-10}-10^{-8}$ at $r\gtrsim10^{3}$ au, and decrease at the inner radii. 
Within $10^{2}$ au, they are $\sim10^{-16}-10^{-13}$ at IRAS 2A and $\sim10^{-17}-10^{-14}$ at IRAS 2A.
The dependance of X-ray fluxes are smaller (within two orders of magnitude) than other dominant carbon-bearing molecules such as CH$_{3}$OH, HCN, and CH$_{4}$.
C$_{2}$H is the representative products of WCCC in star-forming cores (e.g., \citealt{Sakai2013, Higuchi2018, Aikawa2020}, see also Section 4.6), and it is mainly produced by the ion-molecule reaction of CH$_{4}$ gas with C$^{+}$ and the subsequent electron recombination reaction \citep{Aikawa2012}. C$_{2}$H also reacts with C$^{+}$, which links to the formation of longer carbon chain molecules.
\citet{Henning2010} discussed that in the atmospheres of Class II disks, the C$_{2}$H abundance is larger around Herbig Ae stars with stronger X-rays, than that in T Tauri stars.
\section{NH$_{3}$ and N$_{2}$ fractional abundances}
\begin{figure*}
\begin{center}
\includegraphics[scale=0.67]{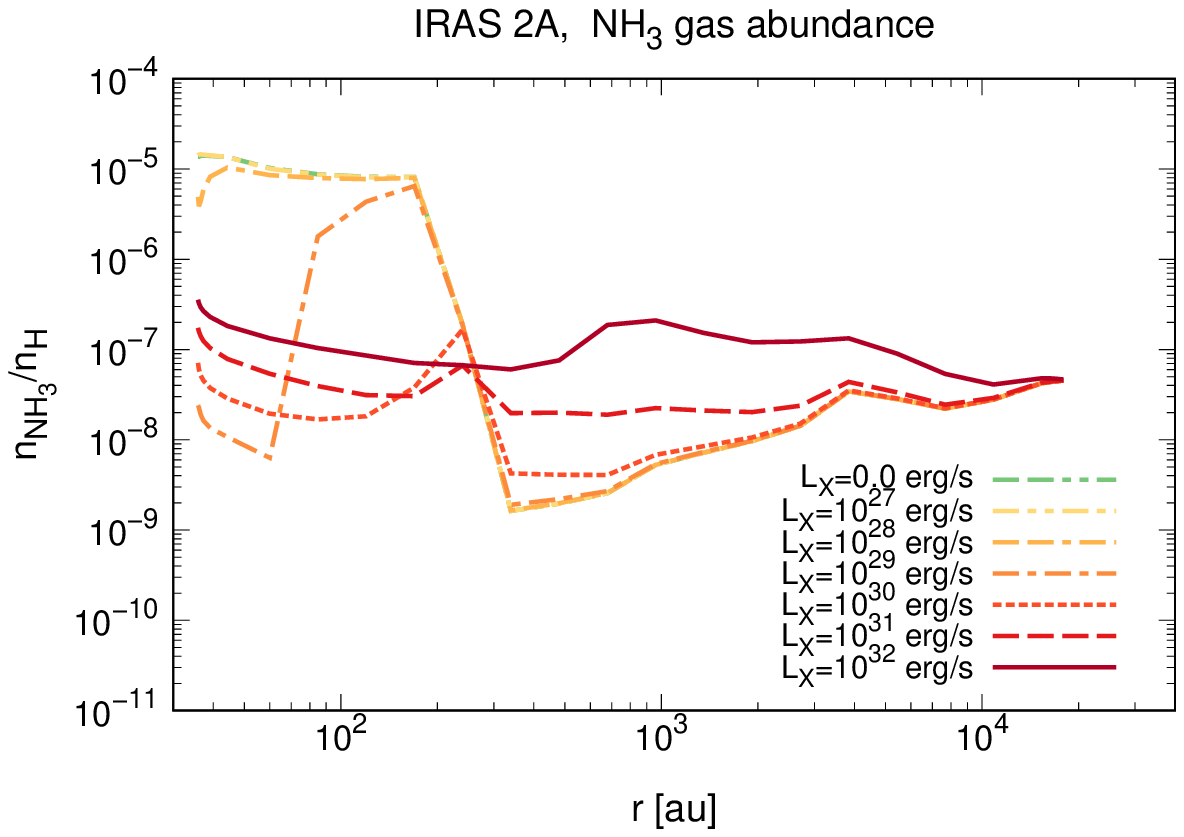}
\includegraphics[scale=0.67]{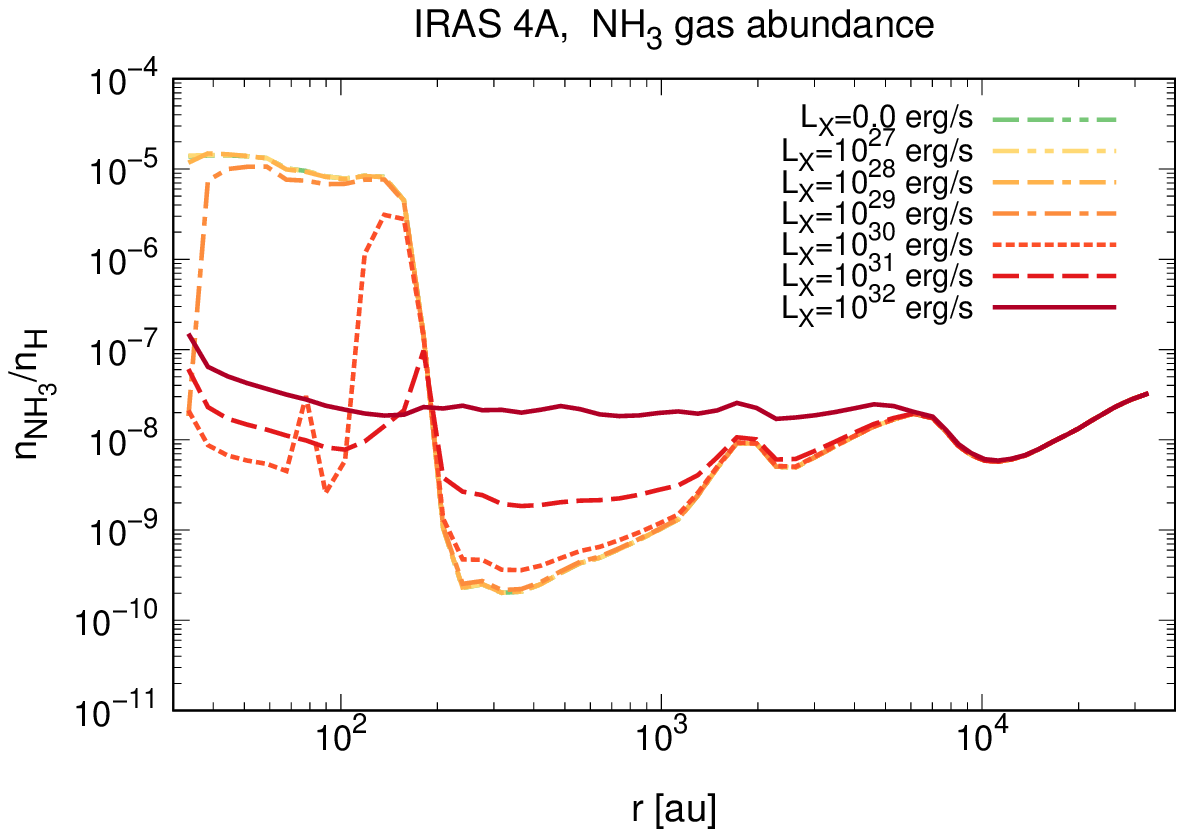}
\includegraphics[scale=0.67]{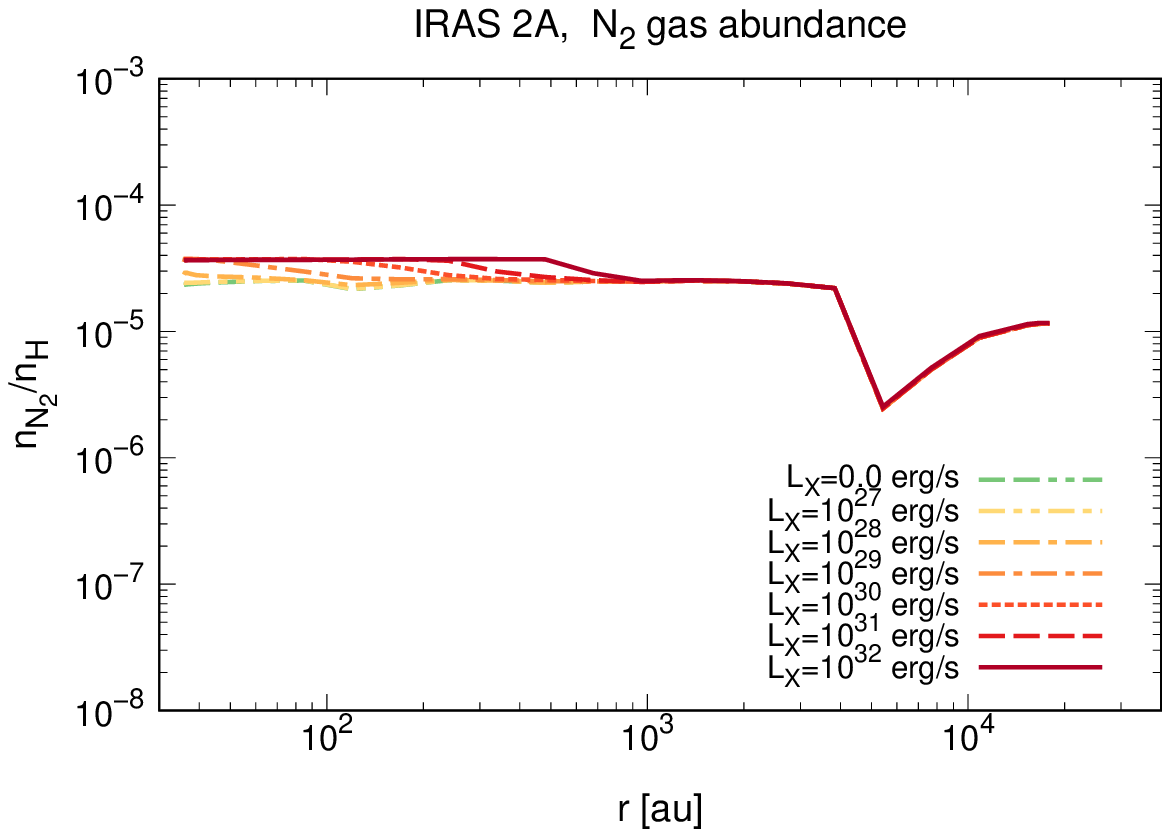}
\includegraphics[scale=0.67]{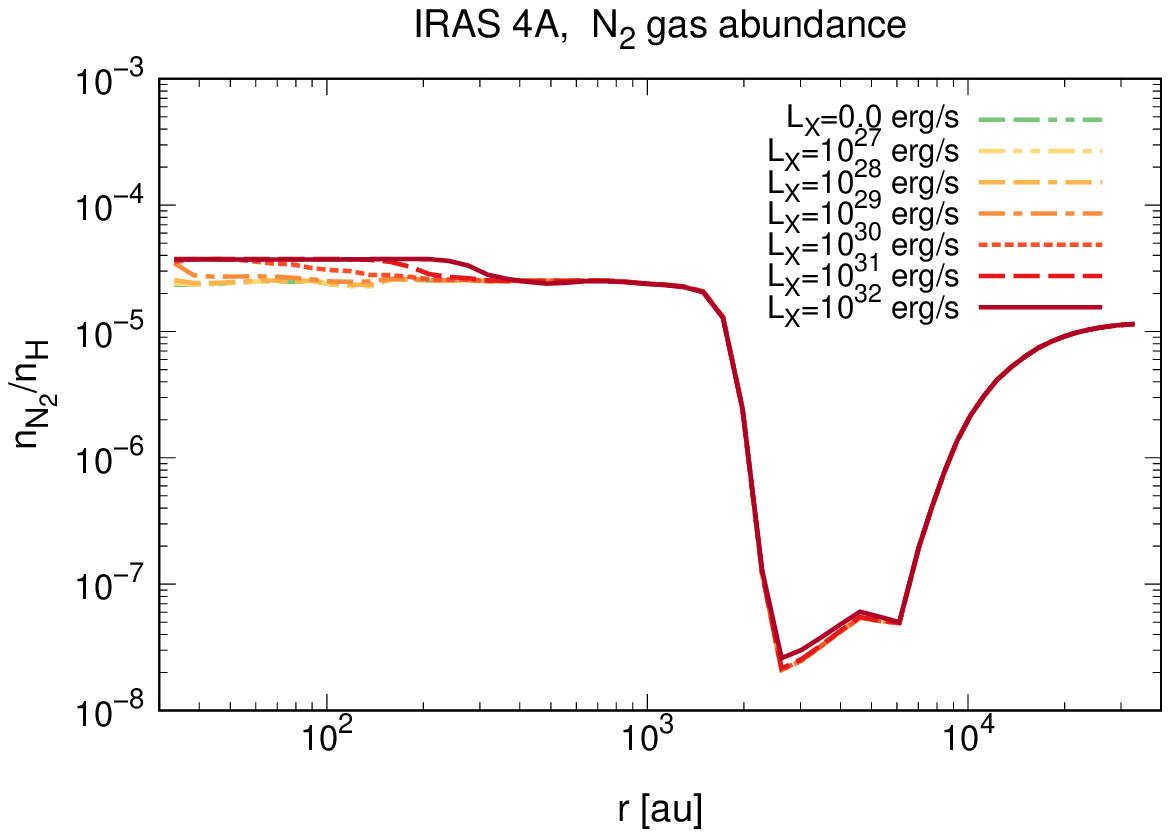}
\end{center}
\caption{
\noindent 
[Top panels]: The radial profiles of NH$_{3}$ fractional abundances $n_{\mathrm{NH}_{3}}$/$n_{\mathrm{H}}$ in NGC 1333-IRAS 2A (left panel) and NGC 1333-IRAS 4A (right panel) envelope models.
[Bottom panels]: 
The radial profiles of N$_{2}$ gas fractional abundances $n_{\mathrm{N}_{2}}$/$n_{\mathrm{H}}$ in the IRAS 2A (left panel) and IRAS 4A (right panel) envelope models.
}\label{FigureE1_NH3&N2gas}
\end{figure*}
Top panels of Figure \ref{FigureE1_NH3&N2gas} show the radial profiles of NH$_{3}$ gas fractional abundances $n_{\mathrm{NH}_{3}}$/$n_{\mathrm{H}}$ in IRAS 2A (left panel) and IRAS 4A (right panel) envelope models, for the various X-ray luminosities.
The binding energy of NH$_{3}$ is smaller than that of H$_{2}$O and HCN, and larger than that of CO$_{2}$ ($E_{\mathrm{des}}$(NH$_{3}$)$=$2715 K, $E_{\mathrm{des}}$(HCN)$=$3610 K, $E_{\mathrm{des}}$(CO$_{2}$)$=$2267 K, and $E_{\mathrm{des}}$(H$_{2}$O)$=$4880 K, see Table \ref{Table:2}), and the NH$_{3}$ snowline position exists outside the water snowline and the HCN snowline ($>2\times10^{2}$ au) and inside the CO$_{2}$ snowline ($<3\times10^{2}$ au).
\citet{Zhang2018} discussed the possibility to observe the NH$_{3}$ line emission (the 23GHz 1,1 and 2,2 lines) with ngVLA as a proxy of the water snowline in disks.
\\ \\
According to Figure \ref{FigureE1_NH3&N2gas}, for $L_{\mathrm{X}}\gtrsim10^{31}$ erg s$^{-1}$, NH$_{3}$ gas abundances become higher (up to $\sim10^{-8}-10^{-7}$) outside the NH$_{3}$ snowline, compared with the values ($\sim10^{-9}-10^{-8}$ in IRAS 2A and $\sim10^{-10}-10^{-9}$ in IRAS 4A) for $L_{\mathrm{X}}\lesssim10^{30}$ erg s$^{-1}$.
The X-ray induced photodesorption is considered to be important in this region. 
Inside the NH$_{3}$ snowline,
for low X-ray luminosities,
the NH$_{3}$ snowline gas abundances are around $\sim10^{-5}$.
As the X-ray fluxes increase, the NH$_{3}$ snowline gas abundances in these radii decrease.
For $L_{\mathrm{X}}\gtrsim10^{30}$ erg s$^{-1}$, the NH$_{3}$ gas abundances are $\lesssim10^{-8}-10^{-7}$ within the NH$_{3}$ snowline.
The X-ray induced photodissociation and ion-molecule reactions are considered to be dominant destruction processes of NH$_{3}$ in the inner envelopes (e.g., \citealt{Gredel1989, Walsh2015, Eistrup2016}).
Thus, with strong X-ray field, the NH$_{3}$ abundance is no longer dominant nitrogen carrier.
In addition, it cannot be used as the tracer of the water snowline position, since NH$_{3}$ gas abundances are similar within and outside the NH$_{3}$ snowline.
\\ \\
The NH$_{3}$ line emission (e.g., the 23GHz (1,1), (2,2), and (3,3) lines) have been observed toward protostar disks and envelopes using VLA (e.g., \citealt{Choi2007, Choi2010}).
However, the spatial resolutions of such VLA observations ($\sim1.0"$) were not sufficient to resolve the inner structures around the protostars at a few hundred pc.
Future ngVLA observations of these NH$_{3}$ lines with much higher resolutions ($\lesssim0.3"$ at around 23 GHz) will resolve the NH$_{3}$ gas emission within its snowline towards many protostars, 
and can also constrain the impact of X-rays on NH$_{3}$ gas abundances. In addition, such observations will be important to trace the chemical history of nitrogen bearing molecules.
\\ \\
Bottom panels of Figure \ref{FigureE1_NH3&N2gas} show the radial profiles of N$_{2}$ gas fractional abundances $n_{\mathrm{N}_{2}}$/$n_{\mathrm{H}}$ in IRAS 2A (left panel) and IRAS 4A (right panel) envelope models, for the various X-ray luminosities.
Like CO, N$_{2}$ gas abundances do not depend on X-ray fluxes.
\section{The fractional abundances and percentage contributions of major oxygen bearing molecules}
Table \ref{Table:3} shows the fractional abundances of major oxygen bearing molecules at $r=60$ au ($T_{\mathrm{gas}}\sim$150 K, inside the water snowline) in the IRAS 2A and IRAS 4A envelope models for the various X-ray luminosities, and their percentage contributions (see also Figures \ref{Figure8_Percentage} and \ref{Figure9_Percentage_PieChart} in Section 4.1). 
The cut-off threshold is 0.01\% for the contributions.
\begin{table*}
\caption{The fractional abundances of major oxygen bearing molecules at $r=60$ au ($T_{\mathrm{gas}}\sim$150 K, inside the water snowline) in the NGC 1333-IRAS 2A and NGC 1333-IRAS 4A envelope models for the various X-ray luminosities, and their percentage contributions relative to the total elemental oxygen abundance ($3.2\times10^{-4}$)}              
\label{Table:3}      
\centering                                      
\begin{tabular}{c c c c c}         
\hline\hline                        
$L_{\mathrm{X}}$ [erg s$^{-1}$]& H$_2$O gas & CO gas & O gas & O$_{2}$ gas \\    
\hline
&&NGC 1333-IRAS 2A&& \\                            
 0 &$2.0\times10^{-4}$, 61\%&$1.1\times10^{-4}$, 34\%&$1.2\times10^{-9}$, $<$$10^{-2}$\%&$3.7\times10^{-7}$, $2.3\times10^{-1}$\% \\
 10$^{27}$&$2.0\times10^{-4}$, 61\%&$1.1\times10^{-4}$, 33\%&$2.5\times10^{-10}$, $<$$10^{-2}$\%&$3.0\times10^{-7}$, $1.9\times10^{-1}$\%\\ 
 10$^{28}$ &$1.9\times10^{-4}$, 60\%&$1.1\times10^{-4}$, 33\%&$1.7\times10^{-10}$, $<$$10^{-2}$\%&$1.9\times10^{-8}$, $1.2\times10^{-2}$\%\\
 10$^{29}$ &$1.5\times10^{-4}$, 48\%&$8.1\times10^{-5}$, 25\%&$4.0\times10^{-8}$, $1.3\times10^{-2}$\%&$1.5\times10^{-6}$, $9.2\times10^{-1}$\%\\
 10$^{30}$ &$1.9\times10^{-8}$, $<$$10^{-2}$\%&$1.4\times10^{-4}$, 43\%&$5.8\times10^{-5}$, 18\%&$6.0\times10^{-5}$, 38\%\\
 10$^{31}$ &$5.3\times10^{-8}$, $1.6\times10^{-2}$\%&$1.4\times10^{-4}$, 44\%&$6.2\times10^{-5}$, 19\%&$5.9\times10^{-5}$, 37\%\\
 10$^{32}$ &$1.4\times10^{-7}$, $4.5\times10^{-2}$\%&$1.4\times10^{-4}$, 44\%&$6.5\times10^{-5}$, 20\%&$5.8\times10^{-5}$, 36\%\\
\hline            
&&NGC 1333-IRAS 4A&& \\                            
 0&$2.0\times10^{-4}$, 62\%&$1.1\times10^{-4}$, 33\%&$1.1\times10^{-10}$, $<$$10^{-2}$\%&$1.0\times10^{-9}$, $<$$10^{-2}$\% \\
 10$^{27}$ &$2.0\times10^{-4}$, 62\%&$1.1\times10^{-4}$, 33\%&$1.1\times10^{-10}$, $<$$10^{-2}$\%&$1.0\times10^{-9}$, $<$$10^{-2}$\%\\
 10$^{28}$ &$2.0\times10^{-4}$, 62\%&$1.0\times10^{-4}$, 33\%&$1.2\times10^{-10}$, $<$$10^{-2}$\%&$1.0\times10^{-9}$, $<$$10^{-2}$\%\\
 10$^{29}$ &$1.9\times10^{-4}$, 60\%&$9.9\times10^{-5}$, 31\%&$2.6\times10^{-10}$, $<$$10^{-2}$\%&$1.5\times10^{-9}$, $<$$10^{-2}$\%\\
 10$^{30}$ &$7.0\times10^{-5}$, 22\%&$7.2\times10^{-5}$, 23\%&$2.0\times10^{-7}$, $6.3\times10^{-2}$\%&$2.6\times10^{-5}$, 16\%\\
 10$^{31}$ &$1.3\times10^{-8}$, $<$$10^{-2}$\%&$1.4\times10^{-4}$, 43\%&$5.3\times10^{-5}$, 17\%&$6.2\times10^{-5}$, 39\%\\
 10$^{32}$ &$3.7\times10^{-8}$, $1.1\times10^{-2}$\%&$1.4\times10^{-4}$, 43\%&$6.2\times10^{-5}$, 19\%&$5.9\times10^{-5}$, 37\%\\
 \hline                                 
%
\end{tabular}
\end{table*}
\section{The additional model calculations for X-ray induced photodesorption rates}
\begin{figure*}
\begin{center}
\includegraphics[scale=0.67]{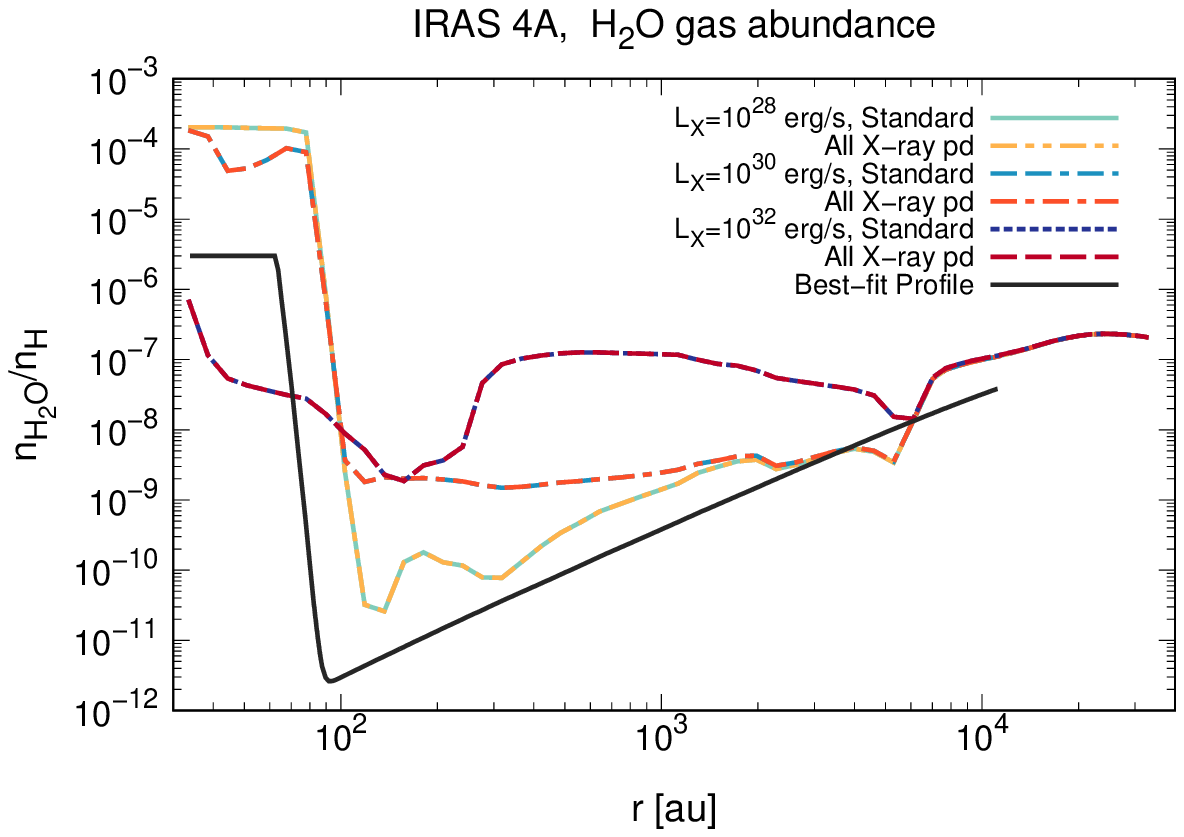}
\includegraphics[scale=0.67]{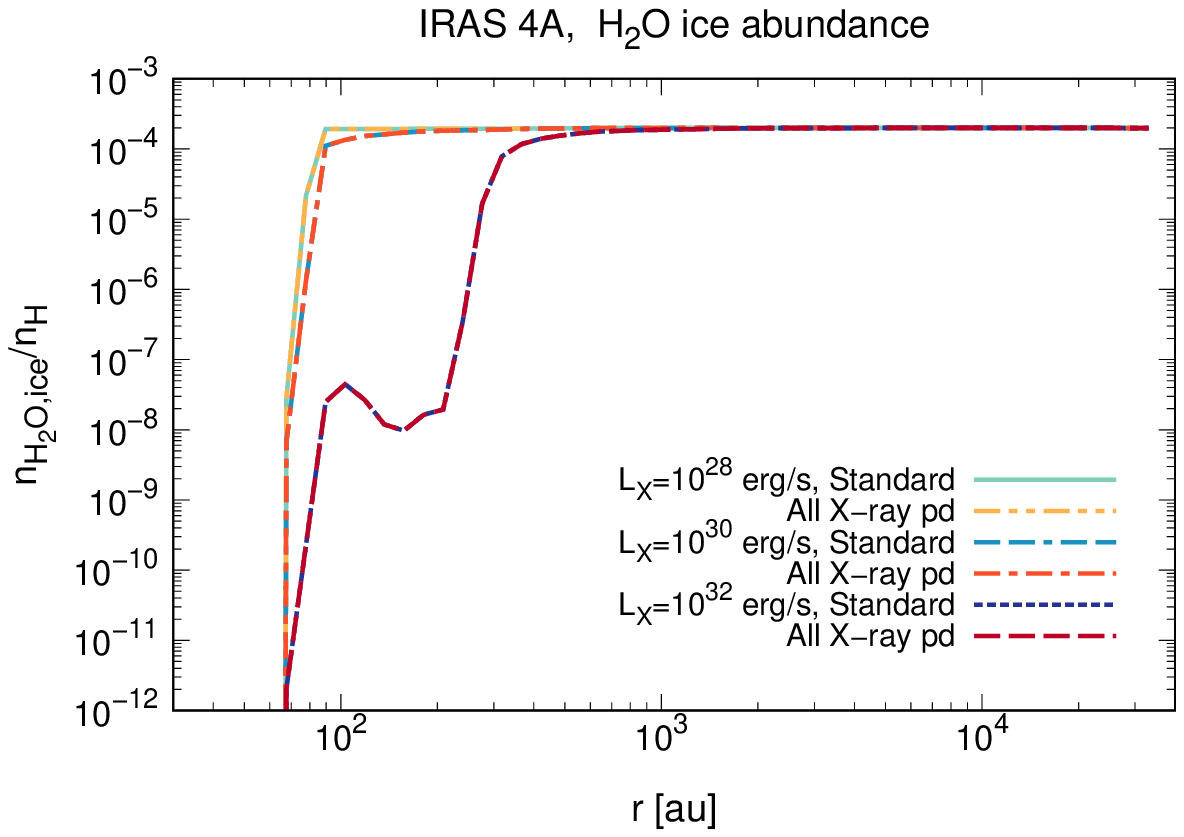}
\includegraphics[scale=0.67]{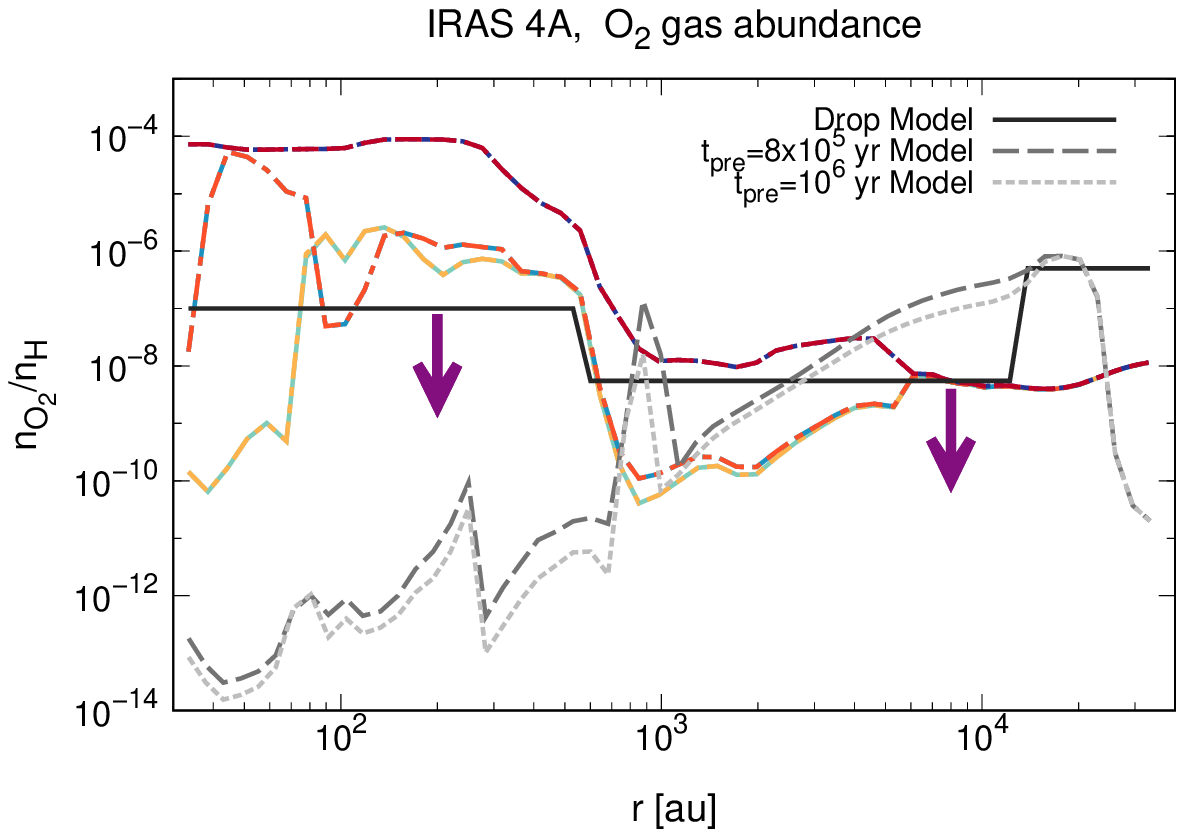}
\includegraphics[scale=0.67]{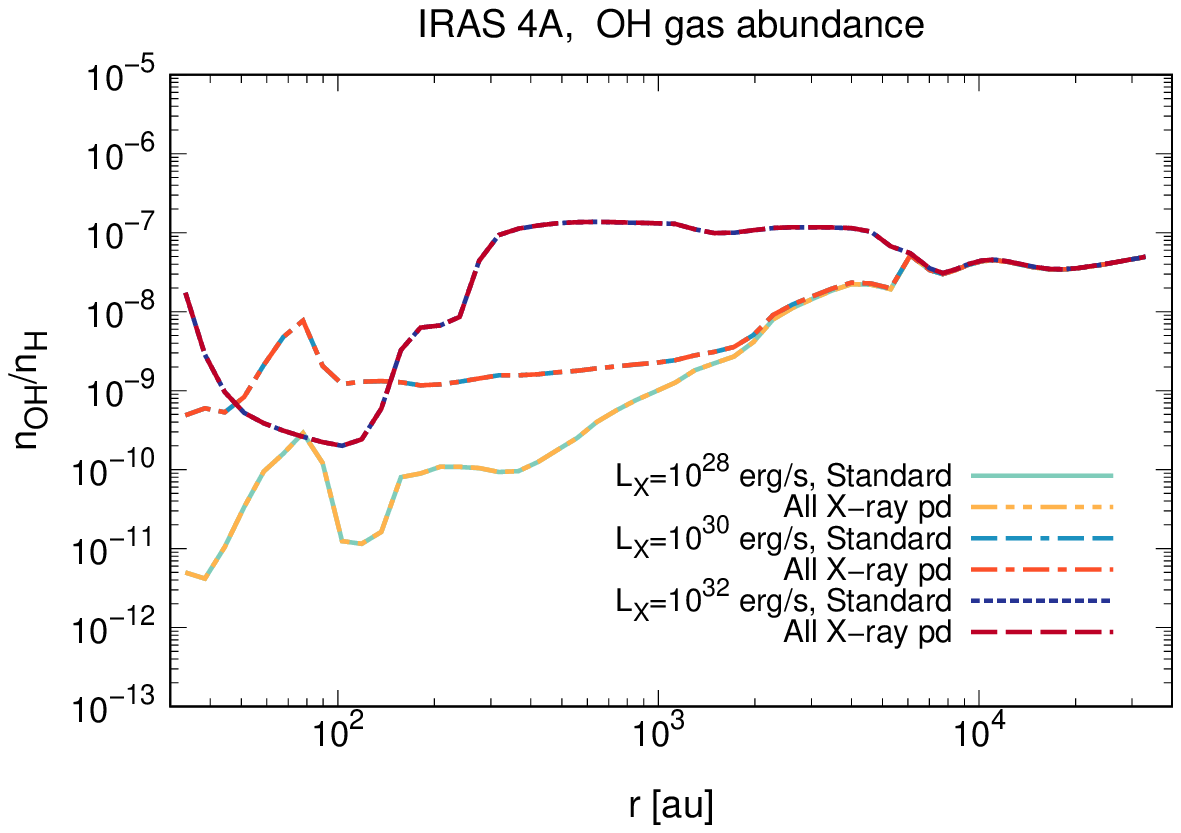}
\end{center}
\caption{
\noindent 
The radial profiles of gaseous fractional abundances 
of H$_{2}$O, O$_{2}$, and OH,
and icy fractional abundances of H$_{2}$O
in the NGC 1333-IRAS 4A envelope models.
The light-green solid lines, the cyan double-dashed dotted lines, and blue dotted lines show the radial profiles of our standard model calculations, for values of central star X-ray luminosities
$L_{\mathrm{X}}$=10$^{28}$, 10$^{30}$, and 10$^{32}$ erg s$^{-1}$, respectively (see also Figures \ref{Figure3_H2Ogas}, \ref{Figure4_H2Oice}, \ref{Figure5_O2&Ogas}, \ref{Figure6_HCO+&OHgas}).
The yellow dashed double-dotted lines, the scarlet dashed dotted lines, and the brown dashed lines show the radial profiles of our additional model calculations for $L_{\mathrm{X}}$=10$^{28}$, 10$^{30}$, and 10$^{32}$ erg s$^{-1}$, respectively. In the additional model calculations, we include the photodesorption by UV photons generated internally via the interaction of secondary electrons produced by X-rays with H$_{2}$ molecules (see also Figure \ref{FigureG2_add_Xp-rev1_model-comparisons}).
In the top left panel, the observational best-fit H$_{2}$O gas abundance profile obtained in \citet{vanDishoeck2021} is over-plotted with the black solid line (see also Figure \ref{Figure10_H2O&O2gas_obs-plot}).
In the bottom left panel, the three model O$_{2}$ gas abundance profiles obtained in \citet{Yildiz2013} are over-plotted (see also Figure \ref{Figure10_H2O&O2gas_obs-plot}).
}\label{FigureG1_add_Xp-rev1_model-comparisons}
\end{figure*} 
\begin{figure*}
\begin{center}
\includegraphics[scale=0.67]{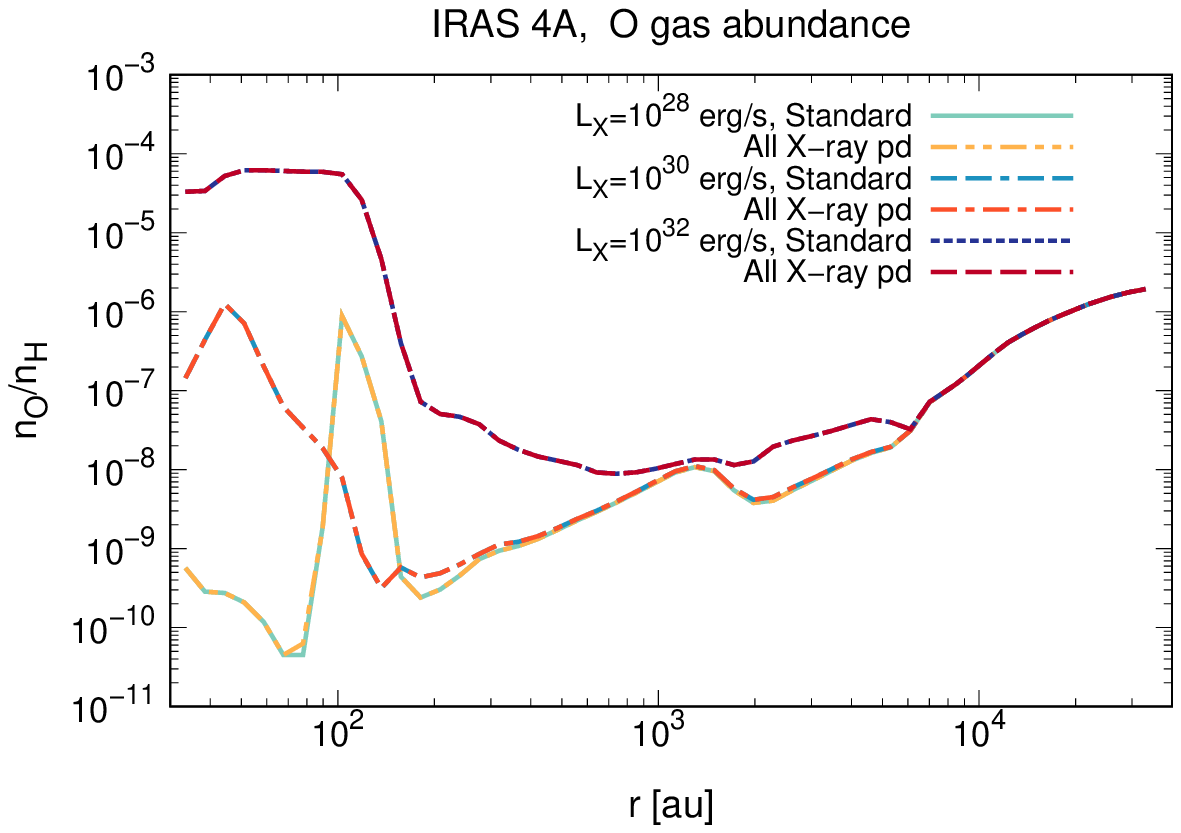}
\includegraphics[scale=0.67]{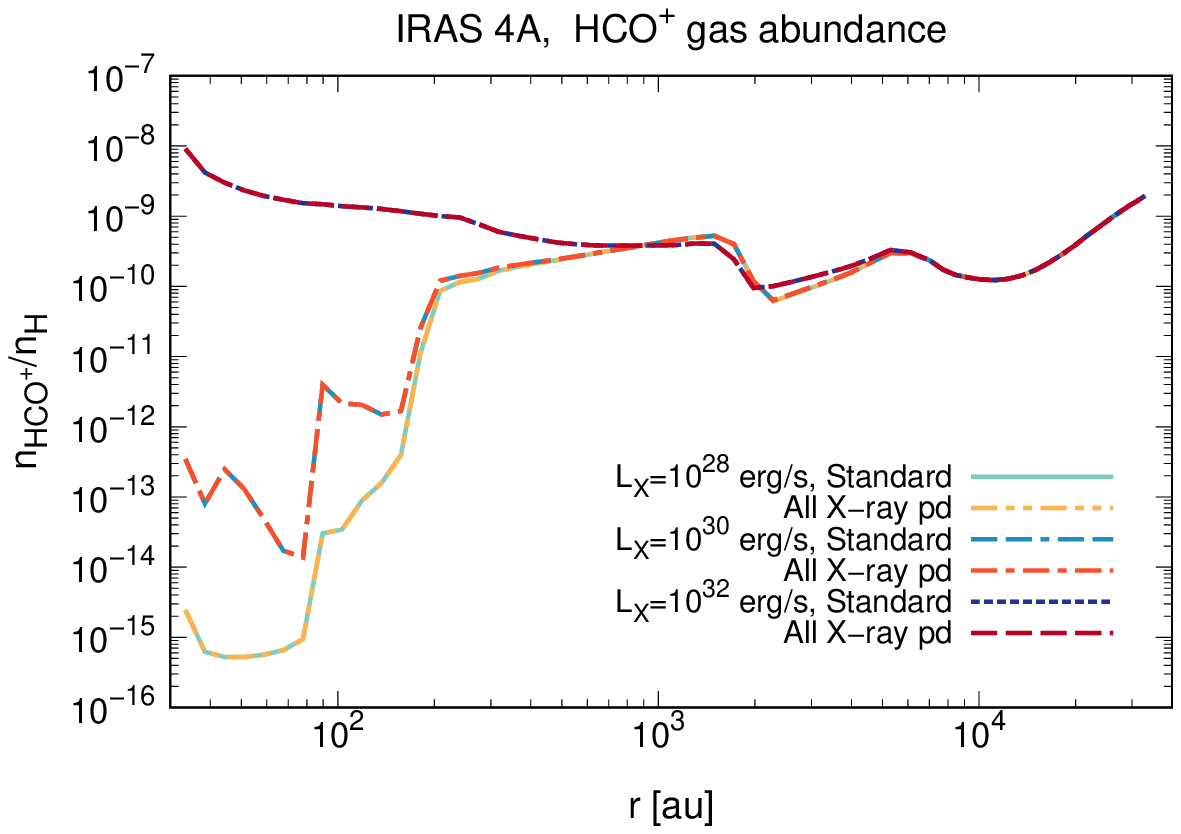}
\includegraphics[scale=0.67]{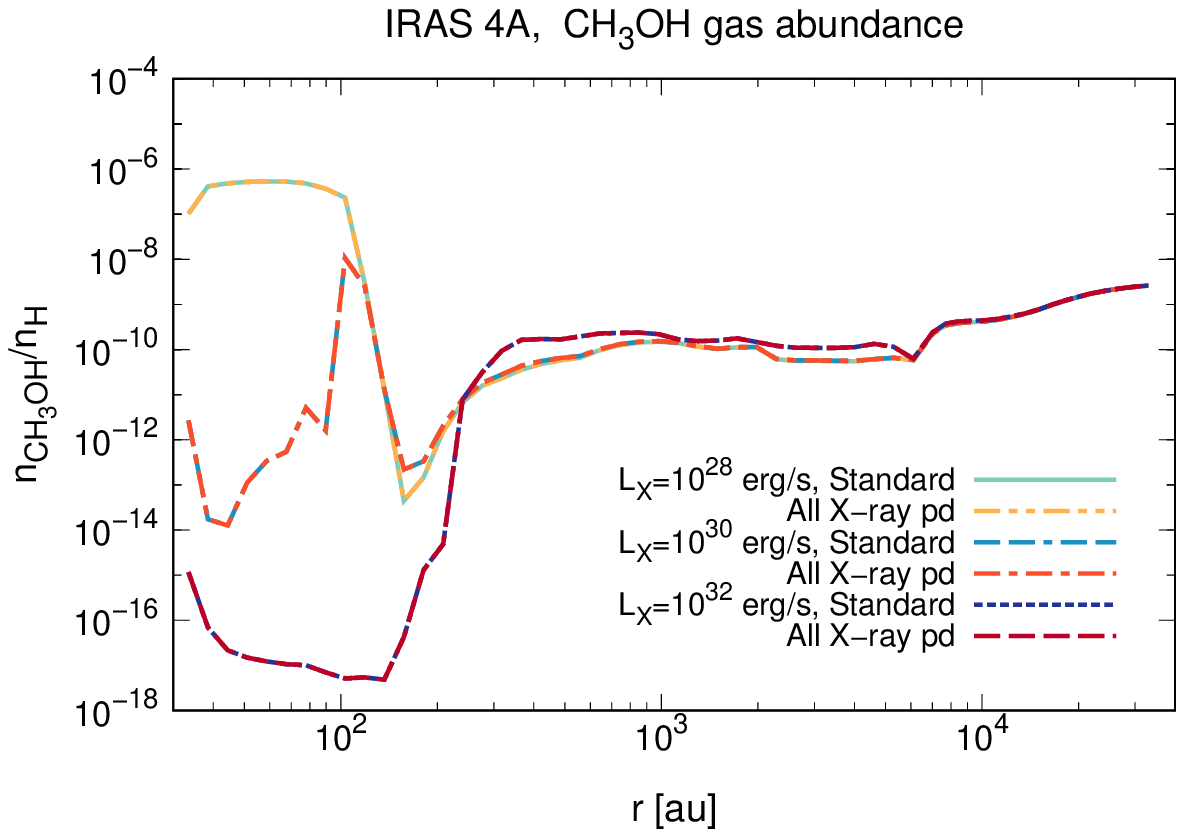}
\includegraphics[scale=0.67]{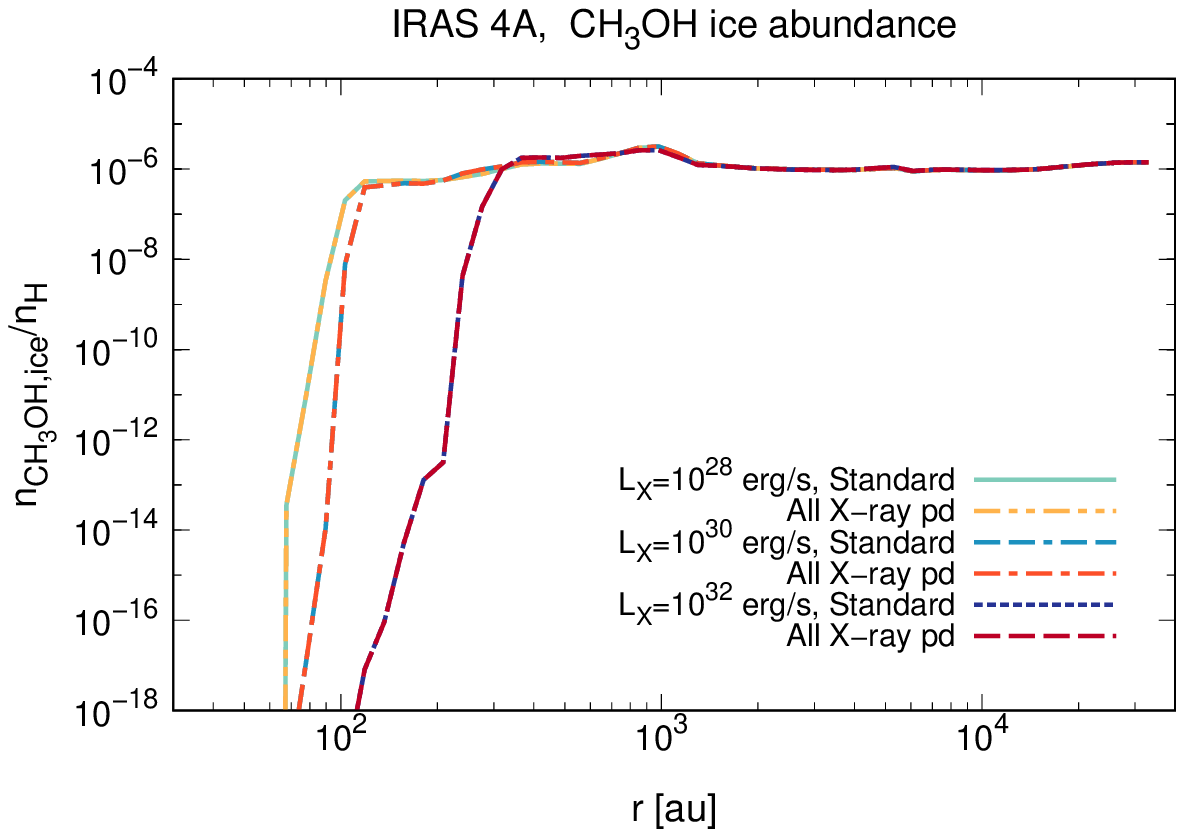}
\end{center}
\caption{
\noindent 
The radial profiles of gaseous fractional abundances 
of 
O, HCO$^{+}$, and CH$_{3}$OH,
and icy fractional abundances of
CH$_{3}$OH
in the NGC 1333-IRAS 4A envelope models.
The line types and color patterns for the radial profiles of our standard model calculations (see also Figures
\ref{Figure5_O2&Ogas}, \ref{Figure6_HCO+&OHgas}, and \ref{Figure7_CH3OHgas&ice}) and additional model calculations are same as Figure \ref{FigureG1_add_Xp-rev1_model-comparisons}.
In the additional model calculations, we include the secondary (indirect) X-ray induced photodesorption (see also Figure \ref{FigureG1_add_Xp-rev1_model-comparisons}).
}\label{FigureG2_add_Xp-rev1_model-comparisons}
\end{figure*} 
\begin{figure*}
\begin{center}
%
\includegraphics[scale=0.67]{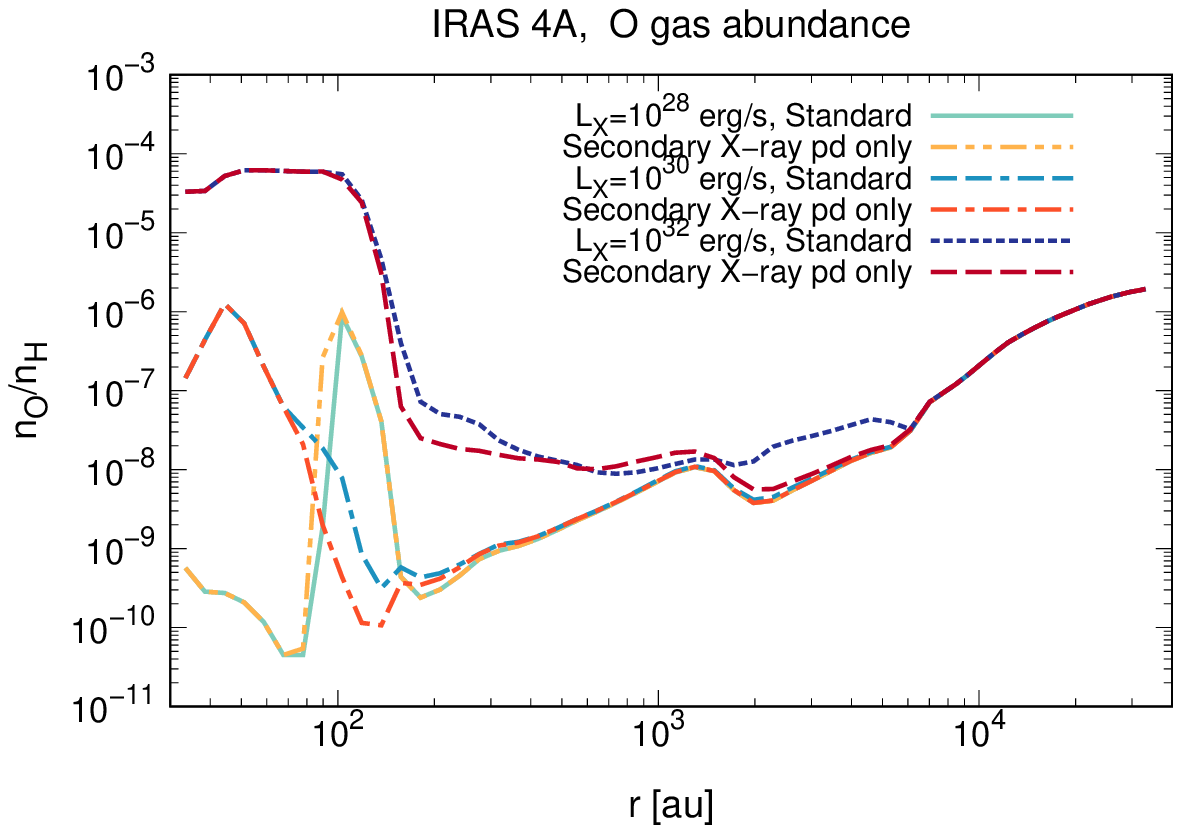}
\includegraphics[scale=0.67]{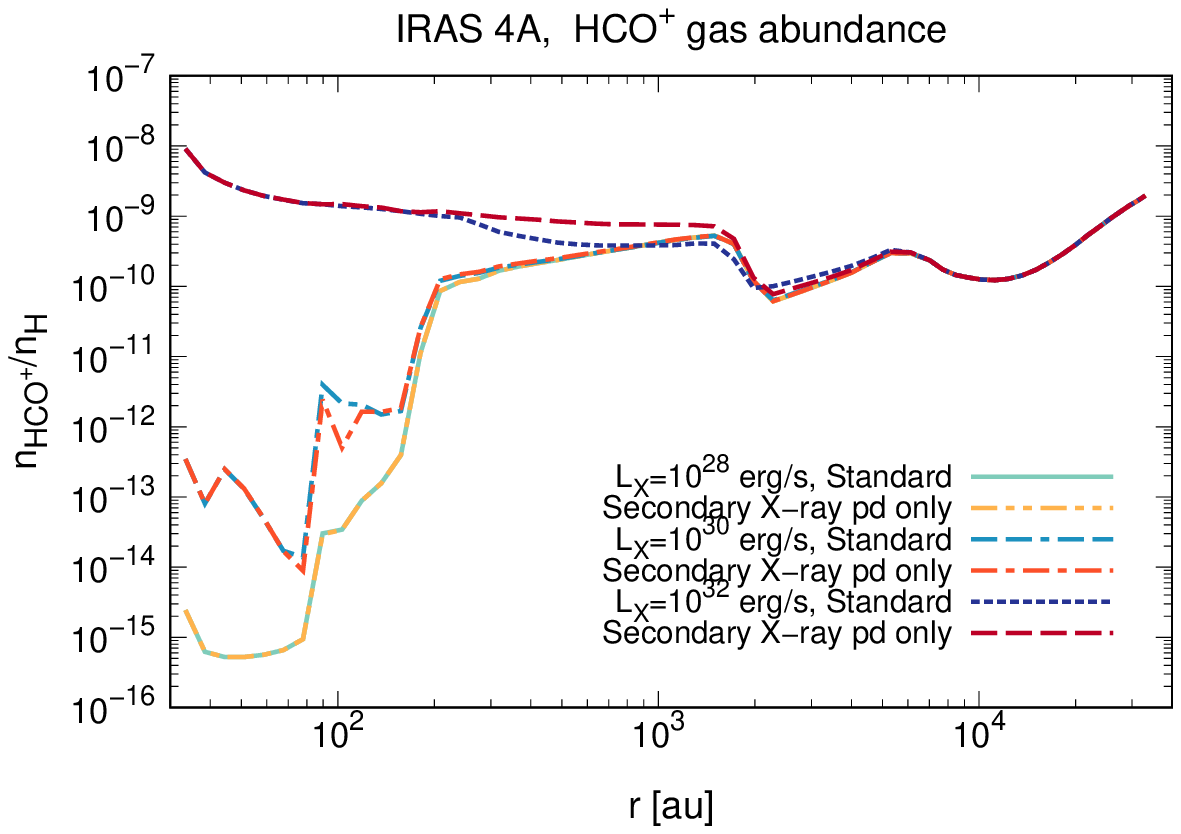}
\includegraphics[scale=0.67]{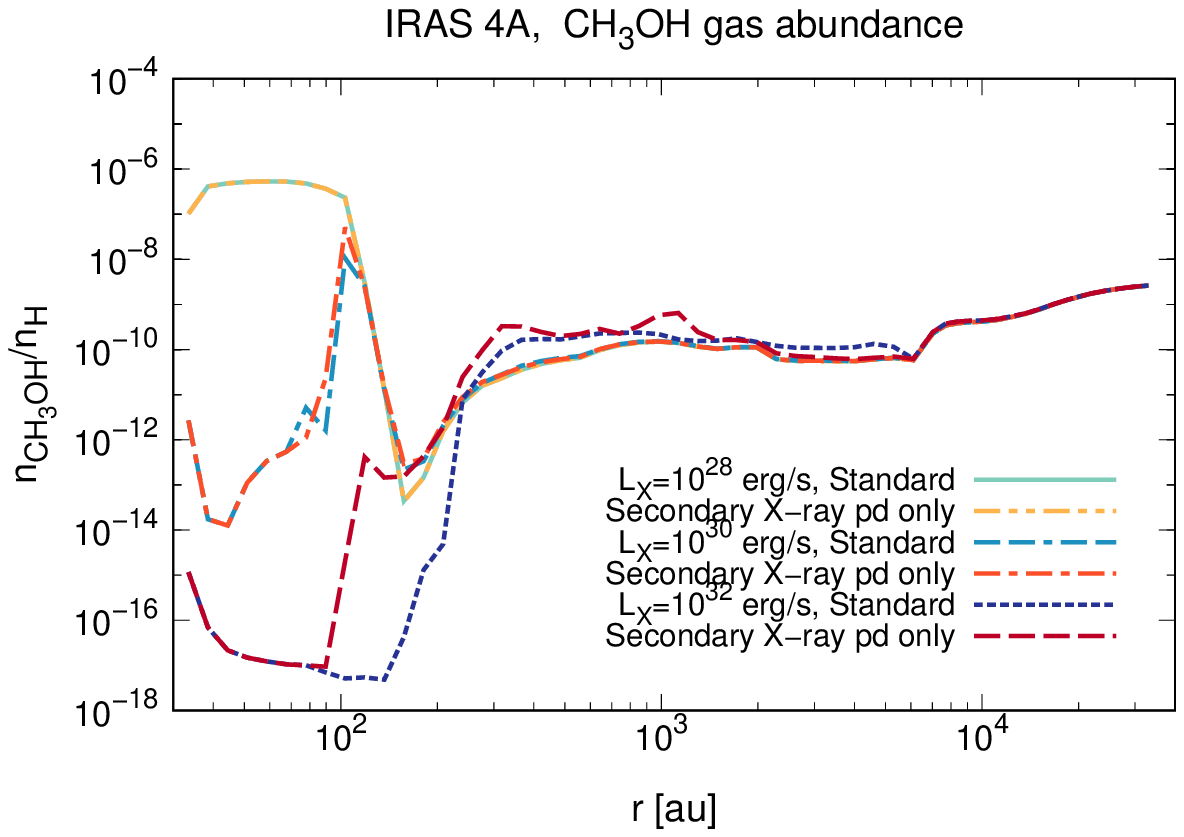}
\includegraphics[scale=0.67]{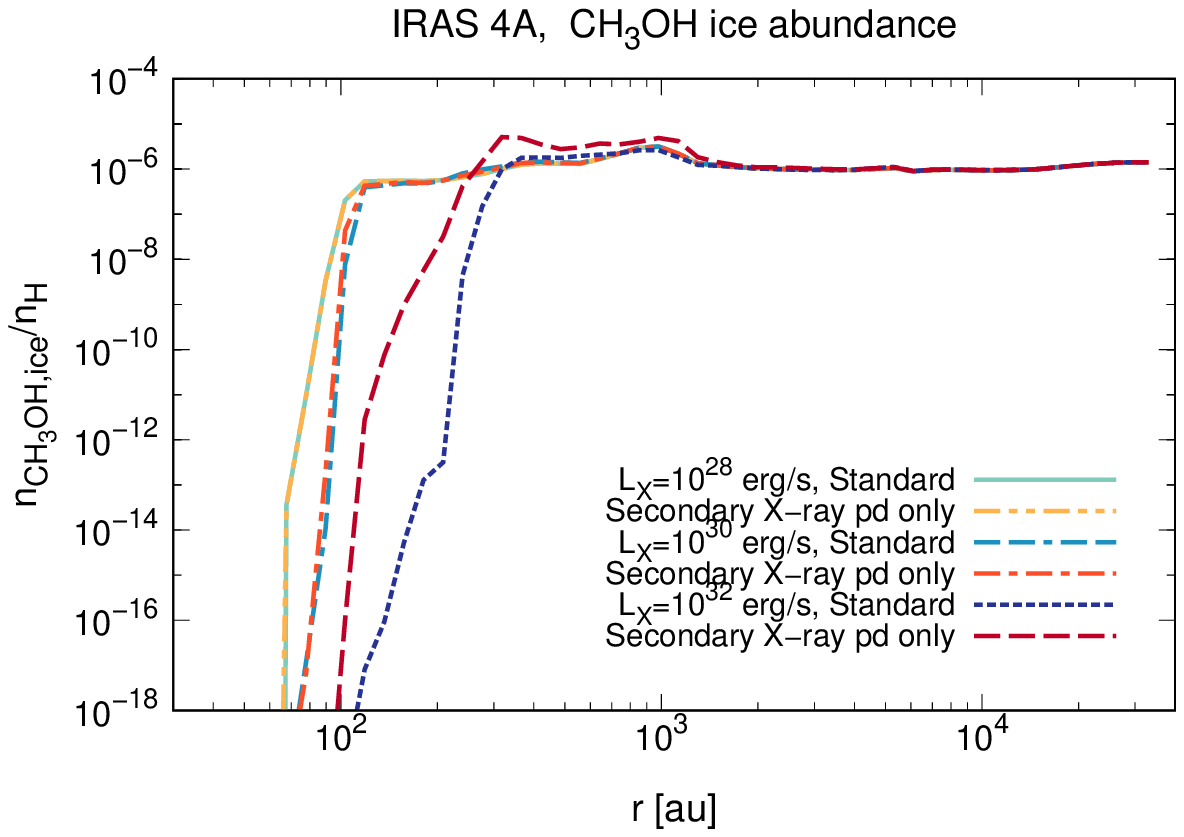}
\end{center}
\caption{
\noindent 
The radial profiles of gaseous fractional abundances 
of
O, HCO$^{+}$, and CH$_{3}$OH,
and icy fractional abundances of 
CH$_{3}$OH
in the NGC 1333-IRAS 4A envelope models.
The profiles are same as Figure \ref{FigureG2_add_Xp-rev1_model-comparisons}, except in the additional model calculations 
we switch off the direct X-ray induced photodesorption and include the secondary (indirect) X-ray induced photodesorption only (see also Figure \ref{Figure11_Xp-rev2_model-comparisons}). 
}\label{FigureG3_add_Xp-rev2_model-comparisons}
\end{figure*} 
In Figure \ref{FigureG1_add_Xp-rev1_model-comparisons} and Figure \ref{Figure11_Xp-rev2_model-comparisons} in Section 4.3, we show the radial profiles of gaseous fractional abundances 
of H$_{2}$O, O$_{2}$, and OH,
and icy fractional abundances of H$_{2}$O
in the IRAS 4A envelope models, for our standard model calculations and additional model calculations.
In Figures \ref{FigureG2_add_Xp-rev1_model-comparisons} and \ref{FigureG3_add_Xp-rev2_model-comparisons}, we show the radial profiles of gaseous fractional abundances 
of O, HCO$^{+}$, and CH$_{3}$OH, and icy fractional abundances of CH$_{3}$OH
in the IRAS 4A envelope models, for our standard model calculations and additional model calculations.
For the additional model calculations in Figures \ref{FigureG1_add_Xp-rev1_model-comparisons} and \ref{FigureG2_add_Xp-rev1_model-comparisons}, we include the photodesorption by UV photons generated internally via the interaction of secondary electrons produced by X-rays with H$_{2}$ molecules.
For the additional model calculations in Figure \ref{FigureG3_add_Xp-rev2_model-comparisons} and Figure \ref{Figure11_Xp-rev2_model-comparisons} in Section 4.3, we switch off the direct X-ray induced photodesorption and include the secondary (indirect) X-ray induced photodesorption only. 
\\ \\
According to Figures \ref{FigureG1_add_Xp-rev1_model-comparisons} and \ref{FigureG2_add_Xp-rev1_model-comparisons}, the effects of such additional secondary (indirect) X-ray induced photodesorption is marginal (the abundances are changed by <1\%, see also Section 4.3).
For molecules shown in Figure \ref{FigureG3_add_Xp-rev2_model-comparisons}, the differences in abundances between the standard model and the second additional model are much smaller than those in H$_{2}$O, OH, and O$_{2}$ (see Figure \ref{Figure11_Xp-rev2_model-comparisons} in Section 4.3).
\section{The impacts of the different photodissociation branching ratio and binding energies on the molecular abundances}
\begin{figure*}
\begin{center}
%
\includegraphics[scale=0.67]{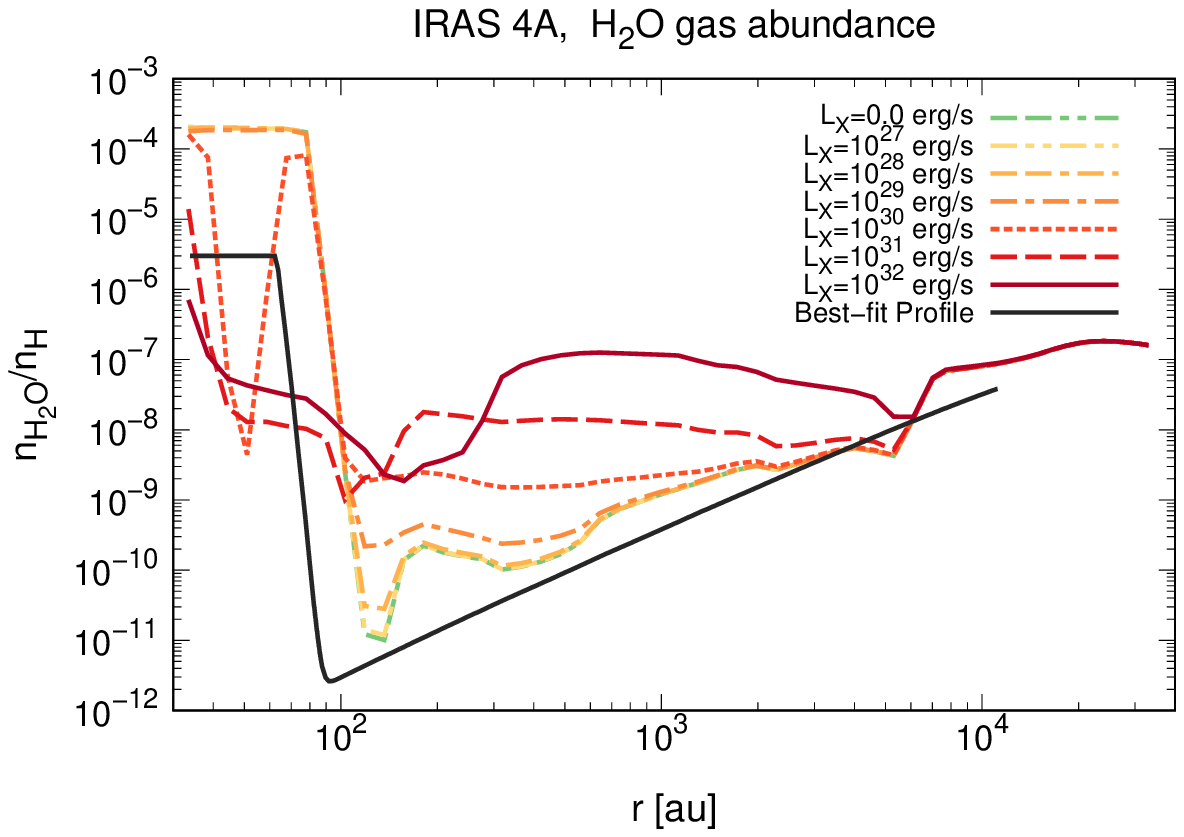}
\includegraphics[scale=0.67]{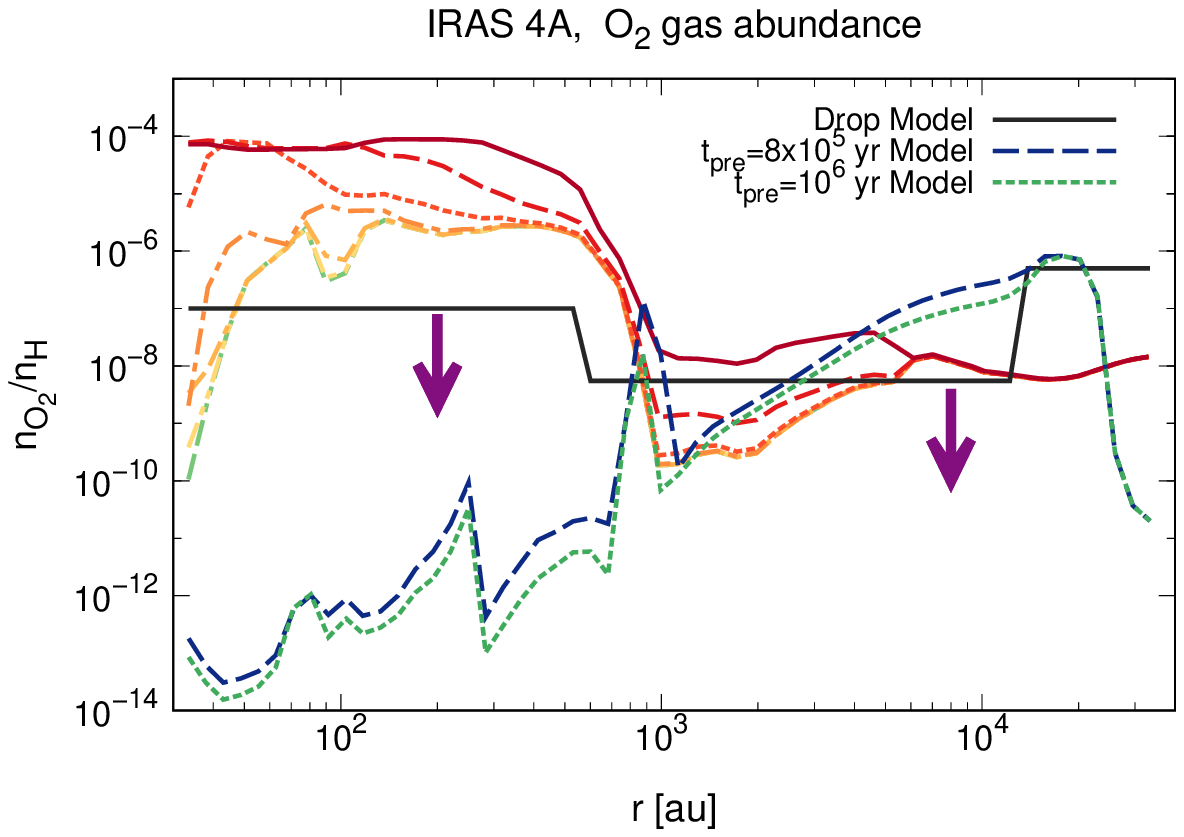}
\includegraphics[scale=0.67]{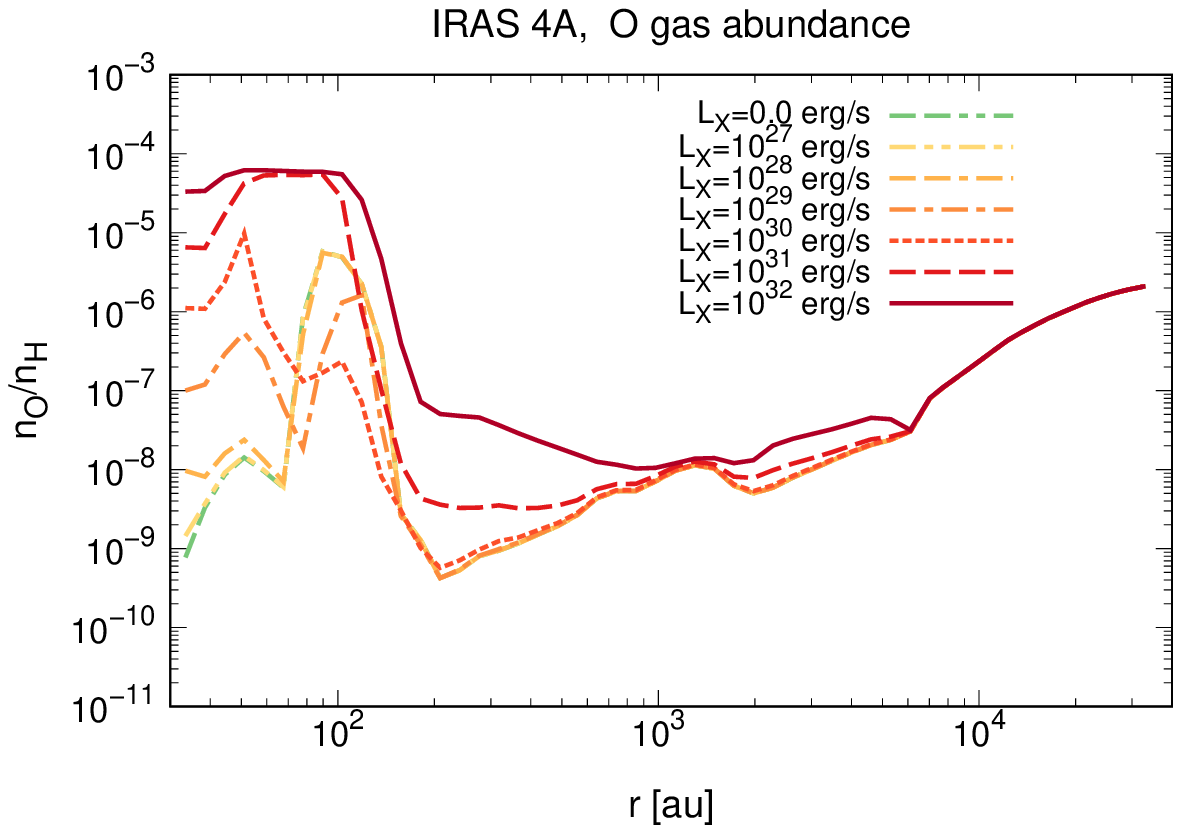}
\includegraphics[scale=0.67]{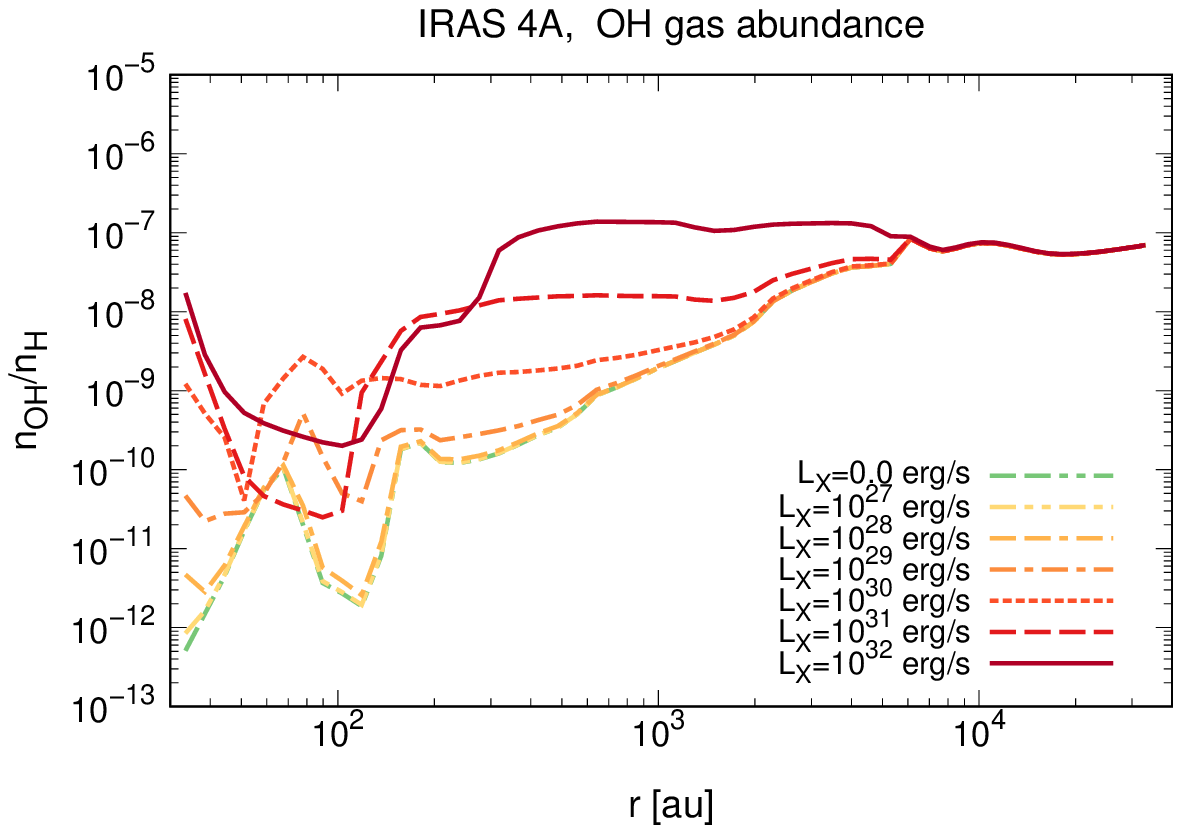}
\end{center}
\caption{
\noindent 
The radial profiles of gaseous fractional abundances 
of H$_{2}$O, O$_{2}$, O, and OH
in NGC 1333-IRAS 4A envelope models.
These profiles are obtained from our test calculations assuming that the product of H$_{2}$O photodissociation is 100\% atomic oxygen, unlike our fiducial model (100\% OH).
In the top left panel, the observational best-fit H$_{2}$O gas abundance profile obtained in \citet{vanDishoeck2021} is over-plotted with the black solid line (see also Figure \ref{Figure10_H2O&O2gas_obs-plot}).
In the top right panel, the three model O$_{2}$ gas abundance profiles obtained in \citet{Yildiz2013} are over-plotted (see also Figure \ref{Figure10_H2O&O2gas_obs-plot}).
}\label{FigureH1_H2O&O2&O&OHgas_reaction-rev}
\end{figure*} 
\begin{figure*}
\begin{center}
\includegraphics[scale=0.67]{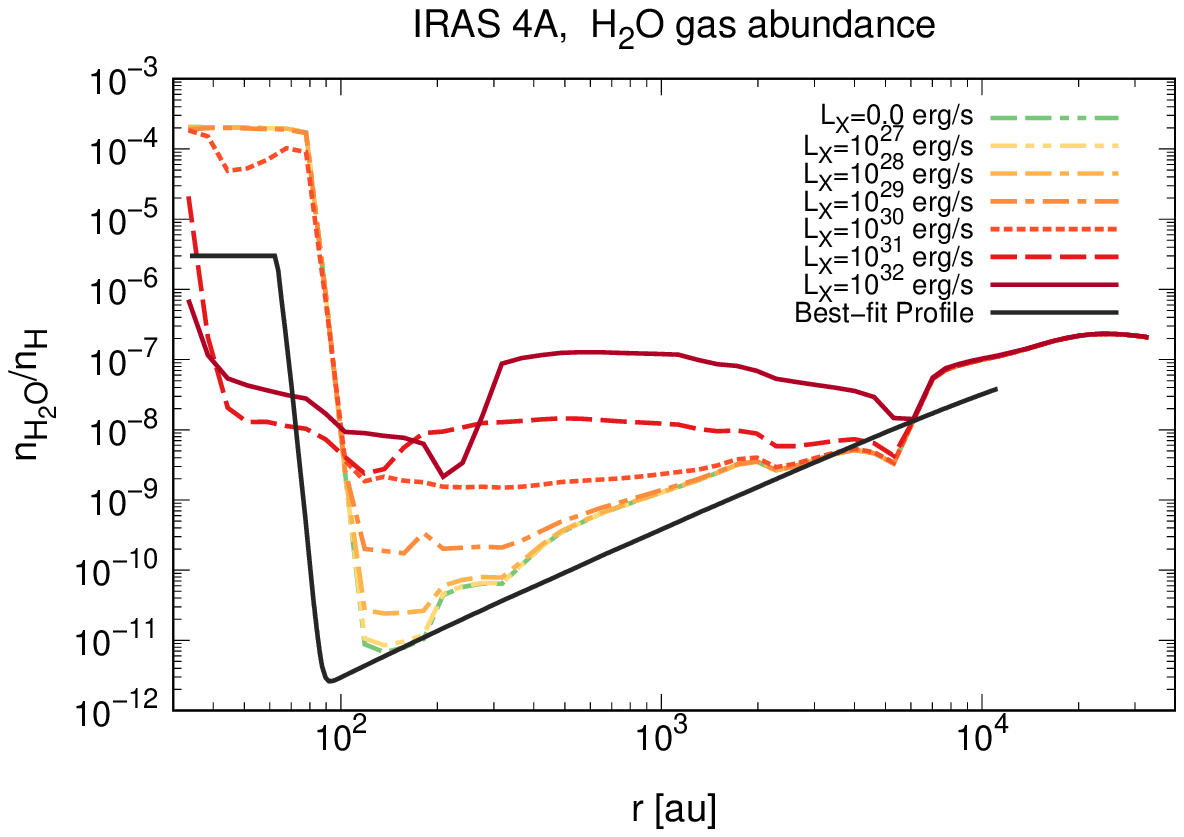}
\includegraphics[scale=0.67]{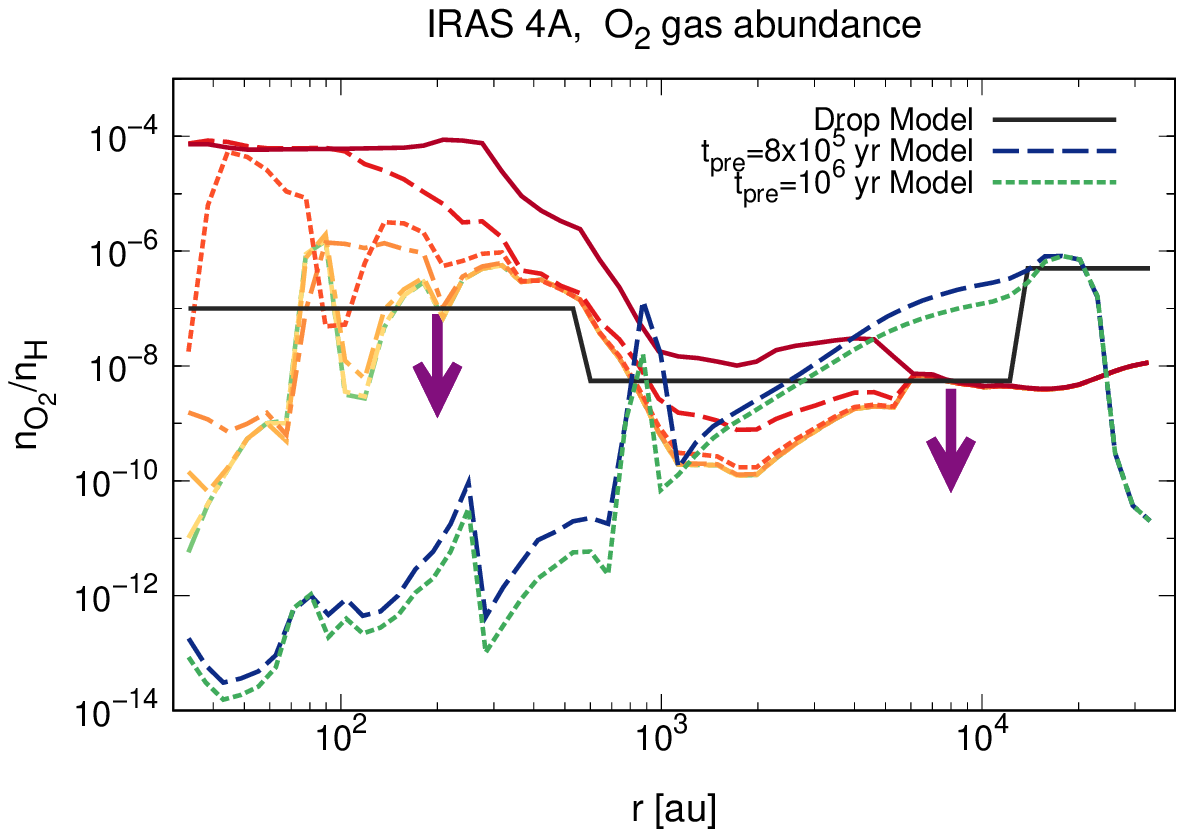}
\includegraphics[scale=0.67]{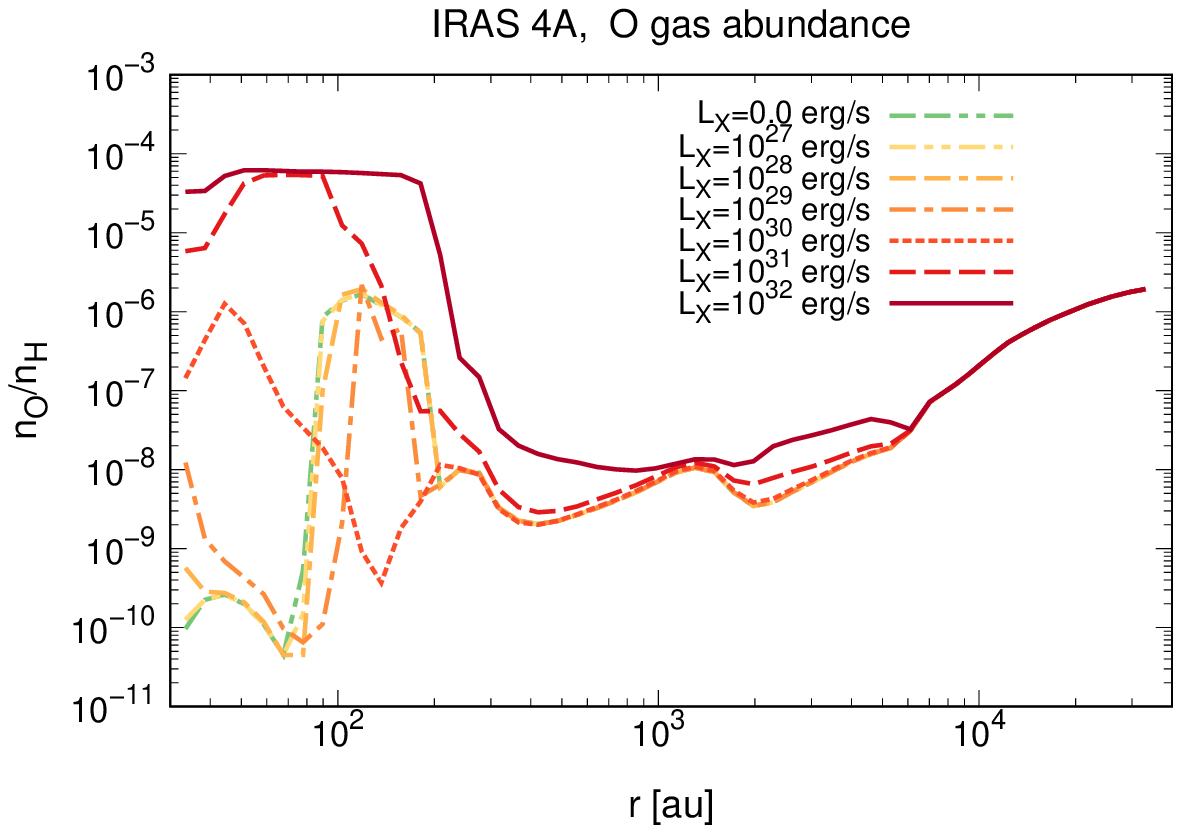}
\includegraphics[scale=0.67]{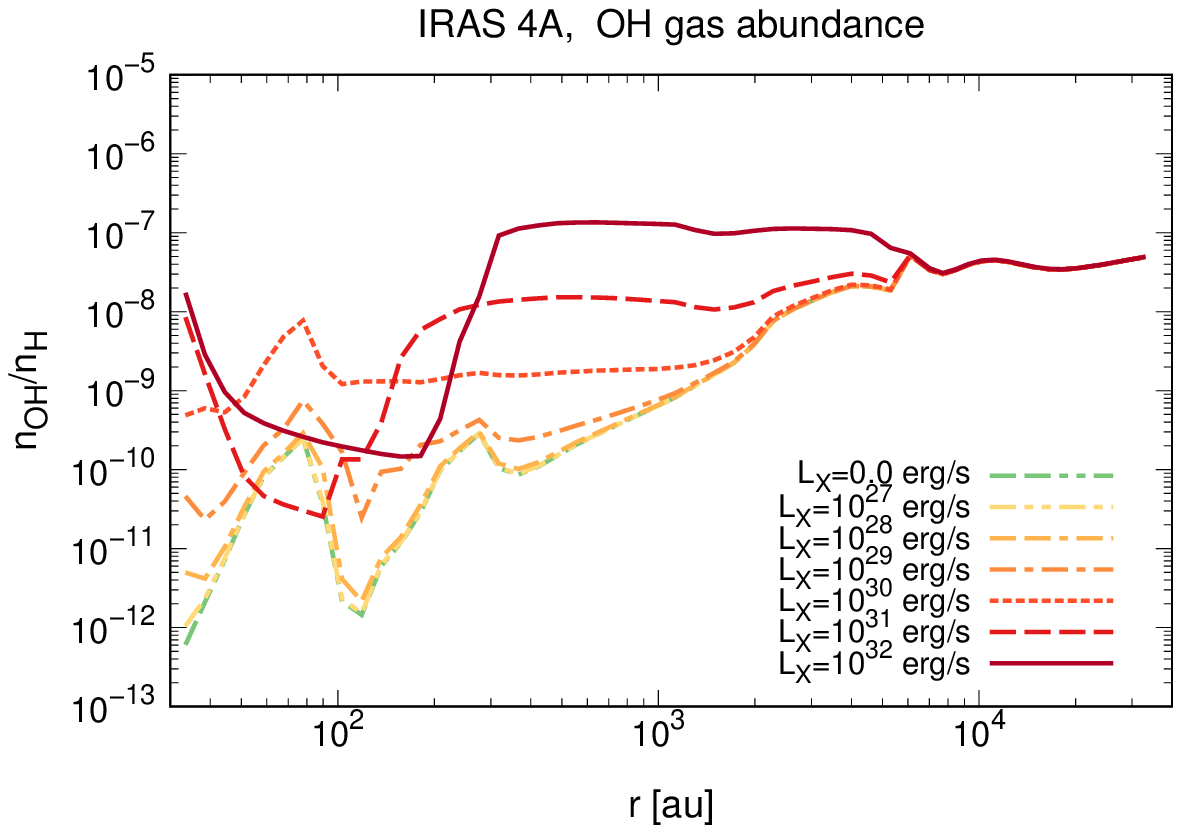}
\end{center}
\caption{
\noindent 
The radial profiles of gaseous fractional abundances 
of H$_{2}$O, O$_{2}$, O, and OH
in NGC 1333-IRAS 4A envelope models.
These profiles are obtained from our test calculations assuming the smaller $E_{\mathrm{des}}(j)$ for atomic oxygen ($=800$ K) than that in our fiducial model (=1660 K). 
In the top left panel, the observational best-fit H$_{2}$O gas abundance profile obtained in \citet{vanDishoeck2021} is over-plotted with the black solid line (see also Figures \ref{Figure10_H2O&O2gas_obs-plot} and \ref{FigureH1_H2O&O2&O&OHgas_reaction-rev}).
In the top right panel, the three model O$_{2}$ gas abundance profiles obtained in \citet{Yildiz2013} are over-plotted (see also Figure \ref{Figure10_H2O&O2gas_obs-plot} and \ref{FigureH1_H2O&O2&O&OHgas_reaction-rev}).
}\label{FigureH2_H2O&O2&O&OHgas_Edes-O-rev}
\end{figure*} 
Figure \ref{FigureH1_H2O&O2&O&OHgas_reaction-rev} shows the gas-phase abundance profiles of H$_{2}$O, O$_{2}$, O, and OH, which are calculated assuming the extreme case that the product of H$_{2}$O photodissociation is 100\% atomic oxygen (Reaction \ref{reaction15} only both in the gas and ice).
%
Figure \ref{FigureH2_H2O&O2&O&OHgas_Edes-O-rev} shows the gas-phase abundance profiles of H$_{2}$O, O$_{2}$, O, and OH, which are obtained from our test calculations assuming the smaller $E_{\mathrm{des}}$(O) ($=800$ K) than that in our fiducial model ($=1660$ K).
\\ \\
We plot these figures in order to investigate the impacts of the different photodissociation branching ratio of H$_{2}$O and different binding energies of O on the abundances of H$_{2}$O and related molecules. 
In Section 4.4, we discuss these impacts in detail (see also Figure \ref{Figure12_H2O&O2&O&OHgas_model-comparisons}).
\end{appendix}

%
%

%
\end{document}